\documentclass[11pt,a4paper]{article}          
\usepackage{jheppub}

\usepackage{enumitem}
\usepackage{mathtools}
\usepackage{mathrsfs}
\usepackage{physics}
\usepackage{braket}
\usepackage[noabbrev]{cleveref}

\DeclareMathOperator{\sgn}{sgn}

\title{Approximate higher-form symmetries and dualities of massive p-forms in the holographic bulk}

\author{Andr\'{e} Oliveira Pinheiro} 

\affiliation{Department of Mathematics and Maxwell Institute for Mathematical Sciences, Heriot-Watt University, Edinburgh EH14 4AS, U.K.}

\emailAdd{ao2012@hw.ac.uk}

\abstract{We develop a holographic framework for continuous higher-form symmetries and their low-energy effective descriptions, based on bulk path integrals, holographic renormalisation and boundary-condition-changing deformations. We show how approximate higher-form symmetries associated with a defect current can be realised holographically through massive antisymmetric tensor fields, either via parametrically small bulk masses or strong deformations associated with Robin boundary conditions. We further study the consequences of massless and massive Hodge dualities in the bulk, deriving the corresponding dualities between boundary theories related by different quantisation choices. Our results provide a unified perspective on approximate symmetries, dualities and strong/weak-coupling relations in holographic theories based on massive $p$-forms. In the self-dual case of an exact symmetry of degree $(d - 3) / 2$, we derive generalised constraints on holographic current-current correlators in the presence of double-trace deformations.}


\arxivnumber{2606.01486}

\begin{document} 	
\maketitle
\flushbottom

\section{Introduction}

\paragraph{}Continuous global symmetries, or simply continuous symmetries, are one of the key elements in describing physical systems. They are tied to the existence of local conservation equations. Usually in QFT one considers local charged operators that are responsible for the creation of pointlike charges. The symmetry is then what gives the notion of a real, physical object to these charges, as it says that (in the absence of charged operators) the total charge is conserved in time. Now one could ask: what about if our system contains extended objects, i.e. whose locus is a higher-dimensional portion of spacetime? A generalisation of ordinary symmetries to include this case is provided by \textit{higher-form symmetries} \cite{Gaiotto:2014kfa}. A continuous $p$-form symmetry arises when the $(p{+}1)$-dimensional worldvolumes along which $p$-dimensional objects move are smooth submanifolds of spacetime. Ordinary and higher-form symmetries correspond to $p = 0$ and $p \geq 1$, respectively.

\paragraph{}Higher-form symmetries (and, more broadly, \textit{generalised symmetries} \cite{Cordova:2022ruw}) have proven useful across areas of physics by unifying different phenomena as manifestations of symmetries, effectively by recognising known gauge theories and lattice models as phases of matter under a generalised Landau paradigm \cite{Chen:2025uno}. Developments regarding these symmetries have been particularly relevant to condensed matter theory \cite{McGreevy:2022oyu}, where they were applied to topological order, Landau Fermi liquids, spin liquids, among other systems.
The most famous example of a higher-form symmetry is Maxwell theory in $d=4$ \cite{Gaiotto:2014kfa}, where a 1-form symmetry is associated with the conservation of magnetic field lines. Electrically charged matter explicitly breaks a second 1-form symmetry associated with conserved electric field lines. In fact, the photon corresponds to the Goldstone of the electric 1-form symmetry, such that our vacuum lives in a (spontaneously) \textit{broken phase}. 

\paragraph{}Higher-form symmetries in real condensed matter systems are either emergent or due to dynamical electromagnetic fields \cite{Wen:2018zux}. Independently of their origin, the low-energy dynamics of such systems is captured by effective field theories as in the 0-form case --- see e.g. the discussion of \cite{Delacretaz:2019brr} on superfluid phases. For continuous symmetries this led in particular to developments in the hydrodynamic description of unconventional fluids \cite{Grozdanov:2016tdf,Grozdanov:2018ewh,Armas:2018ibg}. Examples of this are magnetohydrodynamics \cite{Grozdanov:2016tdf,Armas:2018atq,Glorioso:2018kcp,Armas:2018zbe} and viscoelastic hydrodynamics \cite{Grozdanov:2018ewh,Armas:2019sbe}, which govern long-wavelength fluctuations in polarised plasmas and crystals, respectively. 

\paragraph{}In this paper, we study gravitational systems whose holographic dual possesses continuous higher-form symmetries.\footnote{`Higher-form' is often used as an umbrella term for all $p$-form symmetries.} According to gauge-gravity duality (holography) \cite{Maldacena:1997re,Witten:1998qj,Gubser:1998bc}, the dynamics of certain strongly coupled QFTs are equivalent to those of certain theories of gravity with negative cosmological constant. While the latter lives in the bulk of an asymptotically AdS spacetime, the strongly coupled theory is said to live in its conformal boundary. Interestingly enough, the aforementioned `higher-form hydrodynamics' arose through magnetohydrodynamics alongside a holographic analysis of its gravitational dual \cite{Grozdanov:2017kyl,Hofman:2017vwr}. The motivation in this case was particularly strong, as magnetohydrodynamics is relevant to the physics of the quark-gluon plasma, one of the paradigmatic examples of strongly coupled matter.

\paragraph{}More specifically, we pursue a systematic treatment via holography of systems with a single (non-anomalous) symmetry and then extend it by explicitly breaking the symmetry in a controlled way. Additionally, we consider concrete bulk models for both cases, that are given by Einstein-Maxwell/Proca theories for which the Maxwell/Proca field is a $p$-form. The gravitational actions also include surface terms dual to double-trace deformations of the boundary theory, as these have proved important when extending the fluid-gravity correspondence \cite{Bhattacharyya:2007vjd,Rangamani:2009xk,Hubeny:2011hd} to higher-form charges \cite{Davison:2025sze}. Surface terms of a different type, responsible for a Legendre transformation that exchanges canonical bulk variables, are also allowed. \\ \hspace*{1.75ex}
Starting from bulk antisymmetric tensor fields with arbitrary rank and mass, the aforementioned surface terms lead to an even larger family of holographic duals with higher-form symmetries. Its elements are found to be connected by duality relations that follow from classical electromagnetic-like Hodge dualities in the bulk. In fact, only the mass is duality-invariant. \\ \hspace*{1.75ex}
This work contributes to an improved understanding of approximate higher-form symmetries in holography and of how they are achieved across the different settings within the family of theories under consideration. The second major contribution concerns dualities that interchange quantisation schemes, with a focus on deriving their holographic properties from the bulk. These notably include, in the case of models based on Proca massive $p$-forms, both a Hodge-type duality and a strong/weak-coupling duality.

\paragraph{}This paper is organised as follows. In \cref{boundarytheories}, we introduce the classes of theories that motivate \cref{HologPathInt,WeaklyBroken_sec}. These are broadly defined by the manner in which they realise a continuous higher-form symmetry. In \cref{HologPathInt}, we first discuss how the phenomenology associated with an exact symmetry is implemented holographically, and then focus on massless $p$-forms in the bulk. Having settled the more familiar case, we turn to the main contributions of this work in \cref{WeaklyBroken_sec,sec:duality}. While the former mirrors \cref{HologPathInt} in structure for a weakly broken symmetry, the latter explores how dualities between higher-form bulk fields manifest themselves holographically. In particular, we discuss at the end of \cref{sec:duality} how self-duality (in the massless case) constrains the dynamics of the dual theory. Finally, we conclude in \cref{conclusion}.

\paragraph{Conventions.} $\Omega^p (M)$ denotes the space of smooth differential $p$-forms on a manifold $M$.\footnote{When smoothness is particularly important, we might write $C^\infty \Omega^p (M)$ to emphasise it.} When the manifold is not stated, it is implied that $\Omega^p$ refers to $p$-forms on the $d$-dimensional physical spacetime with metric $g$ (eventually taken to be the Minkowski metric $\eta$). Coordinate indices on this spacetime are denoted by lowercase Greek letters $\mu, \nu, \ldots$. Lowercase Latin letters $a, b, \ldots$ from the beginning of the alphabet are used for indices in the $(d{+}1)$-dimensional bulk spacetime $\mathbb{B}$ with metric $G$, whose boundary is identified with the physical spacetime.\footnote{$\Omega^p$ will eventually refer to $p$-forms in the boundary of the regularised holographic bulk $\mathbb{B}_\varLambda$ (which coincides with $\mathbb{B}$ when $\varLambda \to \infty$).} Greek and Latin indices are raised with the inverse metrics $g^{\mu \nu}$ and $G^{ab}$, respectively. Lastly, antisymmetrisation of indices is denoted with square brackets and it is not normalised, e.g. $X_{[a b]} = X_{a b} - X_{b a}$.

\subsection{Motivation: theories of $p$-form operators} \label{boundarytheories}

\paragraph{} We are interested in theories with antisymmetric fields of arbitrary rank as local observables. Consider first a system with a continuous $(p{-}1)$-form symmetry. The field-theoretical description of this system has a $p$-form Noether current $\mathcal{O}_j$, with conservation equation
\begin{equation} \begin{aligned} \label{Oj_conserv}
	\mathrm{d} \ast \mathcal{O}_j = 0 \qquad \left( \mathcal{O}_j \in \Omega^p \right) .
\end{aligned} \end{equation}
In a quantum setting, the current is promoted to an operator and \eqref{Oj_conserv} to an operator equation or, less drastically, to the conservation of the expectation value $\braket{\mathcal{O}_j}$.
The generating functional $W_j[\psi]$ of $\mathcal{O}_j$'s correlators is a functional of the background gauge field $\psi \in \Omega^p$ that couples to $\mathcal{O}_j$. Hence,
\begin{equation} \begin{aligned} \label{deltaWj}
	\delta W_j [\psi] \sim \int \ast \braket{\mathcal{O}_j} \wedge \delta \psi \, .
\end{aligned} \end{equation}
Note that $\psi$ being a gauge field is consistent with $\mathcal{O}_j$ being a conserved current, since this makes $W_j[\psi]$ invariant (up to surface terms) under gauge transformations that shift $\psi$ by an exact form:
\begin{equation} \begin{aligned} \label{gaugeshift}
	\delta \psi = \mathrm{d} \zeta \, , \qquad \zeta \in C^\infty \Omega^{p-1} .
\end{aligned} \end{equation}
\paragraph{}Now, consider the closely related situation in which the background field $\psi$ is a conserved current. Because of this, $\mathcal{O}^a$ --- the dynamical field that couples to $\psi$ --- is defined up to shifts by arbitrary exact forms. Hence, the physical objects are equivalence classes 
\begin{equation} \begin{aligned} \label{equivclasses}
	[\mathcal{O}^a] = \{ \mathcal{O}^a + \mathrm{d} \zeta \; | \; \zeta \in C^\infty \Omega^{p-1} \} \, .
\end{aligned} \end{equation}
This is useful, for example, when a continuous $p$-form symmetry is spontaneously broken. In this case, the phase of the relevant order parameter (required to characterise equilibrium states and describe near-equilibrium fluctuations) is the integral of a field with the properties of $\mathcal{O}^a$ \cite{Hofman:2018lfz,Lake:2018dqm}. Hence, $\mathcal{O}^a$ is an effective degree of freedom in the low-energy description of `higher-form broken phases' and the generating functional, which we denote $W_a [\psi]$, captures the dynamics of Goldstone modes.

\paragraph{}Let us now address the scenario where the continuous $(p{-}1)$-form symmetry associated with \eqref{Oj_conserv} is no longer exact, but is instead explicitly broken due to a conserved current $\tilde{\mathcal{O}}_j \in \Omega^{p-1}$ \cite{Armas:2023tyx}:
\begin{equation} \begin{aligned} \label{Oj_nonconserv}
	& \mathrm{d} \ast \mathcal{O}_j = \ast \tilde{\mathcal{O}}_j \, .
\end{aligned} \end{equation}
The conservation of $\tilde{\mathcal{O}}_j$ follows from the exterior derivative of the equation above, showing that breaking requires a ${(p{-}2)}$-form continuous symmetry.
The generating functional in this case is $\mathcal{W}_j [\psi , \tilde{\psi}]$ such that
\begin{equation} \begin{aligned}
	\delta \mathcal{W}_j [\psi , \tilde{\psi}] & \sim \int \left( \ast \braket{\mathcal{O}_j} \wedge \delta \psi + \ast \braket{\tilde{\mathcal{O}}_j} \wedge \delta \tilde{\psi} \right) , 
\end{aligned} \end{equation}
where $\tilde{\psi} \in \Omega^{p-1}$ is a background gauge field. In accordance with the (non-)conservation \cref{Oj_nonconserv}, the gauge transformation \eqref{gaugeshift} should be enhanced to 
\begin{equation} \begin{aligned}
	( \delta \psi , \delta \tilde{\psi} ) = \left( \mathrm{d} \zeta , - \zeta \right) 
\end{aligned} \end{equation}
in order to leave the generating functional invariant. 
The fact that $\tilde{\psi}$ transforms via a completely arbitrary shift is compatible with $\tilde{\mathcal{O}}_j$ not being independent from $\mathcal{O}_j$ (it is indeed the divergence of the latter).
Note that by setting $\zeta = \mathrm{d} \tilde{\zeta}$, where $\tilde{\zeta} \in C^\infty \Omega^{p-2}$, we have the usual gauge transformation that can be expected from the conservation of $\tilde{\mathcal{O}}_j$.

\paragraph{}Similar to what we did before, we would like to consider a generating functional $\mathcal{W}_a [\psi , \tilde{\psi}]$ where the background fields obey 
\begin{equation} \begin{aligned} \label{background_nonconserv}
	\mathrm{d} \ast \psi = \ast \tilde{\psi} \, .
\end{aligned} \end{equation}
Before we do that, it is worth mentioning that in the previous case where $\psi$ is conserved, the equivalence classes \eqref{equivclasses} are conveniently captured by the field strength $\mathsf{f} \equiv \mathrm{d} \mathcal{O}^a$, which is the fundamental local, gauge-invariant object that we can build out of $\mathcal{O}^a$. Due to \cref{background_nonconserv}, it is then necessary to extend \eqref{equivclasses} according to
\begin{equation} \begin{aligned} \label{equivclasses_broken}
	[(\mathcal{O}^a , \tilde{\mathcal{O}}^a)] = \{ (\mathcal{O}^a + \mathrm{d} \zeta \, , \, \tilde{\mathcal{O}}^a - \zeta) \; | \; \zeta \in C^\infty \Omega^{p-1} \} \, .
\end{aligned} \end{equation}
Because $\tilde{\psi}$ is conserved, one could think that $\tilde{\mathsf{f}} = \mathrm{d} \tilde{\mathcal{O}}^a$ is the field strength of a `Goldstone field'. However, this is not $\zeta$-invariant. Instead, the fundamental local, gauge-invariant observable in this theory is
\begin{equation} \begin{aligned}
	\tilde{\mathsf{f}} \vcentcolon = \mathrm{d} \tilde{\mathcal{O}}^a + \mathcal{O}^a \, .
\end{aligned} \end{equation}
Note that $\tilde{\mathsf{f}}$ is in general neither exact nor closed. If we want to see $\tilde{\mathsf{f}}$ as the strength of a Goldstone field, then this field has a \textit{multivalued} part that contributes towards $\mathcal{O}^a$.\footnote{\textit{Multivalued Goldstones} differ from {pseudo-Goldstone fields}, as the latter arise upon weakly (explicitly) breaking a spontaneously broken symmetry.} Although the Goldstone field becomes singular, the associated physical observable, $\tilde{\mathsf{f}}$, is still smooth.

\paragraph{}A convenient way to interpret this situation is to introduce the notion of a \textit{magnetic current}. This is a current that is topologically conserved (due to the nilpotency of the exterior derivative), such as $\ast \mathsf{f}$ in the theory $W_a [\psi]$ which therefore possesses a $(d{-}p{-}2)$-form magnetic symmetry. On the other hand, $\mathcal{W}_a [\psi , \tilde{\psi}]$ possesses an explicitly broken $(d{-}p{-}1)$-form magnetic symmetry reflected in the non-conservation of the magnetic current $\ast \tilde{\mathsf{f}}$:
\begin{equation} \begin{aligned} \label{magnetic_nonconserv}
	\mathrm{d} \ast (\ast \tilde{\mathsf{f}}) = \ast \left( (-1)^{d-1} \ast \mathsf{f} \right) \, .
\end{aligned} \end{equation}
Non-magnetic currents are called \textit{electric}, such as $\mathcal{O}_j$ in the theory $W_j[\psi]$. Given the current explicitly broken magnetic symmetry and similarly to its electric counterpart, there is a one-degree-lower symmetry associated with the conservation of $\ast \mathsf{f}$.

\paragraph{}Above, we saw that when a continuous higher-form symmetry is dynamically broken, there is a second symmetry which is exact. Despite not being mentioned above, the second symmetry is missing when the degree of the former is null. This is important to understand the physical implications of the breaking. When a $p$-form symmetry is broken, $p$-dimensional objects can then be created or destroyed. Additionally, if $p>0$ the breaking also allows for the objects themselves to lose their smoothness as submanifolds of spacetime. The $(p{-}1)$-dimensional hypersurfaces where this happens correspond to \textit{defects} and the second symmetry ensures defect conservation.\footnote{The use of the term `defect' differs slightly from \cite{Pinheiro:2025fqg}. (There, it referred to the $p$-dimensional worldvolumes of the defects considered here).} We call $\tilde{\mathcal{O}}_j $ the \textit{defect current} \cite{Armas:2023tyx}. In theories like $\mathcal{W}_a [\psi , \tilde{\psi}]$, the defect current is $\ast \mathsf{f}$ --- cf. \eqref{magnetic_nonconserv}.

\paragraph{}Lastly, we would like to make a brief remark regarding a generic theory with a (non-conserved, gauge-invariant) $p$-form operator, for which the generating functional is
\begin{equation} \begin{aligned} \label{generic_delW}
	\delta W [\psi] \sim \int \ast \braket{\mathcal{O}} \wedge \delta \psi \, .
\end{aligned} \end{equation}
If we consider the divergence of $\mathcal{O}$ as an operator $\ast \tilde{\mathcal{O}} \equiv \mathrm{d} \ast \mathcal{O}$, then in order to introduce a background source that couples to $\tilde{\mathcal{O}}$, we substitute $\psi \to \psi - \mathrm{d} \tilde{\psi}$ such that
\begin{equation} \begin{aligned} 
	\delta W [\psi] \to \delta \mathcal{W}_j [\psi , \tilde{\psi}] \, .
\end{aligned} \end{equation}
On the other hand, we might wish to consider the operator $\tilde{\mathcal{O}}$ that couples to the divergence of the background source, i.e. $\ast \tilde{\psi} \equiv \mathrm{d} \ast \psi$, --- this operator contributes towards the exact part of $\mathcal{O}$. Hence, we let $\mathcal{O} \to \mathcal{O} - \mathrm{d} \tilde{\mathcal{O}}$ such that
\begin{equation} \begin{aligned} 
	\delta W [\psi] \to \delta \mathcal{W}_a [\psi , \tilde{\psi}] \, .
\end{aligned} \end{equation}
The following question then arises: (focusing on the electric case for concreteness) when is it worth introducing (and keeping track of) a defect current? The answer is: when the defect is itself dynamical, e.g. when its mean-field configuration obeys its own PDE. Here, the distinction between a broken symmetry or the simple absence of a conserved charge becomes relevant. Because this work focuses not on dynamics but on the kinematics of holographic models through the analysis of the asymptotics of bulk fields, our position is to include defect currents when we have control over the scale of symmetry breaking, in particular, when we can trace back to the symmetry that was broken and access its weakly broken regime. 

\section{Exact symmetries} \label{HologPathInt}

\paragraph{}In this section, we show how the phenomenology described in \cref{boundarytheories} of systems with exact higher-form symmetries can be reproduced holographically. In particular, we demonstrate that electric and magnetic symmetries arise from different quantisations of a massless $p$-form field in the bulk. Our analysis is carried out within the framework of \textit{bottom-up} holography \cite{Hartnoll:2009sz,Herzog:2009xv,McGreevy:2009xe}.
In contrast to \textit{top-down} constructions \cite{Denef:2009tp,Gubser:2009qm,Gauntlett:2009dn,Gauntlett:2009bh}, our setup lacks a fully fledged boundary theory and the bulk dual offers the only window into it. Nevertheless, guided by AdS-CFT, we assume that the boundary theory admits a fundamental description in terms of matrix-valued fields transforming in some representation of a gauge group of rank $N$. Taking the trace of a local functional of these fields yields a \textit{single-trace} operator \cite{Witten:2001ua}. \\ \hspace*{1.75ex}
In the following, the dynamical fields from \cref{boundarytheories} are taken to be single-trace operators $\mathcal{O}$. Recall that we used $\mathcal{O}_j$ and $\mathcal{O}^a$ to distinguish between $p$-form conserved currents and Goldstone fields associated with electric $(p{-}1)$-form symmetries and magnetic $(d{-}p{-}2)$-form symmetries, respectively.
Quadratic functionals of $\mathcal{O}$ define \textit{double-trace} operators.

\paragraph{}Previously, we introduced generating functionals $W_j$ and $W_a$ for correlators of $\mathcal{O}_j$ and $\mathcal{O}^a$. We now express these as expectation values of the form 
\begin{equation} \begin{aligned} \label{genfuncW0}
	W [\psi] \equiv \expval{1} [\psi] \vcentcolon = \expval{ \exp( i \mathcal{C} (N) \int_{\partial \mathbb{B}} {\ast \mathcal{O} \wedge (\psi - \hat{\psi}) \over (d-p)!})}_{\hat{\psi}}
\end{aligned} \end{equation}
in the respective theory, where the source is constrained by $\mathrm{d} \ast \psi = 0$ in the magnetic case. Note that, according to our notation, an expectation value of an operator in a theory $W [\psi]$ is equivalent to the expectation value of that operator weighted by the exponential above in the theory $W [\hat{\psi}]$ where sources are fixed at $\hat{\psi}$, arbitrary.\footnote{Below, we fix $\hat{\psi} = 0$ for simplicity.}
Lastly, $\partial \mathbb{B}$ is the boundary of the bulk manifold and we have introduced the prefactor $\mathcal{C}(N)$, which is identified with the gravitational coupling $\kappa^{-2}$ expressed in terms of CFT data.\footnote{To be precise, $\mathcal{C}(N)$ is $\kappa^{-2}$ in units of the AdS radius.} According to the standard gauge-gravity duality dictionary, $\mathcal{C}(N)$ is the central charge of the boundary CFT, up to a numerical factor. We assume that a large-$N$ limit exists and it corresponds to classical gravity (i.e. low $\kappa^2$).

\subsection{Overview of $p$-forms in AdS} \label{GKPWmotivation}

\paragraph{}As massless and massive $p$-forms in AdS \cite{lYi:1998trg} constitute the main ingredient for the holographic models in this work, we briefly review them before proceeding. Some statements below are presented without the necessary evidence, but they will be verified later in the paper. \\ \hspace*{1.75ex}
We consider a free field with mass $m^2$ whose components $\phi_{a_1 ... a_p}$ or $\phi_{b_1 ... b_p}$ we denote compactly as $\phi_\mathrm{A}$ or $\phi_\mathrm{B}$. The bulk spacetime is the Poincaré patch of AdS with line element
\begin{equation} \begin{aligned} \label{metric}
	\mathrm{d} s^2 = {L^2 \over r^2} \mathrm{d} r^2 + {r^2 \over L^2} \eta_{\mu \nu} \mathrm{d} x^\mu \mathrm{d} x^\nu ,
\end{aligned} \end{equation}
where $L$ is the AdS radius --- throughout we work in units with $L = 1$.
Although we are actually interested in a bulk $\mathbb{B}$ which is asymptotically AdS, considering pure AdS is sufficient to characterise the leading on-shell behaviour of $\phi_A$ near the conformal boundary at $r \to \infty$. In particular, the non-radial components $\phi_\Xi$ obey
\begin{equation} \begin{aligned} \label{phi_eom}
	r^{1 - d + 2 p} \partial_r \left( r^{d + 1 - 2 p} \partial_r \phi_\Xi \right) + \ldots & = m^2 \phi_\Xi \, .
\end{aligned} \end{equation}
Note that we adopt the notation $\mathrm{A} \to \Xi$ when there is no $r$ index among $\mathrm{A}$ and $\mathrm{A} \to \mathrm{R}$ otherwise. On-shell, $\phi_\Xi$ is the sum of two linearly independent solutions parametrised by a pair of functions, $\phi_\Xi^+$ and $\phi_\Xi^-$, of the transverse coordinates $x^\mu$. From \eqref{phi_eom}, we obtain the leading contribution to each of these:
\begin{equation} \begin{aligned} \label{3.23}
	\phi_\Xi = \left[ r^{- \Delta_-} \phi_\Xi^- (x) + \ldots \right] + \left[ r^{- \Delta_+} \phi_\Xi^+ (x) + \ldots \right] ,
\end{aligned} \end{equation}
where
\begin{equation} \begin{aligned} \label{Deltapm}
	\Delta_\pm = {d - 2 p \pm \sqrt{(d - 2 p)^2 + 4 m^2} \over 2} \, ,
\end{aligned} \end{equation}
for $(d - 2 p)^2 + 4 m^2 > 0$. Note that $\Delta_+ \Delta_- = - m^2$.

\paragraph{}Our bulk field is associated with a \textit{bare} action $S$ and a \textit{regularised} action $S_\varLambda$ obtained by integrating the bare lagrangian over the portion $\mathbb{B}_\varLambda$ of the Poincar\'e patch bounded by $r = \varLambda$. In the massless case, considering that we have fixed radial gauge ($\phi_R = 0$), the \textit{on-shell variations} of these actions are $\delta S = \lim_{\varLambda \to \infty} \delta S_\varLambda$ and
\begin{equation} \begin{aligned} \label{3.24}
	\delta S_\varLambda = \int_{\partial \mathbb{B}_\varLambda} \mathrm{d}^d x \left( r^{d + 1 - 2 p} \partial_r \phi_\Xi \right) \delta \phi^\Xi \, ,
\end{aligned} \end{equation}
where $\phi_\Xi$ and $\delta \phi_\Xi$ are solutions to the equations of motion at $\partial \mathbb{B}_\varLambda$, parametrised by $\phi_\Xi^\pm (x)$ and $\delta \phi_\Xi^\pm (x)$.\footnote{First, as a general assumption, bulk fields and their derivatives are compactly supported in $x^\mu$	(in practice, one would impose sufficiently fast fall-off conditions at large $|x^\mu|$). Second, the indices in $\delta \phi^\Xi \equiv \delta \phi^{\mu_1 \ldots \mu_p}$ are raised with $\eta^{\mu \nu}$.}
In the massive case, there is an additional term involving $\partial_\mu \phi_\mathrm{R}$ inside the parentheses, but this does not affect the current discussion as it contributes only at subleading order. 

\paragraph{}Upon writing \eqref{3.24} explicitly, one sees that $\delta S_\varLambda$ is the integral of:
\begin{enumerate}[label=(\roman*), leftmargin=3.5em, itemsep=-0.2em]
	\item terms that vanish when we take the $\varLambda \to \infty$ limit;
	\item terms of order $O (1)$, which are the important ones and are captured by a renormalised action \cite{Skenderis:2002wp}
	\begin{equation} \begin{aligned} \label{actionren}
		S_{\text{ren}} = {-1 \over d} \lim_{\varLambda \to \infty} \left( S_\varLambda + \int_{\partial \mathbb{B}_\varLambda}	\mathrm{d}^d x \, \mathscr{L}_{\text{counterterms}} [\phi_A] \right) ,
	\end{aligned} \end{equation}
	where the counterterm lagrangian is a local (bulk) functional of $\phi_A$;
	\item terms that diverge when $\varLambda \to \infty$, which are cancelled by the counterterm above.
\end{enumerate}
One can check that, due to $\Delta_- + \Delta_+ = d - 2 p$, (ii) includes $\phi_\Xi^+ \delta \phi^\Xi_-$ and $\phi_\Xi^- \delta \phi^\Xi_+$, except when $m^2 = 0$, in which case only one of these is present.\footnote{When $d > 2p \Rightarrow \Delta_- = 0$, (ii) contains $\phi_\Xi^+ \delta \phi^\Xi_-$ and when $d < 2p \Rightarrow \Delta_+ = 0$, it contains $\phi_\Xi^- \delta \phi^\Xi_+$. One can generalise the latter case to $d \leq 2p$ since, when $d = 2p$, solutions look like $\phi_\Xi = \ln r \phi_\Xi^- + \phi_\Xi^+ + \ldots$} Regarding the presence of terms (iii), let us address the massive and massless cases separately. 

\paragraph{}When $m^2 \neq 0$, the leading term in the regularised on-shell variation of the action is 
\begin{equation} \begin{aligned}
	\delta S_\varLambda & = (- \Delta_-) \varLambda^{\Delta_+ - \Delta_-} \int \mathrm{d}^d x \, \phi_\Xi^- \delta \phi^\Xi_- + \ldots 
\end{aligned} \end{equation}
Recall that we are assuming $\Delta_+ > \Delta_-$. If we consider the remaining case of $\Delta_+ = \Delta_-$, the leading term in $\delta S_\varLambda$ would scale logarithmically with $\varLambda$. Hence, the massive case always requires holographic renormalisation which, besides removing divergences in (iii), also cancels half of the terms (ii), i.e. either $\phi_\Xi^+ \delta \phi^\Xi_-$ or $\phi_\Xi^- \delta \phi^\Xi_+$. 

\paragraph{}The massless case, on the other hand, is a bit more complex. When $d \leq 2 p$, little changes compared to the massive case and the leading term, involving the integral of $\phi_\Xi^- \delta \phi^\Xi_-$, still diverges.\footnote{Note that $d = 2 p$ corresponds to $\Delta_+ = \Delta_-$.} There is no such term, however, when $d > 2 p$ (since $r^{- \Delta_-}$ is constant). In this case, we have to look at the next to leading order term in \eqref{3.23}. This can be written as $r^{- 2} \mathcal{F} [\phi_\Xi^-]$, assuming that $d - 2 p \neq 2$, where $\mathcal{F}$ is a local (boundary) functional, linear in $\phi_\Xi^-$. Hence, the leading term in the regularised on-shell variation of the action, when $m^2 = 0$ and $0 < d - 2 p \neq 2$, is given by
\begin{equation} \begin{aligned}
	\delta S_\varLambda & = - 2 \varLambda^{d - 2 p - 2} \int \mathrm{d}^d x \, \mathcal{F} [\phi_\Xi^-] \delta \phi^\Xi_- + \ldots 
\end{aligned} \end{equation}
Even though we did not consider these cases explicitly, note that the $d - 2 p = 2$ case is similar to $d = 2 p$, in the sense that we have logarithmic scaling of the leading term in $\delta S_\varLambda$. Hence, free massless $p$-forms also require holographic renormalisation, with the case where $d - 2 p = 1$ being the sole exception.

\subsection{Bulk path integrals: the setup} \label{setup}

\paragraph{}Initially, we address the case of an electric higher-form symmetry. Our goal is then to explore holographic realisations of the generating functional $W_j$, i.e. we investigate \textit{bulk generating functionals} $Z_F$ such that\footnote{For simplicity, we consider flat boundaries throughout. However, the portion of this paper where path integrals are discussed abstractly can be generalised with little to no effort. In particular, one can consider a geometry of the Fefferman-Graham type with boundary metric $h \neq \eta$. For the electric and magnetic cases it then suffices to replace $\mathcal{O} \to \sqrt{|h|} \mathcal{O}$ and $\psi \to \sqrt{|h|} \psi$, respectively.}
\begin{equation} \begin{aligned} \label{eq:4.3}
	\ln Z_F (\psi) = \expval{ \exp( i \mathcal{C} (N) \int_{\partial \mathbb{B}} \mathcal{O}^\Xi_j \psi_\Xi)}_0 \, .
\end{aligned} \end{equation}
The label $F$ specifies the choice of boundary conditions in $\mathbb{B}$, as will be clarified in the following sections. We have chosen to omit the boundary volume element and therefore $\int_{\partial \mathbb{B}}$ should be read as $\int_{\partial \mathbb{B}} \mathrm{d}^d {x}$. 
Additionally, we will use the prime symbol and write $\mathrm{A}'$, $\Xi'$, $\mathrm{R}'$, etc., to indicate that the first index has been removed (such that one can write $\psi_\Xi = \psi_{\mu \Xi'}$).\footnote{Note that there is at least one radial index among $\mathrm{R}'$.} 
As a final comment, we would rather have identified the generating functional $W$ with a path integral $Z_F$ for which the integration measure has been rescaled such that $Z_F (\psi) = 1$ when $\psi$ is evaluated at the background of the states with respect to whom we compute expectation values. For the purposes of this paper (where we consider expectation values of single-trace operators, at most, obtained from first derivatives of the generating functional), we can avoid these complications by identifying $W$ with $\ln Z_F$, as we did above.

\paragraph{}We will start by confirming the conventional wisdom that a global symmetry at the boundary corresponds to a local one in the bulk, such that $Z_F$ consists of a path integral weighted by $e^{{i \over \kappa^2} \bar{S}_{\text{ren}}}$ where $\bar{S}_{\text{ren}}$ denotes the renormalised action \eqref{actionren} for a massless $p$-form $\phi$. (An overbar will be common notation to indicate $m^2 = 0$). Initially, we consider \textit{low-degree} fields, i.e. $d > 2 p + 1$. In this case, the equations of motion are solved by $\phi_\Xi = \phi_\Xi^- (x) + O(r^{-1})$ --- cf. \cref{3.23} --- and holographic renormalisation uses a counterterm of the form
\begin{equation} \begin{aligned} \label{counter1}
	\mathscr{L}_{\text{counterterms}} [\phi_A] = \mathscr{L}_{\text{counterterms}} (\phi_\Xi) \, .
\end{aligned} \end{equation}
Note that, when dealing with bulk functionals, we reserve the use of round brackets $(X)$ to denote dependence on $X$ and a finite number of its non-radial derivatives.
The (infinitesimal) variation of the renormalised action can be written as
\begin{equation} \begin{aligned} \label{1.1}
	\delta \bar{S}_{\text{ren}} & = \int_{\mathbb{B}} E^\mathrm{A} \delta \phi_\mathrm{A} + \int_{\partial \mathbb{B}} Y^\Xi \delta \phi_\Xi \, ,
\end{aligned} \end{equation}
up to terms of order $O (\delta \phi)^2$. Same as in the boundary, $\int_\mathbb{B}$ should be read as $\int_\mathbb{B} \mathrm{d} r \wedge \mathrm{d}^d {x}$ since we have chosen to omit the bulk volume element. 
\Cref{1.1} is central to the current section. While $E^\mathrm{A} [\phi]$ are the equations of motion, $Y^\Xi [\phi]$ can be seen, via an abuse of terminology where we regard $r$ as time, as the \textit{renormalised canonical momenta} \cite{Papadimitriou:2007sj} conjugate to $\phi_\Xi$. Note that the canonical momentum is a linear combination of terms of the form $\partial_r \phi_\Xi$ and $\partial_\mu \phi_\mathrm{R}$ (and $\phi_\Xi$, if renormalised). 

\paragraph{}Excluding the scalar case, i.e. for $p \geq 1$, the bare action $\bar{S}$ is invariant under the gauge transformation
\begin{equation} \label{4.6} 
	\delta_\zeta \phi_\mathrm{A} = (\mathrm{d} \zeta)_A \, , 
\end{equation} 
where $\zeta \in C^\infty \Omega^{p-1} (\mathbb{B})$. The boundary counterterm is also gauge invariant since it is designed to cancel exactly the diverging terms in the regularised action $S_\varLambda$. Hence, in conclusion, $\delta_\zeta \bar{S}_{\text{ren}} = 0$. This implies that
\begin{equation} \begin{aligned} \label{eq:5.7}
	\int_{\mathbb{B}} \zeta_{\mathrm{A}'} \partial_a E^{a \mathrm{A}'} + \int_{\partial \mathbb{B}} \zeta_{\Xi'} \left( \partial_\mu Y^{\mu \Xi'} - E^{r \Xi'} \right) = 0 \, .
\end{aligned} \end{equation}
where we have substituted \cref{4.6} in \eqref{1.1} and integrated by parts. Equations like the one above, i.e. $\int X^{\mathrm{A}'} \zeta_{\mathrm{A}'} = 0$, arise frequently during the current section. In order to get rid of this equation's distributional character, we assume that the integral is a non-degenerate bilinear form $\langle X , \zeta \rangle$: since $\langle X , \zeta \rangle = 0$ for all $\zeta$, then $X^{\mathrm{A}'} = 0$. \\ \hspace*{1.75ex}
Since gauge invariance does not depend on the position of the boundary, the two terms in \eqref{eq:5.7} must vanish independently and the following identity is satisfied: $\partial_a E^{a \mathrm{A}'} = 0$ (this follows, upon recalling the equations of motion explicitly, from the commutativity of derivatives). Before addressing the second identity that follows from \eqref{eq:5.7}, note how it is possible, with little effort, to include a wider set of theories in the current section's analysis. 

\paragraph{}Until now, we have assumed $\phi$ to be a $p$-form, whose components are invariant under an even permutation of indices and change sign under odd ones. From now onwards, we consider a more general form of dependence between the components of $\phi$. Denoting by $P(\mathrm{A})$ a general permutation of the indices in $\mathrm{A}$, we are going to allow for $\phi_{P (\mathrm{A})} = e_{P (\mathrm{A})} \phi_\mathrm{A}$ where $e_{P (\mathrm{A})} = \pm 1$. In other words, we are interested in fields transforming in irreducible representations of $\mathrm{GL} (d{+}1 , \mathbb{R})$, which correspond to Young diagrams. We use curly brackets to denote the Young symmetriser such that, given some $\Psi_\mathrm{A}$ whose components are all independent, $\Psi_{ \{P (\mathrm{A})\} } = e_{P (\mathrm{A})} \Psi_{ \{\mathrm{A}\} }$. For example, in the case where $\phi$ is a $p$-form, then $\Psi_{ \{a_1 \ldots a_p\} } = \Psi_{ [a_1 \ldots a_p] }$. \\ \hspace*{1.75ex}
We assume that the current set of theories is associated with an action functional $\bar{\mathcal{S}}$ obeying
\begin{equation} \begin{aligned} \label{eq:4.8}
	\delta_\xi \bar{\mathcal{S}} = \int_{\partial \mathbb{B}} \xi_{\Xi'} Q_\xi^{\Xi'} [\phi] \quad \quad \text{under} \quad \quad \begin{cases}
	\delta_\xi \phi_\Xi = \partial_{ \{ \mu} \xi_{\Xi' \} } \\ 
	\delta_\xi \phi_{r \Xi'} = \Gamma_{\Xi'} (\xi) + \partial_r \xi_{\Xi'} \, ,	
	\end{cases}
\end{aligned} \end{equation}
for some functionals $Q^{\Xi'}_\xi$ and $\Gamma_{\Xi'}$,\footnote{$\Gamma_{\Xi'}$ is linear in $\xi$.} such that we have a {bulk gauge symmetry} that can be either \textit{large} or \textit{small} depending on $Q_\xi$ vanishing or not, respectively.\footnote{Our use of \textit{large gauge transformation} alludes not to ``failure of being continuously connected to the identity" meaning of the term but to the transformation parameter not dying off at the boundary.} While the introduction of $Q_\xi$ allows for anomalous global symmetries in the boundary theory \cite{Witten:1998qj,Bilal:1999ph},\footnote{See also \cite{Gynther:2010ed,Amado:2011zx,Landsteiner:2011iq,Landsteiner:2012kd} in the context of applied holography.} $\Gamma$ is responsible for incorporating theories of massless fields other than the $p$-forms, namely a spin-2 field propagating in spacetimes with cohomogeneity-one metrics. In particular, it is due to $\delta_\xi \phi_{r \Xi'} = \nabla_r \xi_{\Xi'}$, in these cases, that the letter $\Gamma$ --- traditionally used for the Christoffel symbols --- was chosen. 

\paragraph{}The action $\bar{\mathcal{S}}$ is taken to be quadratic in $\partial_r \phi_\Xi$ (with no explicit dependence on $\partial_r^2 \phi_\Xi$) such that its variation can be written exactly as $\delta \bar{S}_{\text{ren}}$ in \cref{1.1}. Then, using \cref{eq:4.8} in \eqref{1.1} and integrating by parts, bulk gauge symmetry implies that
\begin{equation} \begin{aligned} \label{eq:13} 
	\int_{\mathbb{B}} H^{\Xi'} [E] \xi_{\Xi'} + \int_{\partial \mathbb{B}} \left( \partial_\mu Y^{\mu \Xi'} + Q_\xi^{\Xi'} - E^{r \Xi'} \right) \xi_{\Xi'} = 0 \, ,
\end{aligned} \end{equation} 
where $H^{\Xi'}$ are functionals of the equations of motion (that reduce to $\partial_a E^{a \Xi'}$ when $\Gamma_{\Xi'} = 0$.). We assume that the bulk gauge symmetries are independent of the position of the boundary such that the terms in the left-hand side of \cref{eq:13} must vanish independently. Hence, 
\begin{equation} \begin{aligned}
	\left[ E^{r \Xi'} = \partial_\mu Y^{\mu \Xi'} + Q_\xi^{\Xi'} \right]_{\partial \mathbb{B}} ,
\end{aligned} \end{equation} 
so that $Y^{\mu \Xi'} \vert_{\partial \mathbb{B}}$ is classically conserved if $Q_\xi^{\Xi'}$ vanishes on-shell. In fact, that equations of motion normal to the boundary give rise to conservation equations in the boundary theory is well-known in the fluid-gravity correspondence \cite{Bhattacharyya:2007vjd}. 

\subsection{Electric quantisation} \label{ElectricQuant_massless}

\paragraph{}The bulk generating functionals that we consider are given by a path integral over configurations of $\phi$ satisfying the equations of motion at the boundary, i.e. for which $E^\mathrm{A} [\phi] \big \vert_{\partial \mathbb{B}} = 0$. For all the higher derivatives of $\phi$, this will broadly constrain their values at the boundary but the low order ones (including $\phi \vert_{\partial \mathbb{B}}$) are to be fixed instead by the boundary and regularity conditions that we impose on the configurations over which we are integrating. \\ \hspace*{1.75ex}
We assume invariance of the generating functional under infinitesimal shifts of the boundary of the target manifold in which $\phi_A (x)$ takes values. Since the generating functional is trivially invariant under field redefinitions, it must also remain invariant when $\phi$ is varied while keeping the boundary of the target manifold fixed. Such assumption --- which we will simply refer to as \textit{field redefinition invariance} --- is common in derivations of the Schwinger–Dyson equations and is employed here in a similar spirit. \\ \hspace*{1.75ex}
We begin by considering generating functionals with Dirichlet boundary conditions $\phi_\Xi \vert_{\partial \mathbb{B}} = \psi_\Xi$, as originally formulated in the GKPW prescription \cite{Gubser:1998bc,Witten:1998qj}:
\begin{equation} \begin{aligned} 
	Z_\phi \vcentcolon = \int_{\phi \vert_{\partial \mathbb{B}} = \psi} \mathfrak{D} \phi \exp({i \bar{\mathcal{S}} / \kappa^2}) \, .
\end{aligned} \end{equation}
Additionally, we denote (normalised) insertions of functionals, e.g. $X [\phi]$, in the path integral by 
\begin{equation} \begin{aligned} \label{eq:4.12}
	\braket{X}_\phi \vcentcolon = {1 \over Z_\phi} \int_{\phi \vert_{\partial \mathbb{B}} = \psi} \mathfrak{D} \phi \exp({i \bar{\mathcal{S}} / \kappa^2}) X \, .
\end{aligned} \end{equation} 
The label $\phi$ in $Z_\phi$ and $\braket{X}_\phi$ refers to the fact that $\phi_\Xi \vert_{\partial \mathbb{B}}$ is being fixed in the path integral. Note that the dependence of these objects on $\psi$ is implicitly assumed.
Trivially, the insertion of $\phi_\Xi \vert_{\partial \mathbb{B}}$ is fixed according to
\begin{equation} \begin{aligned} \label{112}
	\braket{\phi_\Xi}_\phi = \psi_\Xi \, ,
\end{aligned} \end{equation}
where we have adopted the following convention:
\vspace*{0.4em}

\noindent \hypertarget{conv1}{\textbf{(convention 1)}} any insertion $\braket{X}_\phi$ is tacitly assumed to be localised at the boundary, unless $X$ are the equations of motion $E^\mathrm{A}$. \hangindent=2em \hangafter=0

\vspace*{0.4em} \noindent 
Using field redefinition invariance of $Z_\phi$ under $\phi_\mathrm{A} \to \phi_\mathrm{A} + \delta \phi_\mathrm{A}$ for infinitesimal functions $\delta \phi_\mathrm{A} = \delta \varphi^*_\mathrm{A} (x)$ with $\delta \varphi^*_\Xi$ vanishing in the boundary, we have
\begin{equation} \begin{aligned} \label{1.6}
	0 & = {Z_\phi \over i \kappa^2} \int_{\mathbb{B}} \braket{E^\mathrm{A}}_\phi \delta \varphi^*_\mathrm{A} \, .
\end{aligned} \end{equation}
Since $E^\mathrm{A}$ vanishes at the boundary and $\delta \varphi^*_\mathrm{A}$ is unrestricted in the interior, the equation above implies that $\braket{E^\mathrm{A}}_\phi = 0$.\footnote{Since $\delta \varphi^*_\mathrm{R}$ is everywhere unrestricted, \cref{1.6} implies that $\braket{E^\mathrm{R}}_\phi = 0$ without using that $E^\mathrm{R} \vert_{\partial \mathbb{B}} = 0$ for the configurations we are integrating over. Hence, for the current generating functional, we would still arrive at the same results had we integrated over configurations that satisfy only the equations of motion $E^\Xi$ at the boundary.} Additionally, the linearised perturbation of the generating functional, defined as $\delta Z_\phi (\psi) \vcentcolon = Z_\phi (\psi + \delta \psi) - Z_\phi (\psi)$ up to $O (\delta \psi)^2$, is given by
\begin{equation} \begin{aligned} \label{18}
	- i \kappa^2 {\delta Z_\phi \over Z_\phi} & = \int_\mathbb{B}  \underbrace{\braket{E^\Xi}_\phi}_{= 0} \delta \varphi_\Xi + \int_{\partial \mathbb{B}} \braket{Y^\Xi}_\phi \delta \psi_\Xi \, ,
\end{aligned} \end{equation} 
where $\delta \varphi_\Xi$ is some function that tends to $\delta \psi_\Xi$ at the boundary.\footnote{Invariance under field redefinitions was used to write
	\begin{equation*} \begin{aligned}
		Z_\phi (\psi + \delta \psi) & = \int_{\phi \vert_{\partial \mathbb{B}} = \psi + \delta \psi} \mathfrak{D} \phi \exp(i {\bar{\mathcal{S}} [\phi] / \kappa^2}) = \int_{\phi' \vert_{\partial \mathbb{B}} = \psi} \mathfrak{D} \phi' \exp(i {\bar{\mathcal{S}} [\phi' + \delta \varphi] / \kappa^2}) \, .
	\end{aligned} \end{equation*}
} One can rewrite this as
\begin{equation} \begin{aligned} \label{110a} 
	{\delta \ln Z_\phi \over \delta \psi_\Xi} = {i \over \kappa^2} \braket{Y^\Xi}_\phi \, .
\end{aligned} \end{equation}
Using \cref{eq:4.3}, we have that $\braket{Y^\Xi}_\phi = \braket{\mathcal{O}^\Xi_j}$.\footnote{The compact index notation from \cref{GKPWmotivation} reflects the fact that, at this point, $\mathcal{O}$ and $\psi$ are not necessarily antisymmetric tensor fields but instead transform in the same representation as the corresponding bulk fields.} Hence, insertions of the renormalised momenta in the bulk path integral correspond to expectation values of the boundary operators, i.e. {the holographic dictionary maps between} $\mathcal{O}^\Xi_j$ and $Y^\Xi [\phi] \vert_{\partial \mathbb{B}}$ (where $\phi$ belongs to the set of configurations over which we are integrating). 

\paragraph{}Lastly, we would like to address some consequences of bulk gauge symmetry. Inserting \cref{eq:13} in the path integral, we obtain
\begin{equation} \begin{aligned} \label{17} 
	0 = \int_{\partial \mathbb{B}} \left( \partial_\mu \braket{Y^{\mu \Xi'}}_\phi + \braket{Q_\xi^{\Xi'}}_\phi \right) \xi_{\Xi'} \Rightarrow \partial_\mu \braket{Y^{\mu \Xi'}}_\phi + \braket{Q_\xi^{\Xi'}}_\phi = 0 \, .
\end{aligned} \end{equation}
Hence, assuming $\braket{\delta_\xi \bar{\mathcal{S}}}_\phi = 0$ such that $\braket{Q_\xi^{\Xi'}}_\phi = 0$, we conclude that the bulk generating functional $Z_\phi$ is invariant under transformations $\delta \psi_{\mu \Xi'} = \partial_{ \{ \mu} \zeta_{\Xi' \} }$ and describes a theory in the boundary with conserved currents: $\partial_\mu \braket{\mathcal{O}^{\mu \Xi'}_j} = 0$. Remember that $\delta_\xi \bar{\mathcal{S}}$ does vanish in the case of a Maxwell $p$-form. Here, the boundary operator is an electric higher-form current and we say that the path integral $Z_\phi$ enforces the \textit{electric quantisation} of the theory $\bar{\mathcal{S}}$. 

\subsubsection{High-degree $p$-forms and deformations} \label{high-rank}

\paragraph{}Until now, we have considered low-degree $p$-forms, for which $d > 2 p + 1$. In this case, the on-shell value of $\phi$ is regular at the boundary, while the canonical momentum conjugate to $\phi_\Xi$ diverges. Holographic renormalisation with a counterterm of the form \eqref{counter1} yields the renormalised momenta $Y^\Xi [\phi]$, whose on-shell value at the boundary is finite: $Y^\Xi = \phi^\Xi_+ (x) + O (r^{-1})$. The situation is reversed when $d < 2 p + 1$. For \textit{high-degree} $p$-forms, the equations of motion are solved by $\phi_\Xi = r^{2p-d} \phi_\Xi^- (x) + \ldots + \phi_\Xi^+ (x) + O (r^{-1})$. Hence, the on-shell value of $\phi$ diverges at the boundary, while the canonical momenta (which we keep denoting by $Y^\Xi$) is well behaved: $Y^\Xi = \phi^\Xi_- (x) + O (r^{-1})$. 
This time, holographic renormalisation requires 
\begin{equation} \begin{aligned} \label{counter2}
	\mathscr{L}_{\text{counterterms}} [\phi_A] = \mathscr{L}_{\text{counterterms}} (Y^\Xi) \, ,
\end{aligned} \end{equation}
such that the renormalised action is given by 
\begin{equation} \begin{aligned} 
	\delta \bar{S}_{\text{ren}} & = \int_{\mathbb{B}} E^\mathrm{A} \delta \phi_\mathrm{A} + \int_{\partial \mathbb{B}} Y^\Xi \delta \Phi_\Xi \, .
\end{aligned} \end{equation}
Note that the form of $\Phi_\Xi = \Phi_\Xi [\phi]$ depends on the counterterms used.
The equation above is general in the sense that either $Y^\Xi$ is the renormalised momenta and $\Phi_\Xi = \phi_\Xi$, when $d > 2 p + 1$, or $Y^\Xi$ is the canonical momenta and $\Phi_\Xi$ is a renormalised quantity, when $d < 2 p + 1$. In the $d = 2 p + 1$ intermediate case, $\lim_{\varLambda \to \infty} S_\varLambda$ is well defined such that, since no counterterm is necessary, both $Y^\Xi$ and $\Phi_\Xi$ denote bare quantities. 

\paragraph{}Regarding theories where $\phi$ is not necessarily a $p$-form, we also generalise the action $\bar{\mathcal{S}} [\phi]$ such that its variation is given by
\begin{equation} \begin{aligned} 
	\delta \bar{\mathcal{S}} & = \int_{\mathbb{B}} E^\mathrm{A} \delta \phi_\mathrm{A} + \int_{\partial \mathbb{B}} Y^\Xi \delta \Phi_\Xi \, ,
\end{aligned} \end{equation}
where $\Phi_\Xi$ are functionals of the fundamental field determined by holographic renormalisation. The bulk generating functional is then given by
\begin{equation} \begin{aligned} 
	Z_\Phi \vcentcolon = \int_{\Phi \vert_{\partial \mathbb{B}} = \psi} \mathfrak{D} \phi \exp({i \bar{\mathcal{S}} / \kappa^2}) \, .
\end{aligned} \end{equation}
The boundary conditions $\Phi_\Xi \vert_{\partial \mathbb{B}} = \psi_\Xi$ are the natural generalisation of Dirichlet's when $\phi_\Xi \vert_{\partial \mathbb{B}}$ is ill defined. Because the canonical momentum is gauge-invariant and therefore $\delta_\zeta \Phi_\Xi = \delta_\zeta \phi_\Xi$, one only needs to generalise $\delta \varphi$ in order to guarantee that $Z_\Phi$ satisfies the results derived in the previous section, namely \cref{1.6,18,110a,17}. This generalisation is possible under the assumption that
\vspace*{0.4em}

\noindent \hypertarget{assump1}{\textbf{(assumption 1)}} given $\phi$ and $\psi$ obeying $\Phi [\phi] \vert_{\partial \mathbb{B}} = \psi$, for any infinitesimal $\delta \psi$ there always exists a small shift $\phi_\mathrm{A} \to \phi_\mathrm{A} + \delta \varphi_\mathrm{A}$ under which $\delta \Phi \vert_{\partial \mathbb{B}} = \delta \psi$.\footnote{The fact that we could take $\delta \psi$ to have support at the boundary and still reach the same overall conclusions, shows that field redefinition invariance is never essential. It does, however, provide a controlled framework for deriving such results.} \hangindent=2em \hangafter=0

\vspace*{0.4em} \noindent 
In particular, $\delta \varphi_\mathrm{A}$ corresponding to $\delta \psi_\Xi = 0$ is denoted by $\delta \varphi^*_\mathrm{A}$ (this is consistent with the previous use of $\delta \varphi^*_\mathrm{A}$ when $\Phi_\Xi = \phi_\Xi$). \\ \hspace*{1.75ex}
The fact that $Z_\Phi$ satisfies \cref{17} allows one to conclude that such a generating functional also enforces electric quantisation. At this stage, it is worth clarifying the connection with the conventional terminology \cite{Rivelles:2003ge} --- \textit{standard} and \textit{alternative quantisation}. While the former consists of fixing the leading term $\phi_\Xi^-$ in the near-boundary expansion of $\phi_\Xi$, in the latter, one fixes $\phi_\Xi^+$ as the arbitrary source $\psi_\Xi$. Consequently, in standard (alternative) quantisation, the holographic dictionary maps $\mathcal{O}$ to $\phi^+$ ($\phi^-$). Hence, electric quantisation amounts to standard quantisation when $d > 2 p$ and alternative when $d \leq 2 p$.

\paragraph{}Note that we began by imposing Dirichlet boundary conditions and then extended them to specify $\Phi_\Xi \vert_{\partial \mathbb{B}}$ instead. Let us now consider a more general situation in which (linear) functionals $F_\Xi (\Phi , Y)$ are held fixed at the boundary:
\begin{equation} \begin{aligned} \label{Z_F}
	Z_F (\psi) \vcentcolon = \int_{F \vert_{\partial \mathbb{B}} = \psi} \mathfrak{D} \phi e^{{i \over \kappa^2} \left( \bar{\mathcal{S}} + \int_{\partial \mathbb{B}} D_F \right)} \, .
\end{aligned} \end{equation}
The insertion of a functional $X [\phi]$ is denoted by $\braket{X}_F \vcentcolon = (Z_F)^{-1} \int_{F \vert_{\partial \mathbb{B}} = \psi} \mathfrak{D} \phi e^{{i \over \kappa^2} \left( \bar{\mathcal{S}} + \int_{\partial \mathbb{B}} D_F \right)} X$. We have introduced a surface term $\int_{\partial \mathbb{B}} D_F$ such that
\begin{equation} \begin{aligned} \label{eq:4.23}
	\delta \left( \bar{\mathcal{S}} + \int_{\partial \mathbb{B}} D_F \right) & = \int_{\mathbb{B}} E^\mathrm{A} \delta \phi_\mathrm{A} + \int_{\partial \mathbb{B}} Y^\Xi \delta F_\Xi \, .
\end{aligned} \end{equation}
This requires $D_F = D_F (Y^\Xi)$ to be quadratic in $Y^\Xi$ such that 
\begin{equation} \begin{aligned}
	F_\Xi = \Phi_\Xi + {\delta D_F \over \delta Y^\Xi} \, .
\end{aligned} \end{equation}
Any boundary condition that depends on both $\Phi$ and $Y$ is called \textit{mixed}. A simple example of mixed boundary conditions is when $F_\Xi (\Phi , Y) = \alpha \Phi_\Xi +\beta Y^\Xi$ ($\alpha , \beta \in \mathbb{R}$), such that we have: Dirichlet boundary conditions for $\beta = 0$, Neumann for $\alpha = 0$ and Robin otherwise.\footnote{The use of the denomination `Neumann' comes from the canonical momenta in $p$-form theories being proportional to $\partial_r \phi_\Xi$ when radial gauge is fixed. The same way that we use the term Dirichlet when $\Phi_\Xi$, instead of $\phi_\Xi$, is fixed, we still use Neumann when $Y^\Xi$ is the renormalised canonical momenta.}

\paragraph{}Let us examine the properties of the path integral \eqref{Z_F} as a generating functional. We start by noticing that \cref{112} is to be replaced by
\begin{equation} \begin{aligned} 
	\braket{F_\Xi}_F = \psi_\Xi \; .
\end{aligned} \end{equation} 
It is crucial that we extend assumption \hyperlink{assump1}{1} to $F_\Xi$ instead of $\Phi_\Xi$ such that, using field redefinition invariance of $Z_F (\psi)$ under $\Phi_\mathrm{A} \to \Phi_\mathrm{A} + \delta \varphi^*_\mathrm{A}$, we obtain
\begin{equation} \begin{aligned} 
	0 & = \int_{\mathbb{B}} \braket{E^\mathrm{A}}_F  \delta \varphi^*_\mathrm{A} \, .
\end{aligned} \end{equation}
Since $E^\mathrm{A}$ vanishes at the boundary and $\delta \varphi^*_\mathrm{A}$ is unrestricted in the interior, we conclude that $\braket{E^\mathrm{A}}_F = 0$. The linearised perturbation of the generating functional is given by
\begin{equation} \begin{aligned} \label{1.18a} 
	- i \kappa^2 \delta \ln Z_F (\psi) & = \int_\mathbb{B} \underbrace{\braket{E^\mathrm{A}}_F}_{= 0} \delta \varphi_\mathrm{A} + \int_{\partial \mathbb{B}} \braket{Y^\Xi}_F \delta \psi_\Xi \, ,
\end{aligned} \end{equation}
such that
\begin{equation} \begin{aligned}
	{\delta \ln Z_F \over \delta \psi_\Xi} & = {i \over \kappa^2} \braket{Y^\Xi}_F \, .
\end{aligned} \end{equation}
Hence, the generating functional with non-Dirichlet boundary conditions \eqref{Z_F} enforces electric quantisation. \\ \hspace*{1.75ex}
Because standard and alternative quantisation are typically defined in terms of fixing either $\phi^-$ or $\phi^+$, these terms seem useless with mixed boundary conditions. It is possible to retain their essence through an alternative definition similar to the one above for electric quantisation: standard (alternative) quantisation is such that insertions of the functional of $\phi$ that asymptotes to $\phi^+$ ($\phi^-$) at the boundary correspond to expectation values of operators in the boundary theory. In this case, one generalises the previous conclusion from non-mixed boundary conditions that electric quantisation is equivalent to standard and alternative quantisation when $d > 2 p$ and $d \leq 2 p$, respectively. 

\paragraph{}In agreement with the designation of $Z_F$'s quantisation, the boundary theory dual to our $p$-form model possesses an electric symmetry. This can be verified by inserting \cref{eq:13} in the path integral, which renders a straightforward generalisation of \cref{17}:
\begin{equation} \begin{aligned} \label{119}
	\partial_\mu \braket{Y^{\mu \Xi'}}_F + \braket{Q_\xi^{\Xi'}}_F = 0 \, .
\end{aligned} \end{equation}
Hence, from \cref{1.18a,119}, we recover the result from previous section: assuming that $\braket{\delta_\xi \bar{\mathcal{S}}}_F = 0$, the bulk generating functional is invariant under $\delta \psi_{\mu \Xi'} = \partial_{ \{ \mu} \zeta_{\Xi' \} }$ and describes a theory in the boundary with conserved currents. \\ \hspace*{1.75ex}
We conclude that the choice of boundary conditions is intrinsically tied to the presence of extra surface terms that depend quadratically on the momenta $Y^\Xi$ dual to $\mathcal{O}^\Xi_j$. Such surface terms are mapped via the holographic dictionary to double-trace deformations of the boundary theory \cite{Berkooz:2002ug,Witten:2001ua}.\footnote{See also \cite{Minces:2001zy}. Subsequent developments include \cite{Mueck:2002gm,Minces:2002wp,Sever:2002fk,Aharony:2005sh,Elitzur:2005kz,Gubser:2002vv,Hartman:2006dy,Diaz:2007an,Papadimitriou:2007sj}.} For this reason, we use the term \textit{deformation} in the bulk theory specifically to designate surface terms that depend solely on the functional dual to the single-trace operator.

\subsection{Magnetic quantisation} \label{MagnQuant_massless} 

\paragraph{}Recalling, from \cref{boundarytheories}, the generating functional $W_a$ with a magnetic higher-form symmetry, we propose that 
\begin{equation} \begin{aligned} 
	W_a [\psi] \equiv \expval{ \exp( i \mathcal{C} (N) \int_{\partial \mathbb{B}} \mathcal{O}_\Xi^a \psi^\Xi)}_0 = \ln Z_F (\psi) \, ,
\end{aligned} \end{equation}
where the bulk generating functional is also given by \eqref{Z_F} but $D_F = - \Phi_\Xi Y^\Xi - D (\Phi_\Xi)$ with $D$ quadratic in $\Phi_\Xi$ such that 
\begin{equation} \begin{aligned} \label{2.18}
	\delta \Big( \bar{\mathcal{S}} + \int_{\partial \mathbb{B}} D_F \Big) = \int_{\mathbb{B}} E^\mathrm{A} \delta \phi_\mathrm{A} + \int_{\partial \mathbb{B}} \Phi_\Xi \delta F^\Xi \qquad \text{with} \qquad F^\Xi = - Y^\Xi - {\delta D \over \delta \Phi_\Xi} \, .
\end{aligned} \end{equation}
We see that, in this case, $D_F$ is made up of a term dual to a single-trace operator, responsible for a Legendre transformation \cite{Klebanov:1999tb}, and the double-trace deformation $D$. 

\paragraph{}As with electric quantisation, we now verify the aforementioned proposal, namely that $Z_F$ as defined above exhibits the expected properties of $W_a$. Once again, one can show that insertions of the equations of motion in the path integral vanish, such that
\begin{equation} \begin{aligned} \label{2.22b} 
	{\delta \ln Z_F \over \delta \psi^\Xi} & = {i \over \kappa^2} \braket{\Phi_\Xi}_F \, .
\end{aligned} \end{equation}
Hence, $\braket{\Phi_\Xi}_F = \braket{\mathcal{O}_\Xi^a}$ and, more generally, insertions of $\Phi$ in the path integral correspond to boundary expectation values. 
Given a conserved source such that $\int_{\partial \mathbb{B}} \psi^{\mu \Xi'} \partial_\mu \zeta_{\Xi'}$ vanishes, $\braket{\Phi_\Xi}_F$ and consequently $\braket{\mathcal{O}_\Xi^a}$ are only defined up to pure gauge $\partial_{ \{\mu} \zeta_{\Xi'\} }$. If $\phi$ is a massless $p$-form, we have a magnetic higher-form symmetry at the boundary and we say that the Legendre-transformed path integral enforces \textit{magnetic quantisation}. This is the natural counterpart of electric quantisation in the sense that the holographic dictionary maps $\mathcal{O}_\Xi^a$ to $\Phi_\Xi [\phi] \vert_{\partial \mathbb{B}}$,\footnote{Here, $\phi$ can be any configuration contributing towards the path integral.} instead of the renormalised canonical momenta. As a result, magnetic quantisation amounts to standard quantisation for high-degree $p$-forms ($d \leq 2 p$) and alternative for low/intermediate-degree ones ($d > 2 p$). 

\paragraph{}As before, \cref{119} is obtained by inserting \eqref{eq:13} into the path integral. Using the rightmost equation in \eqref{2.18}, it can be rewritten as
\begin{equation} \begin{aligned} \label{1.22}
	\partial_\mu \braket{F^{\mu \Xi'}}_F & = \braket{Q_\xi^{\Xi'}}_F - \partial_\mu \expval{\delta D \over \delta \Phi_{\mu \Xi'}}_F \\
	& = \partial_\mu \psi^{\mu \Xi'} ,
\end{aligned} \end{equation}
since $\braket{F^\Xi}_F = \psi^\Xi$. Hence, the conservation of $\psi^\Xi$ requires gauge symmetry of the bulk action plus deformation terms:
\begin{equation} \begin{aligned} \label{2.29}
	\expval{\delta_\xi \left( \bar{\mathcal{S}} + \int_{\partial \mathbb{B}} D \right)}_F = 0 \, .
\end{aligned} \end{equation}
In particular, deformations that break gauge invariance at the level of boundary conditions are ruled out when $\expval{\delta_\xi \bar{\mathcal{S}}}_F = 0$. This means that, for Maxwell $p$-forms, Robin boundary conditions are disallowed in magnetic quantisation.

\paragraph{}Let us summarise our results so far regarding the holographic description of higher-form symmetries. In electric quantisation, U(1) higher-form large gauge symmetry of the bulk action corresponds to electric higher-form symmetries at the boundary. In magnetic quantisation, on the other hand, small gauge symmetry is in principle sufficient when the appropriate deformation is included since, in this case, magnetic symmetries come from gauge invariance of the total action minus the Legendre term.

\subsection{Holographic renormalisation of massless $p$-forms} \label{masslessrenorm_new}

\paragraph{}We now specialise to Maxwell theories of higher-form gauge fields. Holographic renormalisation is carried out explicitly and the holographic generating functionals are presented for both electric and magnetic quantisation. Our entire approach is guided by power counting and we therefore restrict to the leading contributions in the bulk model. Moreover, since topological terms are entirely neglected, the resulting generating functionals are parametrised by a single coupling constant associated with a double-trace deformation.

\paragraph{}We consider a massless $p$-form, $\bar{\phi}$, where $p \leq d-1$ is a non-negative integer. Its (regularised) action is given, in terms of the {field strength} $\bar{\mathcal{F}} \equiv \mathrm{d} \bar{\phi}$, by 
\begin{equation} \begin{aligned} \label{actionmassless}
	\bar{S}_\varLambda = {1 \over 2} \int_{\mathbb{B}_\varLambda} \mathrm{d}^{d+1} x \sqrt{|G|} \left( \mathscr{L}_{\text{grav}} + {\bar{\mathcal{F}}_{a_0 ... a_p} \bar{\mathcal{F}}^{a_0 ... a_p} \over (p+1)!} \right) .
\end{aligned} \end{equation} 
The gravitational part of the action is denoted by $\mathscr{L}_{\text{grav}}$, which can be for example the Einstein-Hilbert term with a negative cosmological constant. We only require that the theory possesses asymptotically (locally) AdS solutions. In fact, because we are solely concerned with holographic renormalisation of $\bar{\phi}$, we ignore the coupling between gravitational and matter perturbations, and take $G$ to be the (non-dynamical) metric of pure AdS. \\ \hspace*{1.75ex}
The equations of motion $\mathrm{d}^\dagger \bar{\mathcal{F}} = 0$ and the Bianchi identity $\mathrm{d} \bar{\mathcal{F}} = 0$ can be solved for the field strength, which we do in AdS$_{d+1}$ space with the Poincar\'e metric \eqref{metric}. If, for example, $d-2p+1 = \vcentcolon {\bar{\lambda}}$ is larger than 3 and even, we obtain
\begin{subequations} \begin{align} 
	\bar{\mathcal{F}}_{\mu_0 ... \mu_p} & = \beta_{\mu_0 ... \mu_p} - {r^{-2} \over 2} {\Box \beta_{\mu_0 ... \mu_p} \over 3 - {\bar{\lambda}}} 
	+ \ldots + {r^{1 - {\bar{\lambda}}} \over p! (1 - {\bar{\lambda}})} \partial_{[ \mu_0} J^{{\bar{\lambda}}}_{\mu_1 ... \mu_p]} + \ldots \\
	\bar{\mathcal{F}}_{r \mu_1 ... \mu_p} & = r^{- 3} {\partial^{\mu_0} \beta_{\mu_0 ... \mu_p} \over 3 - {\bar{\lambda}}} + \ldots + r^{- {\bar{\lambda}}} J^{{\bar{\lambda}}}_{\mu_1 ... \mu_p} + \ldots
\end{align} \end{subequations}
where $\beta \in \Omega^{p+1}$ and $J^{{\bar{\lambda}}} \in \Omega^{p}$ obey $\mathrm{d} \beta = 0 = \mathrm{d} \ast J^{{\bar{\lambda}}}$.\footnote{The operator $\Box$ corresponds to $\eta^{\mu \nu} \partial_\mu \partial_\nu$.} Hence, on-shell configurations are given by
\begin{subequations} \begin{align}
	p! \bar{\phi}_{\mu_1 ... \mu_p} & = \partial_{[\mu_1} \boldsymbol{\zeta}_{\mu_2 ... \mu_{p}]} + \alpha_{\mu_1 ... \mu_{p}} - {r^{-2} \over 2} {\partial^{\mu_0} \beta_{\mu_0 ... \mu_p} \over 3 - {\bar{\lambda}}} 
	+ \ldots + {r^{1 - {\bar{\lambda}}} \over 1 - {\bar{\lambda}}} J^{{\bar{\lambda}}}_{\mu_1 ... \mu_p} + \ldots \\
	p! \bar{\phi}_{r \mu_2 ... \mu_{p}} & = \partial_{[r} \boldsymbol{\zeta}_{\mu_2 ... \mu_{p}]} \, ,
\end{align} \end{subequations}
where we have introduced $\boldsymbol{\zeta} \in C^\infty \Omega^{p-1} (\mathbb{B}_\varLambda)$, which is arbitrary, and $\alpha \in \Omega^{p}$ obeys $\mathrm{d} \alpha = p! \beta$. Choosing it to be a general solution to this equation, $\alpha$ is then determined up to closed forms, as if it were a boundary gauge field. Hence, we can fix radial gauge by setting $\boldsymbol{\zeta} = 0$.

\paragraph{}General solutions --- cf. appendix \ref{MaxwellEOM} --- can be presented once we introduce the following notation: $(\bar{\varDelta}_+ , \bar{\varDelta}_-) \vcentcolon = ({\bar{\lambda}} , 3)$ when ${\bar{\lambda}} \geq 3$ and  $(\bar{\varDelta}_+ , \bar{\varDelta}_-) \vcentcolon = (3 , {\bar{\lambda}})$ when ${\bar{\lambda}} \leq 3$. We separate the cases where  $\bar{\varDelta}_+ - \bar{\varDelta}_-$ is odd, even but non-null and null, which correspond to even ${\bar{\lambda}}$, odd ${\bar{\lambda}} \neq 3$ and ${\bar{\lambda}} = 3$, respectively. For even ${\bar{\lambda}}$, we have
\begin{equation} \begin{aligned} \label{A1_new}
	p! \bar{\phi}_{\mu_1 ... \mu_p} & = \alpha_{\mu_1 ... \mu_{p}} + {r^{1 - \bar{\varDelta}_-} \over 1 - \bar{\varDelta}_-} J^{\bar{\varDelta}_-}_{\mu_1 ... \mu_p} + \ldots + {r^{1 - \bar{\varDelta}_+} \over 1 - \bar{\varDelta}_+} J^{\bar{\varDelta}_+}_{\mu_1 ... \mu_p} + \ldots
\end{aligned} \end{equation}
where 
\begin{equation} \begin{aligned} \label{0.12}
	J^{3}_{\mu_1 ... \mu_p} = {\partial^{\mu_0} \beta_{\mu_0 ... \mu_p} \over 3 - {\bar{\lambda}}} \, .
\end{aligned} \end{equation}
Previously (in particular in \cref{GKPWmotivation}), we used ellipses simply as placeholders for subleading terms in a near-boundary expansion. From this point onward, however, they carry additional information about the solutions as we adopt the following convention. Assuming the terms before ``$\ldots$" all involve the same $r$-constant, the ellipsis represents a sum $\sum_{i=1}^j O \left( {\Box \over r^2} \right)^i$ acting on the last of these terms. Such a series has an endpoint $j$ if there is a term after the ellipsis with the same constant of motion. This term contains a power of $\Box$ strictly greater than any term in the series (and this condition determines the endpoint). In fact, in case the term upon which $\sum_{i=1}^j O \left( {\Box \over r^2} \right)^i$ is acting does not obey this condition, then it must be itself set to zero. Lastly, if the terms before ``$\ldots$" involve several $r$-constants, the convention is to be applied separately for each of these --- only $\alpha$ and $\beta$ (when explicitly written) are exempt. \\ \hspace*{1.75ex}
For odd ${\bar{\lambda}} \neq 3$, 
\begin{equation} \begin{aligned} \label{A2_new}
	p! \bar{\phi}_{\mu_1 ... \mu_p} = & \alpha_{\mu_1 ... \mu_{p}} + {r^{1 - \bar{\varDelta}_-} \over 1 - \bar{\varDelta}_-} J^{\bar{\varDelta}_-}_{\mu_1 ... \mu_p} + \ldots + \ln r {(- \Box)^{1 - \bar{\varDelta}_- \over 2} J^{\bar{\varDelta}_-}_{\mu_1 ... \mu_p} \over \varOmega_{\bar{\varDelta}_+ - \bar{\varDelta}_-}} \\
	& + {r^{1-\bar{\varDelta}_+} \over 1-\bar{\varDelta}_+} \left( J^{\bar{\varDelta}_+}_{\mu_1 ... \mu_p} + {(\bar{\varDelta}_+ - 1) \ln r + 1 \over \bar{\varDelta}_- - \bar{\varDelta}_+} {(- \Box)^{\bar{\varDelta}_+ - \bar{\varDelta}_- \over 2} J^{\bar{\varDelta}_-}_{\mu_1 ... \mu_p} \over (\bar{\varDelta}_+ - 1) \varOmega_{\bar{\varDelta}_+ - \bar{\varDelta}_-}} \right) + \ldots
\end{aligned} \end{equation}
where $\varOmega_w = \Pi^{{w \over 2} - 1}_{s=1} 2 s (\bar{\varDelta}_- + 2s - 3)$ for $w > 2$ and $\varOmega_2 = 1$. When ${\bar{\lambda}} > 3$, the equation above should be read without the purely logarithmic term, in which case \cref{0.12} also holds. On the other hand, when ${\bar{\lambda}} < 3$, we obtain
\begin{equation} \begin{aligned} \label{0.13}
	J^{3}_{\mu_1 ... \mu_p} & = {\partial^{\mu_0} \beta_{\mu_0 ... \mu_p} \over 3-{\bar{\lambda}}} - {(- \Box)^{3 - {\bar{\lambda}} \over 2} J^{{\bar{\lambda}}}_{\mu_1 ... \mu_p} \over (3-{\bar{\lambda}})^2 \varOmega_{3-{\bar{\lambda}}}} \, .
\end{aligned} \end{equation}
Finally, for ${\bar{\lambda}} = 3$, we have
\begin{equation} \begin{aligned} \label{A3_new}
	p! \bar{\phi}_{\mu_1 ... \mu_p} & = \alpha_{\mu_1 ... \mu_{p}} - {r^{-2} \over 2} \left( {2 \ln r + 1 \over 2} \hat{J}^3_{\mu_1 ... \mu_p} + J^3_{\mu_1 ... \mu_p} \right) + \ldots
\end{aligned} \end{equation}
where $\hat{J}^3_{\mu_1 ... \mu_p}$ (similar to $J^3_{\mu_1 ... \mu_p}$ when ${\bar{\lambda}} \neq 3$) is given by
\begin{equation} \begin{aligned} \label{0.14}
	\hat{J}^3_{\mu_1 ... \mu_p} = - \partial^{\mu_0} \beta_{\mu_0 ... \mu_p} \, .
\end{aligned} \end{equation}
The role of radial gauge in this section is restricted to \cref{A1_new,A2_new,A3_new}, where it is used simply to avoid clutter.

\subsubsection{Renormalisation counterterms} \label{masslessrenorm}

\paragraph{}We now determine the counterterms that cancel the divergent part (when $\varLambda \to \infty$) of the on-shell variation of the action \eqref{actionmassless}, which can be written as
\begin{equation} \begin{aligned} \label{1.9}
	\delta \bar{S}_\varLambda & = \int_{\partial \mathbb{B}_\varLambda} \mathrm{d}^d x \, r^{{\bar{\lambda}}} \bar{\mathcal{F}}_r {}^{\mu_1 ... \mu_p} \delta \bar{\phi}_{\mu_1 ... \mu_p} \, .
\end{aligned} \end{equation} 
Hence, using the compact index notation from \cref{GKPWmotivation}, the canonical momenta conjugate to $\bar{\phi}_\Xi$ is given by $r^{{\bar{\lambda}}} \bar{\mathcal{F}}_r {}^\Xi$. From \cref{Fr1,Fr2,Fr3}, one immediately sees that when $\bar{\phi}$ is a low-degree $p$-form, i.e. ${\bar{\lambda}} > 2$, the canonical momenta exhibits singular on-shell behaviour near the conformal boundary: 
\begin{equation} \begin{aligned}
	r^{{\bar{\lambda}}} \bar{\mathcal{F}}_r {}^\Xi \vert_{\partial \mathbb{B}_\varLambda} \sim \begin{cases}
		\varLambda^{{\bar{\lambda}} - 3} \, , \quad & {\bar{\lambda}} \neq 3 \\
		\ln \varLambda \, , \quad & {\bar{\lambda}} = 3 \, .
	\end{cases} 
\end{aligned} \end{equation}
The divergent term involves $J^{3}$ (or $\hat{J}^{3}$). An identical behaviour is found for the non-radial components of high-degree $p$-forms:
\begin{equation} \begin{aligned}
	\bar{\phi}_\Xi \vert_{\partial \mathbb{B}_\varLambda} \sim \begin{cases}
		\varLambda^{1 - {\bar{\lambda}}} \, , \quad & {\bar{\lambda}} \neq 1 \\
		\ln \varLambda \, , \quad & {\bar{\lambda}} = 1 \, ,
	\end{cases} 
\end{aligned} \end{equation}
which can be seen from \cref{A1_new,A2_new,A3_new}. The divergent term in this case involves $J^{{\bar{\lambda}}}$. Note that these results imply the need for holographic renormalisation when ${\bar{\lambda}} \neq 2$, in agreement with \cref{GKPWmotivation}.

\paragraph{}In order to derive the necessary counterterms, we start by writing the on-shell variation of the action explicitly in terms of the solutions from previous section. Before we do this, it is useful to introduce some notation: firstly, when $a=b$,
\begin{equation} \begin{aligned} 
	{\varLambda^{a-b} \over a-b} \equiv \ln \varLambda \, ;
\end{aligned} \end{equation}
and, secondly, we are using \cref{0.12,0.13,0.14} to get rid of $J^3$ (or $\hat{J}^3$ when ${\bar{\lambda}} = 3$) in favour of $\beta$, so we can refer to $J^{{\bar{\lambda}}}$ simply by $J$.
Then, \eqref{1.9} becomes
\begin{equation} \begin{aligned} \label{4.4}
	\delta \bar{S}_\varLambda \simeq & {1 \over (p+1)!} \int_{\partial \mathbb{B}_\varLambda} \mathrm{d}^d x \bigg[ (p+1) {\varLambda^{1-{\bar{\lambda}}} \over 1-{\bar{\lambda}}} J^{\mu_1 ... \mu_p} \delta J_{\mu_1 ... \mu_p} + {\varLambda^{-1-{\bar{\lambda}}} \over (-1)-{\bar{\lambda}}} O (J \Box \delta J) \\
	& + {\varLambda^{{\bar{\lambda}}-3} \over {\bar{\lambda}}-3} \beta^{\mu_0 ... \mu_p} \delta \beta_{\mu_0 ... \mu_{p}} + {\varLambda^{{\bar{\lambda}}-5} \over {\bar{\lambda}}-5} O (\beta \Box \delta \beta) + (p+1) J^{\mu_1 ... \mu_p} \delta \alpha_{\mu_1 ... \mu_{p}} \bigg] ,
\end{aligned} \end{equation}
where $\simeq$ means equality up to $O (\varLambda^{-1})$.\footnote{When $O ( J \Box \delta J {) \sim O} \left( J_{\mu_1 ... \mu_p} \Box \delta J^{\mu_1 ... \mu_p} \right)$ appears integrated, it stands for a finite number of $J (\Box^{1+i} / \varLambda^{2i}) \delta J$ terms ($i \geq 0$) and a similar convention applies to $O (\beta \Box \beta)$. Also, note that integration by parts makes the position of $\Box^{1+i}$ irrelevant.} Before we proceed, a brief digression is in order.

\paragraph{}Note that the Maxwell-type action \eqref{actionmassless} is essential in describing holographically the universality class of boundary theories with non-anomalous higher-form symmetries. However, we know that a bulk action is not sufficient for uniquely defining a holographic generating functional --- cf. \cref{ElectricQuant_massless,MagnQuant_massless}. In addition, surface terms corresponding to a choice of quantisation and deformations need to be specified. We are going to consider deformations that are universal (from the low-energy point of view) but we continue to disregard topological contributions. In particular, $\int_{\mathbb{B}_\varLambda} \bar{\mathcal{F}} \wedge \bar{\mathcal{F}} \sim \int_{\partial \mathbb{B}_\varLambda} \bar{\phi} \wedge \bar{\mathcal{F}}$ --- which is a theta term in the bulk and a Chern-Simons term in the boundary (available for ${\bar{\lambda}} = 2$) --- is overall ignored, even though it is invariant under large gauge transformations and therefore keeps the boundary symmetry free of anomalies. Also, we restrict ourselves to quadratic order both in the bulk and in the boundary. \\ \hspace*{1.75ex}
In this section, both acts of renormalisation and deformation are woven together through the use of what we call a ``counterterm$^+$". This is obtained by multiplying renormalisation counterterms by a constant pre-factor according to
\begin{equation} \begin{aligned} \label{4.5}
	\text{counterterm}^+ \vcentcolon = \left( 1 + {\mathcal{M} \over \text{function of } \varLambda} \right) \text{counterterm,}
\end{aligned} \end{equation} 
where ${\text{counterterm} \over \text{function of } \varLambda} \sim O (\varLambda)^0$ and the coupling constant $\mathcal{M}$ can be seen as a scale controlling the deformation's magnitude. The surface terms produced in this way are sufficient for including the desired deformations.

\paragraph{}Let us now return to the derivation of the renormalisation counterterm. This can be written for both cases of ${\bar{\lambda}} < 2$ and ${\bar{\lambda}} > 2$ as
\begin{equation} \begin{aligned} \label{4.8}
	\bar{S}_{\text{ct}} & \vcentcolon = {1 \over 2} \int_{\partial \mathbb{B}_\varLambda} \mathrm{d}^d x \, r^{{\bar{\lambda}}-1} \left[  r^2 {\bar{\mathcal{F}}_{r \mu_1 ... \mu_p} \bar{\mathcal{F}}_r {}^{\mu_1 ... \mu_p} \over \kappa_1 (\varLambda)} + r^{-2} {\bar{\mathcal{F}}_{\mu_0 ... \mu_p} \bar{\mathcal{F}}^{\mu_0 ... \mu_p} \over (p+1) \kappa_2 (\varLambda)} \right] + \ldots 
\end{aligned} \end{equation} 
where $\kappa_{1,2}$ are, at first, defined by 
\begin{equation} \begin{aligned} 
	{\varLambda^{1-{\bar{\lambda}}} \over \kappa_1} & = - {\varLambda^{1-{\bar{\lambda}}} \over 1-{\bar{\lambda}}} \quad \quad \text{and} \quad \quad
	{\varLambda^{{\bar{\lambda}}-3} \over \kappa_2} & = - {\varLambda^{{\bar{\lambda}}-3} \over {\bar{\lambda}}-3} \, .
\end{aligned} \end{equation}
and the ellipsis refers to terms of higher order in $\partial_\mu \bar{\mathcal{F}}_{a_0 ... a_p}$, which we omit. These would scale as $O (\varLambda)^{-1}$ for $0 \leq {\bar{\lambda}} \leq 4$, which is acceptable since they are not required in this case. Taking $\bar{\mathcal{F}}$ to be on-shell in the equation above, the first (second) term also scales as a negative power of $\varLambda$ unless $\bar{\phi}$ is a high-degree (low-degree) $p$-form. Hence, $\bar{S}_{\text{ct}}$ agrees with the counterterms \eqref{counter1} and \eqref{counter2}. \\ \hspace*{1.75ex}
We apply the counterterm$^+$ prescription to the leading terms in \cref{4.8}, such that $\kappa_{1,2}$ are instead given by
\begin{equation} \begin{aligned} 
	{\varLambda^{1-{\bar{\lambda}}} \over \kappa_1 (\varLambda)} & = - {\varLambda^{1-{\bar{\lambda}}} \over 1-{\bar{\lambda}}} + \mathcal{M}_j \quad \quad \text{and} \quad \quad
	{\varLambda^{{\bar{\lambda}}-3} \over \kappa_2 (\varLambda)} & = - {\varLambda^{{\bar{\lambda}}-3} \over {\bar{\lambda}}-3} + \mathcal{M}_a \, .
\end{aligned} \end{equation}
Note that, when ${\bar{\lambda}}=1$ or ${\bar{\lambda}}=3$, the counterterm itself transforms non-homogeneously under a scale transformation due to the presence of a logarithm. Hence, in these cases, $\mathcal{M}_{j,a}$ can be seen as logarithmically running couplings, $\ln \varLambda^*$, corresponding to the shift of $\ln \varLambda$ under a rescaling and the functions $\kappa_{1,2} (\varLambda)$ have indeed been introduced in the past \cite{Faulkner:2012gt,Grozdanov:2017kyl,Hofman:2017vwr} as the most general solution to a renormalisation group equation. \\ \hspace*{1.75ex}
The surface terms in \cref{4.8} will be used for all theories under consideration, including ${\bar{\lambda}} = 2$. Hence, we end up with the ``final action" $\bar{S}_{\text{final}} \vcentcolon = \lim_{\varLambda \to \infty} \left[ p! \bar{S}_\varLambda + \bar{S}_{\text{ct}} \right]$, whose on-shell variation can be written as
\begin{equation} \begin{aligned} \label{4.11}
	\delta \bar{S}_{\text{final}} = \int_{\partial \mathbb{B}} \mathrm{d}^d x \left( \mathcal{M}_j J_{\mu_1 ... \mu_p} \delta J^{\mu_1 ... \mu_p} - \mathcal{M}_a \partial^{\mu_0} \beta_{\mu_0 ... \mu_p} \delta \alpha^{\mu_1 ... \mu_p} + J_{\mu_1 ... \mu_p} \delta \alpha^{\mu_1 ... \mu_p} \right) .
\end{aligned} \end{equation} 

\subsubsection{Generating functionals revisited} \label{revisited}

\paragraph{}The generating functionals $W_j$ and $W_a$ from \cref{boundarytheories} are of the form \eqref{genfuncW0}. We now provide an explicit holographic realisation for both of these in terms of $p$-form Maxwell theories. \\ \hspace*{1.75ex}
\Cref{4.11} implies that, for the deformations to be valid, one must assume that the product $\mathcal{M}_j \mathcal{M}_a$ always vanishes. Hence, introducing $a , j \in \Omega^{p}$, one can rewrite this equation as
\begin{equation} \begin{aligned} \label{4.12}
	\delta \bar{S}_{\text{final}} = \int_{\partial \mathbb{B}} {\ast j \wedge \delta a \over (d-p)!} = \int_{\partial \mathbb{B}} \mathrm{d}^d x \, j_{\mu_1 ... \mu_p} \delta a^{\mu_1 ... \mu_p} \, , 
\end{aligned} \end{equation} 
where
\begin{subequations} \label{4.13} \begin{align} 
	\label{4.13a} j_{\mu_1 ... \mu_p} & = J_{\mu_1 ... \mu_p} - \mathcal{M}_a \partial^{\mu_0} \beta_{\mu_0 ... \mu_p} \\
	\label{4.13b} a_{\mu_1 ... \mu_p} & = \alpha_{\mu_1 ... \mu_p} + \mathcal{M}_j J_{\mu_1 ... \mu_p} \, .
\end{align} \end{subequations}
The on-shell variation of the final action is to be identified with \cref{eq:4.23} when the equations of motion are satisfied. In this case, we are using the electric quantisation scheme such that $\mathcal{M}_a$ should be set to zero since $J_{\mu_1 ... \mu_p}$ is the on-shell value of the renormalised momenta at the boundary. In order to construct the corresponding path integral, it is useful to define $j [\bar{\phi}]$ and $a [\bar{\phi}]$, a pair of form-valued functionals whose components are given by
\begin{subequations} \label{5.31} \begin{align} 
	j_{\mu_1 ... \mu_p} [\bar{\phi}] & = r^{\bar{\lambda}} \bar{\mathcal{F}}_{r \mu_1 ... \mu_p} - {r^{{\bar{\lambda}}-3} \over \kappa_2 (r)} \partial^{\mu_0} \left[ \bar{\mathcal{F}}_{\mu_0 ... \mu_p} + O (\Box \bar{\mathcal{F}}_{\mu_0 ... \mu_p}) \right] \\
	a_{\mu_1 ... \mu_{p}} [\bar{\phi}] & = p! \bar{\phi}_{\mu_1 ... \mu_p} + {r \over \kappa_1 (r)} \left[ \bar{\mathcal{F}}_{r \mu_1 ... \mu_p} + O (\Box \bar{\mathcal{F}}_{r \mu_1 ... \mu_p}) \right] ,
\end{align} \end{subequations}
where $\bar{\phi}$ belongs to the integration domain. Note that the subleading terms inside the square brackets correspond to the unspecified contributions in \cref{4.8} and may be neglected when $0 \leq {\bar{\lambda}} \leq 4$. \\ \hspace*{1.75ex}
Recall that we are considering path integrals over bulk field configurations that satisfy the equations of motion at the boundary. Therefore, the functionals introduced above approach the \textit{renormalised variables} \eqref{4.13} at the boundary: $j [\bar{\phi}] \vert_{\partial \mathbb{B}} = j$ and $a [\bar{\phi}] \vert_{\partial \mathbb{B}} = a$. Importantly, $a [\bar{\phi}]$ carries the same gauge freedom as the pullback of $\bar{\phi}$ into a constant-$r$ submanifold, which is consistent with the ambiguity that $a$ inherited from $\alpha$. \\ \hspace*{1.75ex}
In the present case, describing the deformed electric quantisation of a Maxwell $p$-form, the relation \eqref{eq:4.3} between boundary and bulk generating functionals is
\begin{equation} \begin{aligned} \label{Za} 
	W_j [\psi] = \ln Z^{[{\bar{\lambda}} , \mathcal{M}_j]} (\psi) \qquad \text{where} \qquad Z^{[{\bar{\lambda}} , \mathcal{M}_j]} \vcentcolon = \int_{a [\bar{\phi}] \vert_{\partial \mathbb{B}} = \psi} \mathfrak{D} \bar{\phi} e^{{i \over \kappa^2} \bar{S}_{\text{final}}} .
\end{aligned} \end{equation}
The double-trace deformation is 
\begin{equation} \begin{aligned} \label{defM1}
	{\mathcal{M}_j \over 2} \int_{\partial \mathbb{B}} \mathrm{d}^d x \, \mathcal{O}^j_{\mu_1 ... \mu_p} \mathcal{O}_j^{\mu_1 ... \mu_p} \, , 
\end{aligned} \end{equation} 
which is the sole contribution at leading order in a gradient expansion --- unless ${\bar{\lambda}} = 1$, in which case the topological term $\int_{\partial \mathbb{B}} \mathcal{O}_j \wedge \mathcal{O}_j$ is also available. The coupling constant has energy dimension $[\mathcal{M}_j] = 1 - {\bar{\lambda}}$.

\paragraph{}As discussed in \cref{MagnQuant_massless}, switching quantisation amounts to a Legendre transformation of the deformed action. In this case, the boundary and bulk generating functionals are therefore related via
\begin{equation} \begin{aligned} \label{Zj} 
	W_a [\psi] = \ln Z^{[{\bar{\lambda}} , \mathcal{M}_a]} (\psi) \qquad \text{where} \qquad Z^{[{\bar{\lambda}} , \mathcal{M}_a]} \vcentcolon = \int_{j [\bar{\phi}] \vert_{\partial \mathbb{B}} = \psi} \mathfrak{D} \bar{\phi} e^{{- i \over \kappa^2} \left( \bar{S}_{\text{final}} - \int_{\partial \mathbb{B}} {\ast j [\bar{\phi}] \wedge a [\bar{\phi}] \over (d-p)!} \right)} .
\end{aligned} \end{equation}
Here, it is $\mathcal{M}_j$ that should be set to zero since a proper deformation should depend on the functional $a [\phi]$ dual to the single-trace operator $\mathcal{O}^a$. This time, the double-trace deformation is 
\begin{equation} \begin{aligned} \label{defM2}
	{\mathcal{M}_a \over 2 p!} \int_{\partial \mathbb{B}} \mathrm{d}^d x \, \partial_{[ \mu_0} \mathcal{O}^a_{\mu_1 ... \mu_p]} \partial^{\mu_0} \mathcal{O}_a^{\mu_1 ... \mu_p} \, ,
\end{aligned} \end{equation} 
and the coupling constant has energy dimension $[\mathcal{M}_a] = {\bar{\lambda}} - 3$.
Note that, unlike $\mathcal{O}_j$, the single-trace operator $\mathcal{O}^a$ transforms as a Goldstone $p$-form. Hence, in magnetic quantisation, we discard Robin boundary conditions since gauge invariance forbids double-trace deformations from appearing at zeroeth order in gradients.
Accordingly, we have considered the leading gauge-invariant deformation available for general ${\bar{\lambda}}$.\footnote{A topological contribution of the form $\int_{\partial \mathbb{B}} \mathrm{d} \mathcal{O}^a \wedge \mathrm{d} \mathcal{O}^a$ exists for ${\bar{\lambda}} = 3$. Both this and \eqref{defM2} are subleading in the gradient expansion when compared with $\int_{\partial \mathbb{B}} \mathcal{O}^a \wedge \mathrm{d} \mathcal{O}^a$, available for ${\bar{\lambda}} = 2$, which is also gauge-invariant.}
The attentive reader may have noticed in \cref{Zj} that we are not using the Legendre term $\ast j [\bar{\phi}] \wedge a [\bar{\phi}] \big\vert_{\mathcal{M}_{j,a} = 0}$, as would be expected from \cref{MagnQuant_massless}. This merely results in a sign flip on the deformation, which can be consistently absorbed into $\mathcal{M}_a$.

\paragraph{}Lastly, we point out that the double-trace deformation \eqref{defM1} is relevant when ${\bar{\lambda}} < 1$ and marginal when ${\bar{\lambda}} = 1$.\footnote{Marginally relevant or irrelevant depending on the sign of the coupling \cite{Faulkner:2012gt}.} Similarly, \eqref{defM2} is relevant when ${\bar{\lambda}} > 3$ and marginal when ${\bar{\lambda}} = 3$. These are exactly the instances where the deformations arise from a counterterm$^+$. Note that the only reason the counterterm$^+$ prescription provided all leading non-topological deformations allowed was because we considered the entire family of theories at the same time. Had we focused on a specific ${\bar{\lambda}} \neq 2$, only one of the double-trace deformations in \cref{defM1,defM2} would arise from a genuine counterterm$^+$. In fact, none would if ${\bar{\lambda}} = 2$, precisely because in this case $\bar{S}_{\text{ct}}$ does not represent a holographic counterterm, as such is not necessary. When ${\bar{\lambda}} > 1$ and ${\bar{\lambda}} < 3$, the double-trace deformations \eqref{defM1} and \eqref{defM2} are irrelevant, respectively. 

\section{Approximate symmetries} \label{WeaklyBroken_sec}

\paragraph{}We are now ready to tackle the main topic to which this work contributes new results. Let us then consider the exact symmetries from previous section to be explicitly broken. Note that we borrow most of the structure of that section. One of the key differences is that we eventually restrict to effective holographic models with bulk St\"uckelberg fields. The resulting theory is that of massive higher-form gauge fields subject to Robin boundary conditions. Also, in performing holographic renormalisation, we encounter a two-fold ambiguity which allows us to implement both quantisation schemes without the need for an explicit Legendre transform.

\subsection{Bulk path integrals with defects} \label{LinModels}

 \paragraph{}In \cref{boundarytheories}, we discussed the breaking of higher-form symmetries associated with the introduction of defects in the system. The defect charge is itself a topological operator, associated with a (one-degree-lower) higher-form symmetry. Specifically, we saw how dynamical and background non-conserved currents corresponded to explicitly broken electric and magnetic symmetries, respectively. In the latter case, in particular, the fundamental operators are Goldstone fields, one of which ceased to be gauge-covariant.\footnote{As previously explained, one can also take the point of view where both Goldstone fields transform appropriately but one of them is multivalued.}
Similarly to \cref{genfuncW0} in \cref{HologPathInt}, we start by expressing the generating functionals $\mathcal{W}_j$ and $\mathcal{W}_a$ for correlators of $(\mathcal{O}_j , \tilde{\mathcal{O}}_j)$ and $(\mathcal{O}^a , \tilde{\mathcal{O}}^a)$ as expectation values of the form 
\begin{equation} \begin{aligned} 
	\mathcal{W} [\psi , \tilde{\psi}] \equiv \expval{1} [\psi , \tilde{\psi}] \vcentcolon = \expval{ \exp( i \mathcal{C} (N) \int_{\partial \mathbb{B}} \left[ {\ast \mathcal{O} \wedge \psi \over (d-p)!} + {\ast \tilde{\mathcal{O}} \wedge \tilde{\psi} \over (d-p-1)!} \right])}_0 .
\end{aligned} \end{equation}
Our goal is to, departing from the models discussed in \cref{HologPathInt}, achieve a holographic realisation of these generating functionals. Hence, we begin by considering a second field $\tilde{\phi}_{\mathrm{A}'}$ in the bulk. The coupling between both fields is encoded in the action functional $\mathcal{S} \equiv \bar{\mathcal{S}} [\phi] + \tilde{\mathcal{S}} [\phi , \tilde{\phi}]$, whose variation takes the form
\begin{equation} \begin{aligned} \label{1.50}
	\delta \mathcal{S} & = \int_{\mathbb{B}} \left( \mathcal{E}^\mathrm{A} \delta \phi_\mathrm{A} + \tilde{\mathcal{E}}^{\mathrm{A}'} \delta \tilde{\phi}_{\mathrm{A}'} \right) + \int_{\partial \mathbb{B}} \left( \Upsilon^\Xi \delta \Phi_\Xi + \tilde{\Upsilon}^{\Xi'} \delta \tilde{\Phi}_{\Xi'} \right) .
\end{aligned} \end{equation}
Similarly to \cref{setup}, we extend the range of applicability of this holographic model beyond higher-form symmetries, by considering the bulk fields to be tensors in any irreducible representation of $\mathrm{GL} (d{+}1 , \mathbb{R})$. This assumption is then carried by the compact index notation. \\ \hspace*{1.75ex}
Since the boundary symmetry is broken, we propose to combine \eqref{eq:4.8} with a shift of $\tilde{\phi}$ that depends algebraically on the gauge parameter. Hence, the bulk action $\mathcal{S}$ is to be invariant under
\begin{equation} \begin{aligned} \label{164} 
	\begin{cases}
		\delta_\xi \phi_\Xi = \partial_{ \{ \mu} \xi_{\Xi' \} } \\ 
		\delta_\xi \phi_{r \Xi'} = \Gamma_{\Xi'} (\xi) + \partial_r \xi_{\Xi'} \\
		\delta_\xi \tilde{\phi}_{A'} = - \Theta^{\Xi'}_{\mathrm{A}'} [\phi , \tilde{\phi}] \xi_{\Xi'} \, ,
	\end{cases}
\end{aligned} \end{equation}
where $\Theta$ is a matrix-valued functional.\footnote{Abelian Higgs theory \cite{Englert:1964et,Higgs:1964pj,Guralnik:1964eu} is invariant under \eqref{164} given that the vector and scalar fields correspond to $\phi$ and $\tilde{\phi}$, respectively, and $\Theta = \Theta [\tilde{\phi}]$ is proportional to $\tilde{\phi}$.} To simplify expressions while maintaining the key features that arise when $\delta_\xi \tilde{\phi}_{\mathrm{A}'}$ has some functional dependence, we assume that $\Theta^{\Xi'}_{\Theta'} = \theta [\phi , \tilde{\phi}] \delta^{\Xi'}_{\Theta'}$. Hence, a large gauge symmetry under \eqref{164} implies that
\begin{equation} \begin{aligned} \label{168} 
	\int_{\mathbb{B}} \mathcal{H}^{\Xi'} [\mathcal{E} , \tilde{\mathcal{E}} , \Theta] \xi_{\Xi'} + \int_{\partial \mathbb{B}} \left( \partial_\mu \Upsilon^{\mu \Xi'} + \theta \tilde{\Upsilon}^{\Xi'} - \mathcal{E}^{r \Xi'} \right) \xi_{\Xi'} = 0 \, .
\end{aligned} \end{equation}
Here and below, $\mathcal{H}$ and $\tilde{\mathcal{H}}$ denote functionals whose explicit form plays no role in the present discussion, similarly to $H$ in \eqref{eq:13}. Note that $\tilde{\mathcal{S}}$ introduces a current associated with $\theta \tilde{\Upsilon}^{\Xi'}$ in the boundary theory. For this to be conserved (following \cref{boundarytheories}), the action $\mathcal{S}$ should also remain invariant under
\begin{equation} \begin{aligned} \label{165} 
	\begin{cases}
		\delta_{\tilde{\xi}} \tilde{\phi}_{\Xi'} = - \theta \partial_{ \{ \mu} \tilde{\xi}_{\Xi'' \} } \\ 
		\delta_{\tilde{\xi}} \tilde{\phi}_{r \Xi''} = \tilde{\Gamma}_{\Xi''} (\tilde{\xi}) - \tilde{\Theta}_{\Xi''}^{\Theta''} \partial_r \tilde{\xi}_{\Theta''} \, ,
	\end{cases}
\end{aligned} \end{equation}
where $\tilde{\Theta}$ is also a matrix-valued functional and we assume, for simplicity, that $\tilde{\Theta}^{\Xi''}_{\Theta''} = \tilde{\theta} [\phi , \tilde{\phi}] \delta^{\Xi''}_{\Theta''}$. Note that $\tilde{\Gamma}$ has been introduced by analogy with $\Gamma$ in \eqref{eq:4.8}.\footnote{Not only $\tilde{\Gamma}$, but also $\Gamma$, may exhibit non-trivial $[\phi , \tilde{\phi}]$-dependence without affecting the results of this section. In fact, all such dependence --- but $\theta$'s --- contributes only at order $O (\kappa^2)$.} 
Since \eqref{165} is a large gauge transformation, it follows that
\begin{equation} \begin{aligned} \label{169}
	\int_{\mathbb{B}} \tilde{\mathcal{H}}^{\Xi''} [\tilde{\mathcal{E}} , \Theta , \tilde{\Theta}] \tilde{\xi}_{\Xi''} + \int_{\partial_\mathbb{B}} \left( \partial_\mu \left( \theta \tilde{\Upsilon}^{\mu \Xi''} \right) - \tilde{\theta} \tilde{\mathcal{E}}^{r \Xi''} \right) \tilde{\xi}_{\Xi''} = 0 \, .
\end{aligned} \end{equation}

\paragraph{}Before addressing the electric and magnetic quantisation of $\mathcal{S}$, we would like to consider this theory linearised around a solution $(\phi , \tilde{\phi}) = (\phi_{(0)} , \tilde{\phi}_{(0)})$ to the equations of motion. The linearisation of the bulk theory has a counterpart at the boundary, whose background is dual to $(\phi_{(0)} , \tilde{\phi}_{(0)})$. Since we are ultimately interested in approximate symmetries, we want such a background to display an intact symmetry which will be weakly broken by linearised perturbations. Hence, as a sufficient condition for $\Upsilon^{\mu \Xi'}_{(0)} \big\vert_{\partial \mathbb{B}}$ to be divergenceless, we require
\begin{equation} \begin{aligned} \label{eq:6.19}
	\tilde{\Upsilon}^{\Xi'}_{(0)} \big\vert_{\partial \mathbb{B}} = 0 \, .
\end{aligned} \end{equation}
A subscript $(n)$ was introduced to denote that, after performing the substitution $\phi \to \phi_{(0)} + \phi$ and $\tilde{\phi} \to \tilde{\phi}_{(0)} + \tilde{\phi}$ in the given functional, only terms up to $n^\text{th}$-order in $\phi$ and $\tilde{\phi}$ were retained. 
The variation of the action $\mathcal{S}_{(2)}$ of the linearised theory is given by
\begin{equation} \begin{aligned} \label{1.71}
	\delta \mathcal{S}_{(2)} & = \int_{\mathbb{B}} \left( \mathcal{E}_{(1)}^\mathrm{A} \delta \phi_\mathrm{A} + \tilde{\mathcal{E}}_{(1)}^{\mathrm{A}'} \delta \tilde{\phi}_{\mathrm{A}'} \right) + \int_{\partial \mathbb{B}} \left( \Upsilon_{(1)}^\Xi \delta \Phi_\Xi + \tilde{\Upsilon}_{(1)}^{\Xi'} \delta \tilde{\Phi}_{\Xi'} \right) .
\end{aligned} \end{equation}
In addition to \cref{eq:6.19}, we have $\mathcal{E}_{(0)}^\mathrm{A} = \tilde{\mathcal{E}}_{(0)}^{\mathrm{A}'} = 0$ such that, in the equation above, only $\Upsilon_{(1)}^\Xi$ might contain $0^\text{th}$-order terms through $\Upsilon_{(0)}^\Xi$. 

\paragraph{}Because $\Theta$ and $\tilde{\Theta}$ always multiply either equations of motion $\tilde{\mathcal{E}}_{(0)}^{\mathrm{A}'}$ or the renormalised momenta $\tilde{\Upsilon}^{\Xi'}$, the linearisation of \cref{168,169} features neither $\Theta_{(1)}$ nor $\tilde{\Theta}_{(1)}$ and the action $\mathcal{S}_{(2)}$ is invariant under large gauge transformations which can be obtained by substituting
\begin{equation} \begin{aligned}
	\Theta \to \Theta_{(0)} \quad \quad \text{and} \quad \quad \tilde{\Theta} \to \tilde{\Theta}_{(0)} 
\end{aligned} \end{equation}
in \cref{164,165}. Hence, $\mathcal{S}_{(2)}$ can simply be seen as a subclass of theories $\mathcal{S}$ whose Lagrangian has no terms of order higher than quadratic and for which $\Theta$ and $\tilde{\Theta}$ carry no functional dependence. This will serve as motivation for \cref{MassiveEOM}, where we take the bulk action to be Einstein-Proca. (Below, only when investigating electric quantisation we will for the sake of generality allow $\Theta$, and consequently $\theta$, to depend on the bulk fields.) 

\subsubsection{Electric quantisation} 

\paragraph{}Recalling \cref{boundarytheories}, we propose that the generating functional $\mathcal{W}_j$ corresponds to the electric quantisation of the theory given by $\mathcal{S}$, i.e.
\begin{equation} \begin{aligned} 
	\mathcal{W}_j [\psi , \tilde{\psi}] \equiv \expval{ \exp( i \mathcal{C} (N) \int_{\partial \mathbb{B}} \left( \mathcal{O}^\Xi_j \psi_\Xi + \tilde{\mathcal{O}}^{\Xi'}_j \tilde{\psi}_{\Xi'} \right) )}_0 = \ln Z_\varPhi (\psi , \tilde{\psi}) \, ,
\end{aligned} \end{equation}
where the bulk generating functional is given by the path integral
\begin{equation} \begin{aligned}
	Z_\varPhi = \int_{(\Phi , \tilde{\Phi}) \vert_{\partial \mathbb{B}} = (\psi , \tilde{\psi})} \mathfrak{D} \phi \mathfrak{D} \tilde{\phi} \exp(i \mathcal{S} / \kappa^2) \, .
\end{aligned} \end{equation}
Our proposal does not hold for general $\theta$, as will become clear. We have imposed Dirichlet boundary conditions since, in the interest of clarity, deformations are being neglected. Insertions in the path integral satisfy convention \hyperlink{conv1}{1}, as before, and are denoted by $\braket{X}_\varPhi \vcentcolon = (Z_\varPhi)^{-1} \int_{(\Phi , \tilde{\Phi}) \vert_{\partial \mathbb{B}} = (\psi , \tilde{\psi})} \mathfrak{D} \phi \mathfrak{D} \tilde{\phi} e^{{i \over \kappa^2} \mathcal{S}} X$. 

\paragraph{}Field redefinition invariance of $Z_\varPhi$ under $\Phi_\mathrm{A} \to \Phi_\mathrm{A} + \delta \varphi^*_\mathrm{A} (x)$ and $\tilde{\Phi}_{\mathrm{A}'} \to \tilde{\Phi}_{\mathrm{A}'} + \delta \tilde{\varphi}^*_{\mathrm{A}'} (x)$ implies that $\braket{\mathcal{E}^\mathrm{A}}_\varPhi$ and $\braket{\tilde{\mathcal{E}}^{\mathrm{A}'}}_\varPhi$ must vanish. This time, one also needs to consider insertions of composite functionals involving matrices $\Theta$/$\tilde{\Theta}$ and equations of motion $\tilde{\mathcal{E}}$ in the path integral. All the necessary insertions, such as $\braket{\theta \tilde{\mathcal{E}}^{\Xi'}}_\varPhi$, vanish up to $O (\kappa^2)$ corrections. To see this, apply field redefinition invariance once more, replacing $\delta \tilde{\varphi}^*_{\Xi'}$ with $\theta \delta \tilde{\varphi}^*_{\Xi'}$; one then obtains
\begin{equation} \begin{aligned} 
	{i \over \kappa^2} \int_{\mathbb{B}} \braket{{\theta} \tilde{\mathcal{E}}^{\Xi'}}_\varPhi \delta \tilde{\varphi}^*_{\Xi'} + \braket{\delta \mathbb{J}}_\varPhi = 0 
	\Rightarrow \braket{{\theta} \tilde{\mathcal{E}}^{\Xi'}}_\varPhi = O (\kappa^2) \, ,
\end{aligned} \end{equation}
where $\delta \mathbb{J}$ denotes the linear part (in $\delta \tilde{\varphi}^*_{\Xi'}$) of the Jacobian $\mathbb{J}$, associated with the field redefinition.\footnote{The non-radial components $\delta \varphi^*_\Xi$ and $\delta \tilde{\varphi}^*_{\Xi'}$ vanish at the boundary. Since $\delta \varphi^*_\mathrm{R} \vert_{\partial \mathbb{B}}$ and $\delta \tilde{\varphi}^*_{\mathrm{R}'} \vert_{\partial \mathbb{B}}$ are unconstrained, showing that e.g. $\braket{\tilde{\theta} \tilde{\mathcal{E}}^{\mathrm{R}'}}_\varPhi = O (\kappa^2)$ does not rely on the configurations we are integrating over satisfying the radial equations of motion at the boundary.} It should be noted that, previously, we considered only shifts that were independent of the fields being integrated, implying a unit Jacobian. 

\paragraph{}Inserting \cref{168,169} in the path integral, one obtains
\begin{subequations} \begin{align} 
	\label{1.70a} &	\partial_\mu \braket{\Upsilon^{\mu \Xi'}}_\varPhi + \braket{\theta \tilde{\Upsilon}^{\Xi'}}_\varPhi = O (\kappa^2) \\
	\label{1.70b} & \partial_\mu \braket{\theta \tilde{\Upsilon}^{\mu \Xi''}}_\varPhi = O (\kappa^2) \, .
\end{align} \end{subequations} 
Taken together, the equations above imply that $\partial_\nu \partial_\mu \braket{\Upsilon^{\mu \nu \Xi''}}_\varPhi$ vanishes up to corrections.
The linearised perturbation of $\ln Z_\varPhi$ is given by
\begin{equation} \begin{aligned}
	- i \kappa^2 \delta \ln Z_\varPhi = \int_{\partial \mathbb{B}} \left( \braket{\Upsilon^\Xi}_\varPhi \delta \psi_\Xi + \braket{\tilde{\Upsilon}^{\Xi'}}_\varPhi \delta \tilde{\psi}_{\Xi'} \right) ,
\end{aligned} \end{equation}
such that the generating functional is invariant, in the large-$N$ limit, under $(\delta \psi_\Xi , \delta \tilde{\psi}_{\Xi'}) = (\partial_{ \{ \mu} \zeta_{\Xi' \} } , - \braket{\theta}_\varPhi \zeta_{\Xi'})$ and $\delta \tilde{\psi}_{\Xi'} = - \braket{\theta}_\varPhi \partial_{ \{ \mu} \tilde{\zeta}_{\Xi'' \} }$. Using the relation between boundary and bulk generating functionals, $Z_\varPhi (\psi , \tilde{\psi}) = \mathcal{W}_j [\psi , \tilde{\psi}]$, it follows that
\begin{equation} \begin{aligned}
	\braket{\Upsilon^\Xi}_\varPhi = \braket{\mathcal{O}^\Xi_j} \quad \quad \text{and} \quad \quad \braket{\tilde{\Upsilon}^{\Xi'}}_\varPhi = \braket{\tilde{\mathcal{O}}^{\Xi'}_j} .
\end{aligned} \end{equation}
Hence, focusing on the case of antisymmetric tensor fields, our setup allows for the explicit breaking of the original electric higher-form symmetry. In particular, the (non-)conservation equation is of the form \eqref{Oj_nonconserv} when $\theta$ is a number, in which case the large-$N$ conservation of the defect current signals the expected one-degree-lower symmetry. Moreover, as long as $\theta$ carries no functional dependence, \cref{Oj_nonconserv} can be recovered through a field redefinition. Beyond this, conclusions must be drawn on a case-by-case basis. For instance, if $\theta$ is mapped under the holographic dictionary to a source-dependent quantity, then the resulting theories cannot be accommodated in the class $\mathcal{W}_j$ introduced in \cref{boundarytheories}.

\subsubsection{Magnetic quantisation}

\paragraph{}We now show how the magnetic quantisation of $\mathcal{S}$ provides a holographic realisation of the generating functional $\mathcal{W}_a$ from \cref{boundarytheories}. Following the prescription of \cref{MagnQuant_massless}, we consider a path integral $Z_\varUpsilon$ obtained from $Z_\varPhi$ through a Legendre transformation. Hence,
\begin{equation} \begin{aligned} 
	\mathcal{W}_a [\psi , \tilde{\psi}] \equiv \expval{ \exp( i \mathcal{C} (N) \int_{\partial \mathbb{B}} \left( \mathcal{O}_\Xi^a \psi^\Xi + \tilde{\mathcal{O}}_{\Xi'}^a \tilde{\psi}^{\Xi'} \right) )}_0 = \ln Z_\varUpsilon (\psi , \tilde{\psi}) \, .
\end{aligned} \end{equation}
Deformations are being neglected here so that we impose Neumann boundary conditions with $\Upsilon$ and $\tilde{\Upsilon}$ fixed at the boundary. Here, we consider the magnetic quantisation of $\mathcal{S}$ in the case where $\theta$ carries no functional dependence and should be regarded as an ordinary function.

\paragraph{}As expected, the holographic dictionary maps $\Phi_\Xi$ and $\tilde{\Phi}_{\Xi'}$ respectively to the single-trace operators $\mathcal{O}_\Xi^a$ and $\tilde{\mathcal{O}}_{\Xi'}^a$, which can be obtained via
\begin{equation} \label{2.54} \begin{aligned}
	\braket{\mathcal{O}_\Xi^a} & = - i \kappa^2 {\delta \ln Z_\varUpsilon \over \delta \psi^\Xi} \quad \quad \text{and} \quad \quad
	\braket{\tilde{\mathcal{O}}_{\Xi'}^a} = - i \kappa^2 {\delta \ln Z_\varUpsilon \over \delta \tilde{\psi}^{\Xi'}} \, .
\end{aligned} \end{equation}
The insertions of the renormalised momenta in the path integral are fixed according to $\braket{\Upsilon^\Xi}_\varUpsilon = \psi^\Xi$ and $\braket{\tilde{\Upsilon}^{\Xi'}}_\varUpsilon = \tilde{\psi}^{\Xi'}$. Substituting these in \cref{1.70a,1.70b}, we obtain
\begin{equation} \begin{aligned} 
	\partial_\mu \psi^{\mu \Xi'} + \theta \tilde{\psi}^{\Xi'} = O (\kappa^2) \quad \quad \text{and} \quad \quad \partial_\mu \left( \theta \tilde{\psi}^{\mu \Xi''} \right) = O (\kappa^2) \, .
\end{aligned} \end{equation}
This leads to the boundary observables being equivalence classes, as in \cref{boundarytheories}. In the large-$N$ limit, they consist of
\begin{equation} \begin{aligned} 
	[(\mathcal{O}_\Xi^a , \tilde{\mathcal{O}}_{\Xi'}^a)] & = \{ (\mathcal{O}_\Xi^a + \partial_{ \{ \mu} \zeta_{\Xi' \} } \, , \, \tilde{\mathcal{O}}_{\Xi'}^a - \theta \zeta_{\Xi'} - \theta \partial_{ \{ \mu} \tilde{\zeta}_{\Xi'' \} }) \; | \; \zeta_{\Xi'} , \tilde{\zeta}_{\Xi''} \in C^\infty (\partial \mathbb{B}) \} \, .
\end{aligned} \end{equation}
In fact, if we choose to work with differential forms, $\partial_{ \{ \mu} \tilde{\zeta}_{\Xi'' \} }$ can be absorbed into $\zeta_{\Xi'}$ and we obtain
\begin{equation} \begin{aligned} 
	[(\mathcal{O}^a , \tilde{\mathcal{O}}^a)] & = \{ (\mathcal{O}^a + \mathrm{d} \zeta \, , \, \tilde{\mathcal{O}}^a - \theta \zeta) \; | \; \zeta \in C^\infty \Omega^{p-1} \} \, ,
\end{aligned} \end{equation}
which is equivalent to \eqref{equivclasses_broken} as we can scale $(\tilde{\mathcal{O}}^a , \tilde{\psi}) \to (\theta \tilde{\mathcal{O}}^a , \theta^{-1} \tilde{\psi})$ without altering $\mathcal{W}_a$. 

\subsubsection{St\"uckelberg fields} \label{Stuckelberg}

\paragraph{}Lastly, before proceeding, it is useful to consider theories $\mathcal{S}$ for which $\tilde{\phi}$ is a St\"uckelberg tensor field \cite{Stueckelberg:1938hvi,Ruegg:2003ps}.\footnote{St\"uckelberg scalar fields have been used in holography to describe the chiral anomaly in the boundary \cite{Klebanov:2002gr}. In \cite{Jimenez-Alba:2014iia}, they were studied in the context of anomalous response with non-conserved currents.} For this to be the case, the shift of $\tilde{\phi}$ under \eqref{164} should be enhanced to
\begin{equation} \begin{aligned} 
	\delta_\xi \tilde{\phi}_{A'} = - \Theta^{\mathrm{B}'}_{\mathrm{A}'} \xi_{\mathrm{B}'} \, ,
\end{aligned} \end{equation}
where $\Theta$ is a non-singular matrix of functions. The shift of $\phi$ should also be enhanced so it depends on $\xi_{\mathrm{R}'}$, otherwise the action is simply independent of $(\Theta^{-1})^{\mathrm{A}'}_{\mathrm{R}'} \tilde{\phi}_{A'}$; for concreteness, we are going to consider the large gauge transformation
\begin{equation} \begin{aligned} \label{eq:6.23}
	(\delta_\xi \phi_\mathrm{A} , \delta_\xi \tilde{\phi}_{\mathrm{A}'}) & = (\partial_{ \{a} \xi_{\mathrm{A}'\} } , - \theta \xi_{\mathrm{A}'}) \, .
\end{aligned} \end{equation}
If we use $\xi_{\mathrm{A}'} = \tilde{\phi}_{\mathrm{A}'} / \theta$, this immediately implies that $\mathcal{S} [\phi , \tilde{\phi}] = \mathcal{S} [\hat{\phi} , 0]$ where
\begin{equation} \begin{aligned} \label{2.60}
	\hat{\phi}_{a \mathrm{A}'} = \phi_{a \mathrm{A}'} + \partial_{ \{a} \left( \tilde{\phi}_{\mathrm{A}'\} } / \theta \right) .
\end{aligned} \end{equation}
At least in the large-$N$ limit, one can easily see that integrating over both $\phi$ and $\tilde{\phi}$ is equivalent to integrating solely over $\hat{\phi}$. Note that invariance of the action under \eqref{eq:6.23} implies that
\begin{equation} \begin{aligned} \label{eq:1.72a} 
	\int_{\mathbb{B}} \left( \partial_a \mathcal{E}^{a \mathrm{A}'} + \theta \tilde{\mathcal{E}}^{\mathrm{A}'} \right) \xi_{\mathrm{A}'} + \int_{\partial \mathbb{B}} \left( \delta^{\mathrm{A}'}_{\Xi'} \left( \partial_\mu \Upsilon^{\mu \Xi'} + \theta \tilde{\Upsilon}^{\Xi'} \right) - \mathcal{E}^{r \mathrm{A}'} \right) \xi_{\mathrm{A}'} = 0 \, .
\end{aligned} \end{equation}
Writing $\Phi_\Xi$ and $\tilde{\Phi}_{\Xi'}$ explicitly in \eqref{1.50} generates a boundary counterterm. If we then use the equation above with $\xi_{\mathrm{A}'} = \delta \tilde{\phi}_{\mathrm{A}'} / \theta$, we obtain
\begin{equation} \begin{aligned} \label{2.61}
	\delta \mathcal{S} & = \int_\mathbb{B} \mathcal{E}^\mathrm{A} \delta \hat{\phi}_\mathrm{A} + \int_{\partial \mathbb{B}} \left( \Upsilon^\Xi \delta \hat{\phi}_\Xi + \mathscr{L}_{\text{counterterms}} (\Upsilon^\Xi) \right) \\
	& \equiv \int_\mathbb{B} \mathcal{E}^\mathrm{A} \delta \hat{\phi}_\mathrm{A} + \int_{\partial \mathbb{B}} \Upsilon^\Xi \delta \hat{\Phi}_\Xi \, .
\end{aligned} \end{equation}
Therefore, when $\tilde{\phi}$ is a St\"uckelberg field, we can use the setup from \cref{HologPathInt} with path integrals $Z_F$ over $\hat{\phi}$.\footnote{Deformations are now included once again.} These can be divided into two sets according to the absence or presence of a Legendre transformation, i.e. either
\begin{equation} \begin{aligned} \label{delZ}
	\delta \ln Z_F (\psi) = {i \over \kappa^2} \int_{\partial \mathbb{B}} \braket{\Upsilon^\Xi}_F \delta \psi_\Xi 
	\qquad \text{or} \qquad
	\delta \ln Z_F (\psi) = {i \over \kappa^2} \int_{\partial \mathbb{B}} \braket{\hat{\Phi}_\Xi}_F \delta \psi^\Xi \, .
\end{aligned} \end{equation}
Translating such variations into the boundary theory we see that, at this point, we are in the situation described at the end of \cref{boundarytheories} --- cf. \eqref{generic_delW} --- of a single non-conserved, gauge-invariant operator. In the following, however, we are still going to label path integrals (of this kind) according to quantisation --- electric or magnetic with respect to a certain symmetry --- by choosing this to be the symmetry whose degree of breaking we have control over.

\paragraph{Strong/weak duality for Robin boundary conditions}
Although it will not play a role in the rest of the paper, there is something important one can learn by considering an extra surface term in the path integral such that, instead of \eqref{Z_F}, we have
\begin{equation} \begin{aligned} 
	Z_F (\psi) \vcentcolon = \int_{F \vert_{\partial \mathbb{B}} = \psi} \mathfrak{D} \hat{\phi} e^{{i \over \kappa^2} \left[ \mathcal{S} + \int_{\partial \mathbb{B}} (D_F + \mathcal{D}_F) \right]} \, .
\end{aligned} \end{equation}
where $\mathcal{D}_F = \mathcal{D}_F (F)$.
Since $F \vert_{\partial \mathbb{B}}$ is held fixed in the path integral, this would simply rescale the generating functional by $e^{{i \over \kappa^2} \int_{\partial \mathbb{B}} \mathcal{D}_F (\psi)}$ --- leaving insertions in the path integral unchanged as they are normalised. At the end, it would only contribute contact terms to the correlation functions that are obtained by differentiating $\ln Z_F$ with respect to $\psi$. However, such surface terms can still help us in gaining insight into the relation between different quantisation schemes. 

\paragraph{}Consider a pair of generating functionals in distinct quantisations --- $Z^{[1]}$ and $Z^{[2]}$ --- with Robin boundary conditions $(\hat{\Phi}_\Xi - \mathcal{M} \Upsilon_\Xi) \vert_{\partial \mathbb{B}} = \psi_\Xi$, where $\mathcal{M} \neq 0$ is some constant. One is given by
\begin{equation} \begin{aligned}
	Z^{[1]} = \int_{F^{[1]} \vert_{\partial \mathbb{B}} = \psi} \mathfrak{D} \hat{\phi} e^{{i \over \kappa^2} \left( \bar{\mathcal{S}} +\int_{\partial \mathbb{B}} D^{[1]}_F \right)} ,
\end{aligned} \end{equation}
where $D^{[1]}_F = - \mathcal{M} \Upsilon_\Xi \Upsilon^\Xi / 2$ such that $F^{[1]}_\Xi = \hat{\Phi}_\Xi - \mathcal{M} \Upsilon_\Xi$, while for the other we have
\begin{equation} \begin{aligned}
	Z^{[2]} = \int_{F_{[2]} \vert_{\partial \mathbb{B}} = \psi / \mathcal{M}} \mathfrak{D} \hat{\phi} e^{{i \over \kappa^2} \left( \bar{\mathcal{S}} + \int_{\partial \mathbb{B}} D^{[2]}_F \right)} ,
\end{aligned} \end{equation}
where $D^{[2]}_F = \mathcal{M}^{-1} \hat{\Phi}_\Xi \hat{\Phi}^\Xi / 2 - \hat{\Phi}_\Xi \Upsilon^\Xi$ such that $F_{[2]}^\Xi = \hat{\Phi}^\Xi / \mathcal{M} - \Upsilon^\Xi$. Note that $D^{[2]}_F - D^{[1]}_F = \mathcal{D}_F$ where 
\begin{equation} \begin{aligned}
	\mathcal{D}_F = {\left( \hat{\Phi}_\Xi - \mathcal{M} \Upsilon_\Xi \right) \left( \hat{\Phi}^\Xi - \mathcal{M} \Upsilon^\Xi \right) \over 2 \mathcal{M}} \, ,
\end{aligned} \end{equation}
such that adding a term $\mathcal{D}_F = \mathcal{D}_F (F^{[1]})$ to the first generating functional yields the second one. 
Hence, under identical Robin boundary conditions, correlation functions in different quantisations differ only by contact terms. This can be interpreted as a strong/weak-coupling duality since the couplings of the double-trace deformations that implement such boundary conditions in $Z^{[1]}$ and $Z^{[2]}$ are $- \mathcal{M}$ and $\mathcal{M}^{-1}$.\footnote{These dualities are closely related to those of \cite{Leigh:2003ez}. Recently, they have been discussed in \cite{Karch:2025hof,Grozdanov:2025ulc,Pinheiro:2025fqg}.}
As a final remark, this duality does not apply to the $p$-form Maxwell theories encoding an exact symmetry, for which Robin boundary conditions are ruled out in magnetic quantisation. 

\subsection{Higher St\"uckelberg model} \label{MassiveEOM}

\paragraph{}Having identified the bulk gauge symmetries that bring about broken continuous global symmetries at the boundary, we wish to consider a concrete $p$-form theory with the appropriate gauge-invariant action.\footnote{As before, we neglect topological contributions.}
In the case of exact symmetries (cf. \cref{HologPathInt}), Maxwell theories were ready to be used as the dual to a system with U(1) higher-form symmetries. However it is instructive to consider the process that leads to such theories: starting with the transformations in \cref{eq:4.8} for $\bar{\phi} \in \Omega^p (\mathbb{B})$, covariantise them such that $\delta_\xi \phi = \mathrm{d} \xi$ and enumerate the gauge invariant combinations involving the minimum number of derivatives --- in this case, it is only $\mathrm{d} \bar{\phi}$. The Maxwell lagrangian consists of the quadratic non-topological terms made out of $\mathrm{d} \bar{\phi}$, i.e. $\star \mathrm{d} \bar{\phi} \wedge \mathrm{d} \bar{\phi}$. The same procedure can be used for broken symmetries. Motivated by the linearised theories $\mathcal{S}_{(2)}$ from \cref{LinModels}, we start from transformations \eqref{164} and \eqref{165} for $\phi \in \Omega^p (\mathbb{B})$ and $\tilde{\phi} \in \Omega^{p-1} (\mathbb{B})$, in the case where $\Theta$ and $\tilde{\Theta}$ have no functional dependence. Covariantising these, we obtain \cref{eq:6.23}, which we rewrite as
\begin{equation} \begin{aligned} \label{eq:6.28} 
	(\delta_\xi \phi , \delta_\xi \tilde{\phi}) & = (\mathrm{d} \xi , - \theta \xi) \, ,
\end{aligned} \end{equation}
together with $\delta_\xi \tilde{\phi} = - \theta \mathrm{d} \tilde{\xi}$, which is actually included in \eqref{eq:6.28} as can be seen by setting $\xi = \mathrm{d} \tilde{\xi}$. The gauge invariant combinations (involving the minimum number of derivatives) in this case are 
\begin{equation} \begin{aligned}
	\mathrm{d} \phi \quad \quad \text{and} \quad \quad \hat{\phi} \equiv \phi + \mathrm{d} \left( {\tilde{\phi} / \theta} \right) ,
\end{aligned} \end{equation}
which are not independent since $\mathrm{d} \hat{\phi} = \mathrm{d} \phi$. Hence, one reaches a lagrangian made of $\star \mathrm{d} \phi \wedge \mathrm{d} \phi$ and $\star \hat{\phi} \wedge \hat{\phi}$ which are the quadratic non-topological terms available.
Before writing this explicitly, we fix some notation. We define the field strengths $\mathcal{H} \equiv \mathrm{d} \phi$ and $\mathcal{F} \equiv p! \hat{\phi}$, which satisfy the Bianchi identities
\begin{equation} \begin{aligned}
	\mathrm{d} \mathcal{H} = d^2 \phi = 0 \quad \quad \quad \text{and} \quad \quad \quad \mathrm{d} \mathcal{F} - p! \mathcal{H} = \mathrm{d}^2 \theta^{-1} \tilde{\phi} = 0 \, .
\end{aligned} \end{equation}
Note that the former is the exterior derivative of the latter. We also introduce $n \equiv p - 1$, so that $\phi$ is an $(n{+}1)$-form and $\tilde{\phi}$ is an $n$-form, and assume $0 \leq n \leq d-2$. 

\paragraph{}Finally, the bulk theory we consider --- dual to a boundary system in which a U(1) higher-form symmetry is broken --- is the \textit{Higher St\"uckelberg model} \cite{Aurilia:1980jz,Rajeev:1986id,Allen:1990gb,Bizdadea:1996np},\footnote{We have adopted the designation used recently by \cite{Borsten:2024gox}.} given in terms of the regularised action
\begin{equation} \begin{aligned} \label{action}
	S_\varLambda & = - {1 \over 2} \int_{\mathbb{B}_\varLambda} \mathrm{d}^{d+1} x \sqrt{|G|} \left[ {\mathcal{H}_{a_0 ... a_{n+1}} \mathcal{H}^{a_0 ... a_{n+1}} \over n+2} + m^2 \mathcal{F}_{a_0 ... a_n} \mathcal{F}^{a_0 ... a_n} \right] ,
\end{aligned} \end{equation}
which is equivalent to the Proca action for a massive Abelian gauge field.\footnote{Following the discussion below \cref{actionmassless}, we have ignored the gravitational Lagrangian $\mathscr{L}_{\text{grav}}$.} In fact, the coupling constant $m^2$ was denoted this way since, as we are about to see, it corresponds to the mass squared of $\mathcal{F}$.

\paragraph{}Varying $S_\varLambda$, one sees the equations of motion for $\phi$ and $\tilde{\phi}$ are respectively $\mathrm{d}^\dagger \mathcal{H} = m^2 \mathcal{F}$ and $\mathrm{d}^\dagger \mathcal{F} = 0$. Due to $\tilde{\phi}$ being a St\"uckelberg field, it is not surprising that the latter is simply the adjoint exterior derivative of the former. As in \cref{masslessrenorm_new}, we wish to solve $\mathrm{d}^\dagger \mathcal{H} = m^2 \mathcal{F}$ for the field strengths. Hence, we require the Bianchi identity $\mathrm{d} \mathcal{F} = (n+1)! \mathcal{H}$. In components, these are written
\begin{subequations} \label{1.4} \begin{align} 
	\label{1.4b} \partial_{a_0} \left( \sqrt{|G|} \mathcal{H}^{a_0 ... a_{n+1}} \right) & = m^2 \sqrt{|G|} \mathcal{F}^{a_1 ... a_{n+1}} \\
	\label{1.5} \partial_{[ a_0} \mathcal{F}_{a_1 ... a_{n+1} ]} & = (n+1)! \mathcal{H}_{a_0 ... a_{n+1}} \, .
\end{align} \end{subequations}
We refer to them as the \textit{massive equations}.\footnote{We refrain from using the term \textit{Proca equations} since these would more accurately refer to \cref{2.7a}.} Substituting the Bianchi identity in the equation of motion yields
\begin{equation} \begin{aligned} \label{2.7a} 
	{\mathrm{d}^\dagger \mathrm{d} \mathcal{F} \over (n+1)!} = m^2 \mathcal{F} \, ,
\end{aligned} \end{equation}
showing that $\mathcal{F}$ is indeed massive. In fact, the same substitution in \eqref{action} generates the action for a field $\mathcal{F}$ with \eqref{2.7a} as the equation of motion. 

\subsubsection{Near-boundary massive solutions} \label{massivesolutions}

\paragraph{}Our goal now is to solve the equations of motion near the conformal boundary, which constitutes a preliminary step towards holographic renormalisation. Using the AdS-Poincar\'e metric $G$, given by \eqref{metric}, the components of \cref{2.7a} can after some manipulation be written as
\begin{subequations} \begin{align}  
	\label{4.1}	r^2 \left( \Delta_+ + 3 + r \partial_r \right) \left( \Delta_- + 3 + r \partial_r \right) \mathcal{F}_{r \mu_1 ... \mu_n} + \Box \mathcal{F}_{r \mu_1 ... \mu_n} & = 0 \\
	\label{1.13} r^2 \left( \Delta_+ + r \partial_r \right) \left( \Delta_- + r \partial_r \right) \mathcal{F}_{\mu_0 ... \mu_n} + \Box \mathcal{F}_{\mu_0 ... \mu_n} + {2 r^3 \over n!} \partial_{[ \mu_0} \mathcal{F}_{r | \mu_1 ... \mu_n ]} & = 0 \, ,
\end{align} \end{subequations}
where $\lambda \equiv d - 2 n + 1$ and
\begin{equation} \begin{aligned} 
	\Delta_\pm = {\lambda - 3 \pm \sqrt{(\lambda - 3)^2 + 4 m^2} \over 2} \, ,
\end{aligned} \end{equation}
from \cref{Deltapm}. In \cref{masslessrenorm_new}, we distinguished three cases based on the difference $\bar{\varDelta}_+ - \bar{\varDelta}_-$ being odd, even (non-null) or null. This classification remains useful for $\Delta_\pm$ but, because these are not necessarily integers, the first case should be generalised to $\Delta_+ - \Delta_-$ not being an even integer. This is in fact the only regime we are interested in, since we work in the limit $|m^2| \ll 1$, which as we are about to show renders the symmetry approximate. 

\paragraph{}Note that \cref{4.1} coincides with \cref{4.2} if $m^2 = 0$ and $n = p$, since in this case $\Delta_\pm = \bar{\varDelta}_\pm - 3$. Hence, their solutions take the same form, such that
\begin{equation} \begin{aligned} \label{1.16}
	\mathcal{F}_{r \mu_1 ... \mu_n} = r^{-\Delta_- - 3} X^{-}_{\mu_1 ... \mu_n} + \ldots + r^{-\Delta_+ - 3} X^{+}_{\mu_1 ... \mu_n} + \ldots
\end{aligned} \end{equation}
where we have introduced $X^{-} , X^{+} \in \Omega^{n}$ as arbitrary constants.\footnote{In this context, the term constant refers to the lack of dependence on the radial coordinate.} Having solved for $\mathcal{F}_{r \mu_1 ... \mu_n}$, we plug the result into \cref{1.13} and solve for $\mathcal{F}_{\mu_0 ... \mu_n}$. Using a polyhomogeneous ansatz of the form \eqref{poly}, we obtain
\begin{equation} \begin{aligned} \label{eq:6.38}
	\mathcal{F}_{\mu_0 ... \mu_n} = & r^{-\Delta_-} K^{-}_{\mu_0 ... \mu_n} + {r^{-\Delta_- -2} / 2 \over \Delta_+ - \Delta_- - 2} \left( \Box K^{-}_{\mu_0 ... \mu_n} + {2 \over n!} \partial_{[ \mu_0} X^{-}_{\mu_1 ... \mu_n ]} \right) + \ldots \\
	& + r^{-\Delta_+} K^{+}_{\mu_0 ... \mu_n} + {r^{-\Delta_+ -2} / 2 \over \Delta_- - \Delta_+ - 2} \left( \Box K^{+}_{\mu_0 ... \mu_n} + {2 \over n!} \partial_{[ \mu_0} X^{+}_{\mu_1 ... \mu_n ]} \right) + \ldots
\end{aligned} \end{equation}
where we have introduced $K^{-} , K^{+} \in \Omega^{n+1}$ as arbitrary constants.

\subsubsection{Approximate conservation equation}

\paragraph{}From this point onwards we focus on $\lambda \neq 3$. We are going to show how the mass $m^2$ controls the degree of symmetry breaking. We start off by introducing and adopting 
\vspace*{0.4em}

\noindent \hypertarget{conv3}{\textbf{(convention 2):}} when dealing with massive theories, the labels $\pm$ and $\mp$ are to be read respectively as $+$ and $-$, when $\lambda < 3$, or as $-$ and $+$, when $\lambda > 3$. \hangindent=2em \hangafter=0

\vspace*{0.4em} \noindent 
In the small-$m^2$ regime,
\begin{equation} \begin{aligned}
	\Delta_\pm = {m^2 \over 3-\lambda} + O (m^4) \qquad \text{and} \qquad \Delta_\mp =	\lambda-3 + O (m^2) \, .
\end{aligned} \end{equation}
Hence, substituting the solutions we have previously obtained into $(\mathrm{d}^\dagger \mathcal{F})^{\mu_1 ... \mu_n} = 0$, yields an approximate conservation equation 
\begin{equation} \begin{aligned} \label{3.16a} 
	\partial^{\mu_0} K^{\mp}_{\mu_0 ... \mu_n} + \Delta_\pm X^{\mp}_{\mu_1 ... \mu_n} = 0 \, ,
\end{aligned} \end{equation}
as well as
\begin{equation} \begin{aligned} \label{3.16b} 
	\partial^{\mu_0} K^{\pm}_{\mu_0 ... \mu_n} + \Delta_\mp X^{\pm}_{\mu_1 ... \mu_n} = 0 \, .
\end{aligned} \end{equation}
This last equation immediately clarifies how the arbitrary constants parametrising $\mathcal{F}$ are related to $J$ and $\beta$, which parametrise $\bar{\mathcal{F}}$, in the case of a massless $p$-form. In \cref{masslessrenorm_new},\footnote{At that stage, we had not yet adopted the simplified notation $J$ in place of $J^{{\bar{\lambda}}}$.} $J$ was a conserved current and $\beta$ failed to be co-closed due to \cref{0.12} (when $\bar{\varDelta}_+ - \bar{\varDelta}_-$ was not even). When $m^2 \to 0$ , this equation does in fact agree with \cref{3.16b} under the appropriate identification: $K^{\pm} \to \beta$, $\lambda \to {\bar{\lambda}}$ and $X^{\pm} \to J^{3}$.

\paragraph{}$K^{-}_{\mu_0 ... \mu_n} \equiv K^{-}_\Xi$ and $K^{+}_{\mu_0 ... \mu_n} \equiv K^{+}_\Xi$ are the fundamental fields that parametrise solutions to the equations of motion. They correspond to the on-shell boundary values of $\Upsilon^\Xi$ and $\hat{\Phi}_\Xi$ --- cf. \cref{Stuckelberg}: 
\begin{equation} \label{mapK} \{ \braket{\Upsilon^\Xi}_F , \braket{\hat{\Phi}_\Xi}_F \} \sim \{ K^{-}_\Xi , K^{+}_\Xi \} \, . \end{equation}
Taking \cref{3.16a} into account, electric quantisation is enforced by the non-Legendre transformed path integral in \eqref{delZ} if $\braket{\Upsilon^\Xi}_F \sim K^\mp_\Xi$  or by the Legendre transformed one if $\braket{\hat{\Phi}_\Xi}_F \sim K^\mp_\Xi$. One would then expect that, whatever bulk functional corresponds to $K^\mp$, magnetic quantisation can be achieved by doing or undoing the Legendre transformation. However, as we are about to see in the next section, the map \eqref{mapK} is not unambiguously defined and one can actually achieve both quantisations, without Legendre transforming, via different renormalisation schemes.

\subsection{Holographic renormalisation of massive $p$-forms} \label{massiverenorm}

\paragraph{}The on-shell variation of the regularised action \eqref{action} can be written in a form that makes use of $\tilde{\phi}$ being a St\"uckelberg field --- cf. \eqref{2.61}:
\begin{equation} \begin{aligned} \label{1.12}
	\delta S_\varLambda & = - \int_{\partial \mathbb{B}_\varLambda} \mathrm{d}^d x \, r^{1 + \Delta_+ + \Delta_-} \mathcal{H}_r {}^{\mu_0 ... \mu_n} \delta \mathcal{F}_{\mu_0 ... \mu_n} \, .
\end{aligned} \end{equation} 
The momentum conjugate to $\mathcal{F}_\Xi$ consists of the radial components of the field strength, just as in the massless case. None of the arbitrary constants $K^{-}$ and $K^{+}$ appears at order $r^0$ in the solution \eqref{eq:6.38} for $\mathcal{F}_\Xi$ and the same is true for the canonical momenta conjugate to $\mathcal{F}_\Xi$. Hence, the canonical variables must be rescaled and this can be done in two distinct ways:
\[
\left( \mathcal{F}_\Xi , r^{1 + \Delta_+ + \Delta_-} \mathcal{H}_r {}^\Xi \right) \; \;
\begin{aligned}
	\raisebox{-1\height}{$\nearrow$\quad} & \text{(a) \quad} \left( r^{\Delta_+} \mathcal{F}_\Xi , r^{1 + \Delta_-} \mathcal{H}_r {}^\Xi \right) \\[-0.3ex]
	\raisebox{1\height}{$\searrow$\quad} & \text{(b) \quad} \left( r^{\Delta_-} \mathcal{F}_\Xi , r^{1 + \Delta_+} \mathcal{H}_r {}^\Xi \right) 
\end{aligned}
\]
Option (a) leads to a well-defined canonical momentum, whereas its conjugate field must be renormalised:\footnote{Here, $\simeq$ means equality up to $O (r^{-1})$.}
\begin{subequations} \begin{align} 
	r^{\Delta_+} \mathcal{F}_\Xi & \simeq r^{\Delta_+ - \Delta_-} K^{-}_\Xi + \ldots + K^{+}_\Xi \\
	r^{\Delta_- + 1} \mathcal{H}_{r \Xi} & \simeq -\Delta_- K^{-}_\Xi \, .
\end{align} \end{subequations}
Option (b), on the other hand, gives a finite canonical field, but a divergent momentum:
\begin{subequations} \begin{align} 
	r^{\Delta_-} \mathcal{F}_\Xi & \simeq K^{-}_\Xi \\
	r^{\Delta_+ + 1} \mathcal{H}_{r \Xi} & \simeq -\Delta_- r^{\Delta_+ - \Delta_-} K^{-}_\Xi + \ldots -\Delta_+ K^{+}_\Xi \, .
\end{align} \end{subequations}
These two alternatives interchange the canonical variables that each coefficient, $K^{+}$ or $K^{-}$, is associated with. Note, however, that the divergent behaviour is always tied to $K^{-}$. 

\paragraph{}With these preliminaries in place, we now perform holographic renormalisation at the level of \eqref{1.12}, which can be written as
\begin{equation} \begin{aligned} \label{4.24}
	\delta S_\varLambda \simeq - m^2 \int_{\partial \mathbb{B}_\varLambda} \bigg[ \varLambda^{\Delta_+ - \Delta_-} {K^{-} \cdot \delta K^{-} \over \Delta_+} + \varLambda^{\Delta_+ - \Delta_- - 2} \Box O \left( K^{-} \cdot \delta K^{-} \right) & + {K^{-} \cdot \delta K^{+} \over \Delta_+} \\ & + {K^{+} \cdot \delta K^{-} \over \Delta_-} \bigg] ,
\end{aligned} \end{equation} 
where $K \cdot K \equiv \ast K \wedge K / (d-n-1)!$.\footnote{We write $\Box$ before $O \left( K^{-} \cdot \delta K^{-} \right)$ to show that this stands for the action of $\Box^i$ ($i \geq 0$) on both $K^{-}_{\mu_0 ... \mu_n} \Box \delta K_{-}^{\mu_0 ... \mu_n}$ and $\partial^\nu K^{-}_{\nu \mu_1 ... \mu_n} \partial_\rho \delta K_{-}^{\rho \mu_1 ... \mu_n}$.} 
Following option (a) at first, the counterterm is given by
\begin{equation} \begin{aligned} \label{ct2} 
	S_{\text{ct}} & \vcentcolon = {\chi (\varLambda) \over 2 \Delta_-} \int_{\partial \mathbb{B}_\varLambda} \mathrm{d}^d x \, r^{\lambda-1} \mathcal{H}_{r \mu_0 ... \mu_n} \mathcal{H}_r {}^{\mu_0 ... \mu_n} + \ldots
\end{aligned} \end{equation}
where $\chi = -1$ and the ellipsis denotes terms of higher order in $\partial_\mu \left( r^{\Delta_- + 1} \mathcal{H}_{r \Xi} \right)$ --- these are necessary unless $2 \leq \lambda \leq 4$, when $m^2 > 0$, or $1 \leq \lambda \leq 5$, when $m^2 < 0$. Like in \cref{masslessrenorm}, we are going to apply the counterterm$^+$ prescription to the leading term in \eqref{ct2}, such that
\begin{equation} \begin{aligned} 
	\chi (\varLambda) = 1 - {\Delta_+ - \Delta_- \over \Delta_-} {\varLambda^{\Delta_- - \Delta_+}} \mathcal{M}_- \, .
\end{aligned} \end{equation}
Hence, we define the renormalised and deformed action as
\begin{equation} \begin{aligned} 
	S_{\text{final},-} \vcentcolon = \lim_{\varLambda \to \infty} {S_\varLambda - S_{\text{ct}} \over \Delta_- - \Delta_+} \quad \quad \text{such that} \quad \quad
	\delta S_{\text{final},-} & = \int K^{-} \cdot \delta \left( K^{+} - \mathcal{M}_{-} K^{-} \right) .
\end{aligned} \end{equation}
Alternatively, if we follow option (b) then the counterterm is
\begin{equation} \begin{aligned} \label{ct1} 
	S'_{\text{ct}} & \vcentcolon = {\Delta_- \over 2} \int_{\partial \mathbb{B}_\varLambda} \mathrm{d}^d x \, r^{\lambda-3} \mathcal{F}_{\mu_0 ... \mu_n} \mathcal{F}^{\mu_0 ... \mu_n} + \ldots
\end{aligned} \end{equation}
where the ellipsis refers to terms of higher order in $\partial_\mu \left( r^{\Delta_-} \mathcal{F}_\Xi \right)$. Naturally, the counterterms $S_{\text{ct}}$ and $S'_{\text{ct}}$ agree on-shell up to finite pieces at large $\varLambda$. The renormalised action is given by
\begin{equation} \begin{aligned} 
	S_{\text{ren},+} = \lim_{\varLambda \to \infty} {S_\varLambda - S'_{\text{ct}} \over \Delta_+ - \Delta_-} \quad \quad \text{such that} \quad \quad
	\delta S_{\text{ren},+} & = \int K^{+} \cdot \delta K^{-} \, .
\end{aligned} \end{equation}
\Cref{4.24} was ill-defined for two reasons: the presence of divergent terms and the simultaneous appearance of $K^{-} \cdot \delta K^{+}$ and $K^{+} \cdot \delta K^{-}$. However, the counterterms required to cancel the former also remove one of the latter, thus making the on-shell variation of the action fully well defined.

\paragraph{}We conclude that the counterterms $S_{\text{ct}}$ and $S'_{\text{ct}}$ lead to different quantisations, thereby removing the need for Legendre terms in the path integral. Note that, under choice (b) of canonical variables, the renormalised momentum is associated with $K^{+}$. Hence, the counterterm$^+$ prescription was not considered in \eqref{ct1} as it cannot produce deformations of the dual boundary theory. The deformation that we wish to consider in this quantisation is therefore achieved through a surface term that has to be put in by hand, using the fact that $K^{+}$ is given by the renormalised momentum at the boundary $\partial \mathbb{B}$:
\begin{equation} \begin{aligned} 
	K^{+}_{\mu_0 ... \mu_n}  & = r^{\Delta_+} {r \mathcal{H}_{r \mu_0 ... \mu_n} + \Delta_- \mathcal{F}_{\mu_0 ... \mu_n} + \ldots \over \Delta_- - \Delta_+} \bigg \vert_{\partial \mathbb{B}} \, ,
\end{aligned} \end{equation}
where the ellipsis relates to the one in \cref{ct1}. The end result is that we have two renormalised and deformed actions $S_{\text{final},-}$ and $S_{\text{final},+}$, such that
\begin{equation} \begin{aligned} \label{eq:6.52}
	\delta S_{\text{final},-} = \int_{\partial \mathbb{B}} \mathcal{K}^{-} \cdot \delta \mathcal{K}^{+} \qquad \quad \text{and} \qquad \quad
	\delta S_{\text{final},+} = \int_{\partial \mathbb{B}} \mathcal{K}^{+} \cdot \delta \mathcal{K}^{-} \, ,
\end{aligned} \end{equation}
where we have introduced $\mathcal{K}^{+} , \mathcal{K}^{-} \in \Omega^{n+1}$ given by
\begin{equation} \label{4.28} \begin{aligned} 
	\mathcal{K}^+ & \vcentcolon = K^{+} - \mathcal{M}_{-} K^{-} \qquad \quad \text{and} \qquad \quad 
	\mathcal{K}^- \vcentcolon = K^{-} - \mathcal{M}_{+} K^{+} \, .
\end{aligned} \end{equation}
While $\mathcal{M}_{-}$ vanishes in the $S_{\text{final},+}$ theory, $\mathcal{M}_{+}$ vanishes with $S_{\text{final},-}$.

\subsubsection{Generating functionals re-revisited} \label{rerevisited}

\paragraph{}Now that we have a well-defined bulk theory, it is time to establish the holographic dictionary that connects to its boundary counterpart. Since $K^{\mp}$ is approximately conserved, according to \cref{3.16a}, the boundary generating functional in electric quantisation is given by
\begin{equation} \begin{aligned} \label{Zpm}
	W [\psi] = \ln Z^{[\lambda , \mathcal{M}_\mp]} (\psi) \qquad \text{with} \qquad Z^{[\lambda , \mathcal{M}_\mp]} \vcentcolon = \int_{\mathcal{K}^\pm [\mathcal{F}] \vert_{\partial \mathbb{B}} = \psi} \mathfrak{D} \mathcal{F} e^{{i \over \kappa^2} S_{\text{final},\mp}} \, ,
\end{aligned} \end{equation}
where the generating functional $W$ is given by \eqref{genfuncW0} and it is of the type discussed around \cref{generic_delW}. Building on the construction from \cref{revisited}, we have introduced $\mathcal{K}^{+} [\mathcal{F}]$ and $\mathcal{K}^{-} [\mathcal{F}]$, a pair of form-valued functionals that approach variables \eqref{4.28} at the boundary, where their argument $\mathcal{F}$ satisfies the equations of motion.\footnote{Since the argument of $\mathcal{K}^{+} [\mathcal{F}]$ and $\mathcal{K}^{-} [\mathcal{F}]$ belongs to the domain of integration for \eqref{Zpm}.} The Robin boundary conditions in the path integral above correspond to the following double-trace deformation
\begin{equation} \begin{aligned} \label{eq:6.55}
	{\mathcal{M}_\mp \over 2} \int_{\partial \mathbb{B}} \mathrm{d}^d x \, \mathcal{O}_{\mu_0 ... \mu_n} \mathcal{O}^{\mu_0 ... \mu_n} \, .
\end{aligned} \end{equation} 
This is the only contribution at leading order in the gradient expansion, because the topological term $\int_{\partial \mathbb{B}} \mathcal{O} \wedge \mathcal{O}$ requires $\lambda = 3$, which lies outside the scope of our analysis. The energy dimension of the coupling constants $\mathcal{M}_-$ and $\mathcal{M}_+$ is given by
\begin{equation} \begin{aligned} 
	[\mathcal{M}_-] = - [\mathcal{M}_+] = \Delta_+ - \Delta_- = \sqrt{(\lambda - 3)^2 + 4 m^2} \, .
\end{aligned} \end{equation} 
The deformations parametrised by $\mathcal{M}_-$ are relevant (these ones come from a counterterm$^+$), whereas those parametrised by $\mathcal{M}_+$ are irrelevant. 

\paragraph{}Lastly, we turn to the quantisation attained by exchanging the labels \[\mp \leftrightarrow \pm\] in the preceding discussion, including \cref{Zpm,eq:6.55}. As the reader might already expect, we refer to it as the magnetic quantisation, yet an approximate magnetic symmetry can only be identified in the weakly deformed case. In particular, when $m^2 \ll 1$ and $\mathcal{M}_{\pm} = 0$, the approximate conservation \cref{3.16a} constrains the sources as in \eqref{background_nonconserv}. However, turning on the double-trace deformation, this equation becomes
\begin{equation} \begin{aligned} 
	\partial^{\mu_0} K^{\pm}_{\mu_0 ... \mu_n} + {\partial^{\mu_0} \mathcal{K}^{\mp}_{\mu_0 ... \mu_n} + \Delta_\pm X^{\mp}_{\mu_1 ... \mu_n} \over \mathcal{M}_{\pm}} = 0 \, ,
\end{aligned} \end{equation}
where $\mathcal{K}^{\mp}$ is fixed as the source due to boundary conditions. Hence, for large ${\mathcal{M}_{\pm} / \Delta_\pm}$ there is instead an electric higher-form symmetry associated with the approximate conservation of the expectation value dual to $K^{\pm}$.
Note that with electric symmetries we always assume that $X^{\mp}$ is given by a constitutive relation possessing the appropriate analytic properties (that they have a small-$m^2$ or large-${\mathcal{M}_{\pm} \over \Delta_\pm}$ limit, according to what makes the symmetry exact, is sufficient).
This assumption was verified for thermal states dual to asymptotically AdS black branes \cite{Pinheiro:2025fqg}, in the hydrodynamic, low-$m^2$ regime. More importantly, there is no constitutive relation for $X^{\pm}$ such that \cref{3.16b} plays no role in the dynamics, in contrast to \eqref{3.16a} (see the discussion at the end of \cref{boundarytheories}).

\section{Dualities of higher-form fields} \label{sec:duality}

\paragraph{}In this section, we start from dualities between the bulk dynamics of higher-form fields of different degree. In particular, we consider electric-magnetic Hodge duality in massless theories, as well as its massive counterpart \cite{Kawai:1980qq,Quevedo:1996uu,Quevedo:1997jb},\footnote{See \cite{Hell:2021wzm,Burgess:2025geh} and references therein for more recent discussions of this topic.} and show how these reflect in the holographic lower-dimensional theories. 
Discussions of electric-magnetic duality in holography include \cite{Witten:2003ya,Leigh:2003ez,Yee:2004ju,deHaro:2007eg,deHaro:2007fg} and also \cite{Herzog:2007ij,Hartnoll:2007ip}, in the AdS/CMT context.

\subsection{Massless and massive Hodge duality} \label{duality_subsection} 

\paragraph{}The $p$-form Maxwell equations are a pair of geometric constraints imposing that the field strength $\bar{\mathcal{F}}$ be both closed and co-closed. We are interested in a change of variables given by
\begin{equation} \begin{aligned} \label{Hodgemap1}
	\bar{\mathcal{F}} \to \bar{\mathcal{F}}' = u \star \bar{\mathcal{F}}'' \, , \quad \quad \text{where} \quad \quad \bar{\mathcal{F}}' \in \Omega^{p+1} (\mathbb{B}) \, , \; u \in \mathbb{R} \quad \text{and} \quad \bar{\mathcal{F}}'' \in \Omega^{d-p} (\mathbb{B}) \, .
\end{aligned} \end{equation}
Crucially, \eqref{Hodgemap1} maps the $p$-form Maxwell equations according to 
\begin{equation} \begin{aligned} \label{Hodgemap1_eqs}
		\mathrm{d}^\dagger \bar{\mathcal{F}} & = 0 \\
		\mathrm{d} \bar{\mathcal{F}} & = 0
	\end{aligned} \quad \makebox[-3pt][l]{$\nearrow$}\searrow \;
	\begin{aligned}
		\mathrm{d}^\dagger \bar{\mathcal{F}}'' & = 0 \\
		\mathrm{d} \bar{\mathcal{F}}'' & = 0 \, ,
\end{aligned} \end{equation}
with the right-hand side reproducing the $(d{-}p{-}1)$-form Maxwell equations for a field strength $\bar{\mathcal{F}}''$. Thus, at least classically, the two theories are in fact equivalent. The existence of dual formulations in terms of field strengths related via the Hodge map will be termed \textit{massless Hodge duality} in the case of Maxwell theories.
The analogue of this duality for the Higher St\"uckelberg model describing massive $(n{+}1)$-forms will be referred to as \textit{massive Hodge duality} and corresponds to the change of variables
\begin{equation} \label{Hodgemap2}
	\begin{cases}
		\mathcal{F} \to \mathcal{F}' = v \star \mathcal{H}'' \\ 
		\mathcal{H} \to \mathcal{H}' = v (-1)^{n+1} m^2 \star \mathcal{F}'' 
	\end{cases}
	\begin{aligned}
		& \text{where} \quad \mathcal{F}' \in \Omega^{n+1} (\mathbb{B}) \, , \; v \in \mathbb{R} \quad \text{and} \quad \mathcal{H}'' \in \Omega^{d-n} (\mathbb{B}) \, ; \\
		& \text{where} \quad \mathcal{H}' \in \Omega^{n+2} (\mathbb{B}) \, , \; m^2 \in \mathbb{R} \quad \text{and} \quad \mathcal{F}'' \in \Omega^{d-n-1} (\mathbb{B}) \, . 
	\end{aligned}
\end{equation}
These map the (first-order) equations for a $(n{+}1)$-form $\mathcal{F}$ with mass-squared $m^2$ to the equations for a $(d{-}n{-}1)$-form $\mathcal{F}''$ with the same mass-squared:
\begin{equation} \begin{aligned} \label{Hodgemap2_eqs}
		\mathrm{d}^\dagger \mathcal{H} & = m^2 \mathcal{F} \\
		\mathrm{d} \mathcal{F} & = (n+1)! \mathcal{H}
	\end{aligned} \; \; \makebox[-3pt][l]{$\nearrow$}\searrow \;
	\begin{aligned}
		\mathrm{d}^\dagger \mathcal{H}'' & = m^2 \mathcal{F}'' \\
		\mathrm{d} \mathcal{F}'' & = (d-n-1)! \mathcal{H}'' \, .
\end{aligned} \end{equation}
Note that both massless and massive Hodge duality act within a single spacetime, relating theories defined on it.

\paragraph{}Our goal is to determine how such bulk dualities reflect on the lower-dimensional holographic theories.
For this, we view the massless and massive dualities as automorphisms of the sets $\mathsf{P}$ and $\mathsf{N}$, respectively, of equations in a certain dimension $d{+}1$. (All the massive equations in $\mathsf{N}$ have mass-squared $m^2 \in \mathbb{R}$). Each Maxwell equations in $\mathsf{P}$ is characterised by 
\begin{equation} \begin{aligned}
	{\bar{\lambda}} \in \mathsf{P} \cong \{ d - 2 p + 1 | p \in \mathsf{P}' \}
	\, , \quad \quad \text{where} \quad \quad \mathsf{P}' = \{ p \in \mathbb{N}_0 | p \leq d-1 \} \, ,
\end{aligned} \end{equation}
such that massless Hodge duality acts as a reflection ${\bar{\lambda}} \to 4 - {\bar{\lambda}}$ around ${\bar{\lambda}} = 2$.\footnote{In $\mathsf{P}' \cong \mathsf{P}$, the duality generates a reflection $p \to d - p - 1$ around $p = {d-1 \over 2}$.} Equivalently, each equation in $\mathsf{N}$ is associated with
\begin{equation} \begin{aligned}
	\lambda \in \mathsf{N} \cong \{ d - 2 n + 1 | n \in \mathsf{N}' \}
	\, , \quad \quad \text{where} \quad \quad \mathsf{N}' = \{ n \in \mathbb{N}_0 | n \leq d-2 \} \, ,
\end{aligned} \end{equation}
such that massive Hodge duality acts as a reflection $\lambda \to 6 - \lambda$ around $\lambda = 3$.\footnote{In $\mathsf{N}' \cong \mathsf{N}$, the duality generates a reflection $n \to d - n - 2$ around $n = {d - 2 \over 2}$.} Hence, both dualities generate a $\mathbb{Z}_2$ group of automorphisms.
We set aside the self-dual cases. For massless theories, this will be taken up in \cref{self_duality}.

\paragraph{}In broad terms, we have described how a duality between theories operates at the level of the equations. Naturally, this extends to their solutions as well. 
Recall from \cref{masslessrenorm_new} that on-shell configurations of $\bar{\mathcal{F}}$ are specified by the pair $(J , \beta)$.\footnote{At that time we wrote $J^{{\bar{\lambda}}}$ since the superscript was only dropped in \cref{masslessrenorm}.}
Thus, for each theory in $\mathsf{P}$ we consider an arbitrary pair $(J^{({\bar{\lambda}})} , \beta^{({\bar{\lambda}})})$, thereby defining a solution $\bar{\mathcal{F}}^{({\bar{\lambda}})}$. Massless Hodge duality induces automorphisms of $\{ \bar{\mathcal{F}}^{({\bar{\lambda}})} | {\bar{\lambda}} \in \mathsf{P} \}$ with a $\mathbb{Z}_2$ group structure, generated by
\begin{equation} \begin{aligned} \label{1.34} 
	\bar{\mathcal{F}}^{({\bar{\lambda}})} \to \bar{\mathcal{F}}'^{({\bar{\lambda}})} & = U (4-{\bar{\lambda}}) \star \bar{\mathcal{F}}^{(4-{\bar{\lambda}})} \, .
\end{aligned} \end{equation} 
Similarly, $(K^{-} , K^{+})$ parametrise on-shell configurations of $\mathcal{F}$ and $\mathcal{H}$. Let $(K^{(\lambda) -} , K^{(\lambda) +})$ be, for each theory in $\mathsf{N}$, an arbitrary pair of fields defining solutions $\mathcal{F}^{(\lambda)}$ and $\mathcal{H}^{(\lambda)}$ of the massive equations. Massive Hodge duality induces automorphisms of $\{ (\mathcal{F}^{(\lambda)} , \mathcal{H}^{(\lambda)}) | \lambda \in \mathsf{N} \}$ forming as before a $\mathbb{Z}_2$ group, generated by
\begin{equation} \label{5.13} \begin{cases}
	\mathcal{F}^{(\lambda)} \to \mathcal{F}'^{(\lambda)} = V (6-\lambda) \star \mathcal{H}^{(6-\lambda)} \\
	\mathcal{H}^{(\lambda)} \to \mathcal{H}'^{(\lambda)} = (-1)^{n+1} m^2 V (6-\lambda) \star \mathcal{F}^{(6-\lambda)} \, .
\end{cases} \end{equation}
The proportionality factors $u$ and $v$ in \cref{Hodgemap1,Hodgemap2} must at this stage be respectively promoted to a function $U$ of ${\bar{\lambda}} \in \mathsf{P}$ and a function $V$ of $\lambda \in \mathsf{N}$, required to obey
\begin{equation} \begin{aligned} \label{eq:7.9}
	U ({\bar{\lambda}}) U (4-{\bar{\lambda}}) = (-1)^{(d-p) (p+1) +1} \quad \quad \text{and} \quad \quad V (6-\lambda) V (\lambda) = {(-1)^{(n+2)(d-n)} \over m^2} \, ,
\end{aligned} \end{equation} 
such that the actions of \eqref{1.34} and \eqref{5.13} are involutive. This can be attained for example by choosing 
\begin{equation} \begin{aligned} 
	U ({\bar{\lambda}}) = \begin{cases}
		1 \, , & {\bar{\lambda}} < 2 \\
		(-1)^{(d-p) (p+1) +1} \, , & {\bar{\lambda}} > 2
	\end{cases} \quad \quad \text{and} \quad \quad V (\lambda) = \begin{cases}
		{\sgn (m^2) \over \sqrt{|m^2|}} \, , & \lambda < 3 \\
		{(-1)^{(n+2)(d-n)} \over \sqrt{|m^2|}} \, , & \lambda > 3 \, .
	\end{cases} 
\end{aligned} \end{equation}
Note that, whenever the rank $p$ appears, it should be seen as a function of the integer ${\bar{\lambda}}$ identifying each theory. The same is true for $n$ and $\lambda$.

\paragraph{}One should note that what turns the automorphisms of $\mathsf{P}$ and $\mathsf{N}$ into dualities is the observation that we are merely performing a change of variables. Hence, two systems of equations related by \eqref{Hodgemap1} and \eqref{Hodgemap2} are physically equivalent. In this sense, the automorphism group acts trivially and, since it is isomorphic to $\mathbb{Z}_2$ , we refer to it as a \textit{duality}. 
The same principle can be applied at the level of solutions. We then wish to consider the set $\{ \bar{\mathcal{F}}^{({\bar{\lambda}})} | {\bar{\lambda}} \in \mathsf{P} \}$ and $\{ (\mathcal{F}^{(\lambda)} , \mathcal{H}^{(\lambda)}) | \lambda \in \mathsf{N} \}$ that trivialise the action of \eqref{1.34} and \eqref{5.13}. The former corresponds to
\begin{subequations} \label{1.38} \begin{align} 
	\label{1.38a} \beta^{({\bar{\lambda}})} & = (-1)^{p+1} U (4-{\bar{\lambda}}) \ast J^{(4-{\bar{\lambda}})} \\
	\label{1.38b} J^{({\bar{\lambda}})} & = U (4-{\bar{\lambda}}) \ast \beta^{(4-{\bar{\lambda}})} \, ,
\end{align} \end{subequations}
found by solving $\bar{\mathcal{F}}^{({\bar{\lambda}})} = \bar{\mathcal{F}}'^{({\bar{\lambda}})}$ for all ${\bar{\lambda}} \in \mathsf{P}$, while the latter follows from imposing either $\mathcal{F}^{(\lambda)} = \mathcal{F}'^{(\lambda)}$ or $\mathcal{H}^{(\lambda)} = \mathcal{H}'^{(\lambda)}$ for all $\lambda \in \mathsf{N}$, which leads to
\begin{subequations} \label{5.18} \begin{align}
	\label{5.18a} K^{(\lambda) +} & = \Delta_- (-1)^{n+1} V (6-\lambda) \ast K^{(6-\lambda) +} \\
	\label{5.18b} K^{(\lambda) -} & = \Delta_+ (-1)^{n+1} V (6-\lambda) \ast K^{(6-\lambda) -} \, .
\end{align} \end{subequations}
Equivalence between equations \eqref{1.38} evaluated at ${\bar{\lambda}} = {\bar{\lambda}}'$ and ${\bar{\lambda}} = 4 - {\bar{\lambda}}'$ and between equations \eqref{5.18} evaluated at $\lambda = \lambda'$ and $\lambda = 6 - \lambda'$ follows from the transformations \eqref{1.34} and \eqref{5.13} being involutions.

\subsection{Dual path integrals}

\paragraph{}We are finally in position to derive the effect of Hodge dualities in the boundary theory. Starting with the massless generating functionals $Z^{[{\bar{\lambda}} , \mathcal{M}_j]}$ and $Z^{[{\bar{\lambda}} , \mathcal{M}_a]}$ of \cref{revisited}, we denote the on-shell variations around $\bar{\mathcal{F}}^{({\bar{\lambda}})}$ of the respective actions by $\delta \bar{S}_j^{({\bar{\lambda}} , \mathcal{M}_j)}$ and $\delta \bar{S}_a^{({\bar{\lambda}} , \mathcal{M}_a)}$ such that
\begin{equation} \begin{aligned}
	\delta \bar{S}_j^{({\bar{\lambda}} , \mathcal{M}_j)} & \equiv \delta \bar{S}_{\text{final}} \big\vert_{\mathcal{M}_a = 0, J = J^{({\bar{\lambda}})} , \beta = \beta^{({\bar{\lambda}})}} \\
	\delta \bar{S}_a^{({\bar{\lambda}} , \mathcal{M}_a)} & \equiv \delta \left( \int_{\partial \mathbb{B}} j [\bar{\phi}] \cdot a [\bar{\phi}] - \bar{S}_{\text{final}} \right) \bigg\vert_{\mathcal{M}_j = 0, J = J^{({\bar{\lambda}})} , \beta = \beta^{({\bar{\lambda}})}} \, ,
\end{aligned} \end{equation}
where the variation of the final action is given by \cref{4.12}. 
Moving on to the massive generating functionals $Z^{[\lambda , \mathcal{M}_-]}$ and $Z^{[\lambda , \mathcal{M}_+]}$ of \cref{rerevisited}, the on-shell variations around $\mathcal{F}^{(\lambda)}$ of the respective actions are denoted $\delta S_-^{(\lambda , \mathcal{M}_-)}$ and $\delta S_+^{(\lambda , \mathcal{M}_+)}$ such that
\begin{equation} \begin{aligned}
	\delta S_-^{(\lambda , \mathcal{M}_-)} & \equiv \delta S_{\text{final},-} \big\vert_{K^{+} = K^{(\lambda) +} , K^{-} = K^{(\lambda) -}} \\
	\delta S_+^{(\lambda , \mathcal{M}_+)} & \equiv \delta S_{\text{final},+}\big\vert_{K^{+} = K^{(\lambda) +} , K^{-} = K^{(\lambda) -}} \, ,
\end{aligned} \end{equation}
where the variations of the final actions are given in \eqref{eq:6.52}. We proceed by substituting equations \eqref{1.38} and  \eqref{5.18} in the final actions \eqref{4.12} and \eqref{eq:6.52}, respectively. This yields, in first place,
\begin{equation} \begin{aligned} 
	\delta \bar{S}_{\text{final}} \vert_{J = J^{({\bar{\lambda}})} , \beta = \beta^{({\bar{\lambda}})}} & = {- p! \over (d - p - 1)!} \int \left( \alpha^{(4-{\bar{\lambda}})}_{\mu_1 ... \mu_p} + \mathcal{M}_a J^{(4-{\bar{\lambda}})}_{\mu_1 ... \mu_p} \right) \cdot \delta \left( J^{(4-{\bar{\lambda}})}_{\mu_1 ... \mu_p} - \mathcal{M}_j \partial^{\mu_0} \beta^{(4-{\bar{\lambda}})}_{\mu_0 ... \mu_p}  \right) ,
\end{aligned} \end{equation} 
showing for all ${\bar{\lambda}}$ that $\delta \bar{S}_j^{({\bar{\lambda}} , \mathcal{M})}$ equals $\delta \bar{S}_a^{(4-{\bar{\lambda}} , \mathcal{M})}$ up to an overall factor and, similarly,
\begin{subequations} \begin{align} 
	\delta S^{(\lambda , \mathcal{M})}_{+} & = \sgn (m^2) {(n+1)! \over (d-n-1)!} \delta S^{(6-\lambda , \mathcal{M}')}_{+} \Big\vert_{\mathcal{M}' = {\Delta_+ \over \Delta_-} \mathcal{M}} \\
	\delta S^{(\lambda , \mathcal{M})}_{-} & = \sgn (m^2) {(n+1)! \over (d-n-1)!} \delta S^{(6-\lambda , \mathcal{M}')}_{-} \Big\vert_{\mathcal{M}' = {\Delta_- \over \Delta_+} \mathcal{M}} \, .
\end{align} \end{subequations}
Hence, in conclusion, we found that massless and massive Hodge dualities match bulk path integrals in a way that exchanges electric and magnetic quantisation, amongst other details:
\begin{subequations} \label{HodgeZ} \begin{align}
	\label{HodgeZ_massless} Z^{[{\bar{\lambda}}_1 , \mathcal{M}_j]} \leftrightarrow Z^{[{\bar{\lambda}}_2 , \mathcal{M}_a]} , \quad \qquad \mathcal{M}_j & = \mathcal{M}_a \\[1ex]
	\label{HodgeZ_massive} Z^{[\lambda_1 , \mathcal{M}_\mp]} \leftrightarrow Z^{[\lambda_2 , \mathcal{M}_\pm]} , \; \qquad {\mathcal{M}_\mp \over \mathcal{M}_\pm} & = - {\Delta_\pm (\lambda_1) \over \Delta_\mp (\lambda_2)} \, ,
\end{align} \end{subequations}
where ${\bar{\lambda}}_1 + {\bar{\lambda}}_2 = 4$ and $\lambda_1 + \lambda_2 = 6$.\footnote{$\Delta_+$ and $\Delta_-$ are functions of $\lambda$ and, in particular, $\Delta_+ (6-\lambda) = - \Delta_- (\lambda)$. Since the definition of e.g. $\Delta_\pm$, through convention \hyperlink{conv3}{2}, depends on its argument, we have $\Delta_\pm (6-\lambda) = - \Delta_\pm (\lambda)$.} Assuming that the regularity conditions, satisfied in the bulk interior by the configurations over which we integrate, are duality-covariant, then \eqref{HodgeZ} implies equivalence at the level of correlation functions. This assumption was verified explicitly in the context of \cite{Davison:2025sze}. 

\paragraph{The lowest and highest degree cases}
Having discussed Hodge dualities, we are in position to address a detail that might have previously caught the attention of some readers. In \cref{masslessrenorm_new}, we have considered the case of  $p = 0$ that corresponds to a massless scalar. However, this theory is solely invariant under global shifts $\delta \phi = \text{constant}$ and so does not belong to the class discussed in \cref{setup}, such that its standard quantisation does not possess an electric symmetry.\footnote{Even when a symmetry (exact or approximate) is missing in a certain quantisation of these extreme cases, we still use the terms ``electric and magnetic quantisations" by natural extension of the general case. For instance, the denominations for the scalar case come from the fact that standard and alternative quantisations of \textit{low-degree} $p$-forms correspond to the electric and magnetic ones, respectively.} \\ \hspace*{1.75ex} 
The scalar is dual to a massless $(d{-}1)$-form, whose electric quantisation possesses the corresponding symmetry. In magnetic quantisation, on the other hand, the boundary theory admits a $(d{-}1)$-form-valued operator corresponding to a Goldstone $\mathcal{O}^a$ whose field strength $\mathsf{f}$ is a top form, such that the topological conservation equation of $\ast \mathsf{f}$ is missing. The lack of a magnetic symmetry in this case is dual to the electric symmetry found to be absent for the scalar. \\ \hspace*{1.75ex}
Most importantly, through massless Hodge duality, the electric symmetry of the $(d{-}1)$-form field shows that the alternative quantisation of the massless scalar possesses a magnetic symmetry associated with a scalar Goldstone $\mathcal{O}^a \sim \mathcal{O}^a + \text{constant}$ such that its field strength is closed. Note that our apparatus from \cref{MagnQuant_massless} alone was not prepared to derive such symmetry.

\paragraph{}The extreme cases of massive theories are the scalar and the $d$-form, for which $n = -1$ and $n = d-1$, respectively. Both were excluded from \cref{MassiveEOM}: the latter just so we could display the Bianchi identity $\mathrm{d} \mathcal{H} = 0$; in fact, the massive $d$-form can be automatically integrated in that section and its on-shell configurations are part of the general solutions found. The massive scalar, on the other hand, involves some conceptual differences compared to when $n \geq 0$ (although technically simpler to solve). First and foremost, such a theory does not belong to the class of theories discussed in \cref{WeaklyBroken_sec} and therefore one would not expect it to possess a broken symmetry at the boundary. However, the duality between massive equations according to \eqref{Hodgemap2_eqs} is also true for the cases where $n = -1$ and $n = d-1$. In particular, the comments in the previous paragraphs hold for broken electric and magnetic symmetries --- one of the conclusions is that massive Hodge duality implies the alternative quantisation of the massive scalar possesses a broken magnetic symmetry.

\subsection{Self-duality constraints} \label{self_duality}

\paragraph{}The discussion in \cref{duality_subsection} can be extended to the self-dual point ${\bar{\lambda}} = 2$. For \eqref{1.34} to act involutively, one must then have $U (2)^2 = (-1)^{(p+1)^2+1}$, which would require that we complexify our solutions when $p$ is odd. However, this does not impact the derivation below. \\ \hspace*{1.75ex}
Self-duality of the equations of motion implies that self-dual solutions exist. By extending \cref{duality_subsection} to ${\bar{\lambda}} = 2$, we are then considering the solution $\bar{\mathcal{F}}^{(2)}$ to be self-dual, so its parameters obey \cref{1.38a,1.38b}, which are now both equivalent to $J^{(2)} = U (2) \ast \beta^{(2)}$. Using this, we write
\begin{equation} \begin{aligned} \label{product}
	J^{(2)}_{\mu_1 ... \mu_p} J^{(2)}_{\nu_1 ... \nu_p} = {(-1)^{p+1} \over p!^2} \left( \varepsilon_{\mu_1 ... \mu_p} {}^{\alpha_0 ... \alpha_p} \partial_{\alpha_0} \alpha^{(2)}_{\alpha_1 ... \alpha_p} \right) \left( \varepsilon_{\nu_1 ... \nu_p} {}^{\beta_0 ... \beta_p} \partial_{\beta_0} \alpha^{(2)}_{\beta_1 ... \beta_p} \right) .
\end{aligned} \end{equation} 
Dropping the label $(2)$ for simplicity, we express this equation in terms of the renormalised variables in momentum space, i.e. $j (x) \to \int \mathrm{d}^d k e^{i k_\mu x^\mu} j (k)$ and $a (x) \to \int \mathrm{d}^d k e^{i k_\mu x^\mu} a (k)$. Note that we take the background to be Minkowski, for concreteness, which is homogeneous in the $x^\mu \equiv (t, x, y^A)$ directions.
Furthermore, we use rotational symmetry to align the wavevector with the $x$ direction, without loss of generality. Hence, the momentum vanishes along the $y^A$ directions and the Maxwell equations decouple into \textit{scalar, vector and tensor sectors} (the latter of which is trivial) \cite{Pinheiro:2025fqg}. The scalar and vector sectors involve \[(j^{A_1 ... A_p} , a_{B_1 ... B_p}) \qquad \text{and} \qquad (j^{\vec{\mu} A_2 ... A_p} , a_{\vec{\nu} B_2 ... B_p}) , \quad \text{where} \quad \vec{\mu} , \vec{\nu} \in \{ t , x \} ,\] respectively. Lastly, we consider the electric quantisation such that $j$, which is an expectation value in the boundary theory, should be seen as a functional of the source $a$.

\paragraph{}Now we return to \cref{product}, written in terms of the renormalised variables, and consider the mixed scalar-vector product $j_{A'_1 ... A'_p} j_{\vec{\mu} A_2 ... A_p}$. Differentiating this, we obtain
\begin{equation} \begin{aligned} \label{general_constraint}
	{\delta j^{A'_1 ... A'_p} \over \delta a_{C'_1 ... C'_p}} {\delta j^{\vec{\mu} A_2 ... A_p} \over \delta a_{\vec{\nu} C_2 ... C_p}} = \; & \epsilon^{t x A'_1 ... A'_p B'_2 ... B'_p} \epsilon^{t x A_2 ... A_p B_1 ... B_p} \left( \delta_{B_1 ... B_p}^{C'_1 ... C'_p} - {\mathcal{M}_j \over p!} {\delta j_{B_1 ... B_p} \over \delta a_{C'_1 ... C'_p}} \right) \\ 
	& k^2 \Pi^{\vec{\mu} \vec{\sigma}} \left( \delta_{\vec{\sigma}}^{\vec{\nu}} \delta_{B'_2 ... B'_p}^{C_2 ... C_p} - {\mathcal{M}_j \over p!} {\delta  j_{\vec{\sigma} B'_2 ... B'_p} \over \delta a_{\vec{\nu} C_2 ... C_p}} \right) ,
\end{aligned} \end{equation}
where $\delta_{A_1 ... A_q}^{B_1 ... B_q}$ is the generalised Kronecker delta \cite{Frankel_2011}, which is 1 (-1) if the bottom indices are distinct and an even (odd) permutation of the ones on top, or null in all other cases. Additionally, we have introduced the transverse projector $\Pi^{\mu \nu}$ such that $k^2 \Pi^{\mu \nu} = \eta^{\mu \nu} k^2 - k^\mu k^\nu$, where $k^2 \equiv k^\rho k_\rho$. The simplest case to consider is a vector gauge field in the bulk ($p=1$), corresponding to the ordinary U(1) 0-form symmetry at the boundary. \Cref{general_constraint} then becomes
\begin{equation} \begin{aligned} 
	{\delta j^{A} \over \delta a_{A}} {\delta j^{\vec{\mu}} \over \delta a_{\vec{\nu}}} & = \left( 1 - \mathcal{M}_j {\delta j_{A} \over \delta a_{A}} \right) k^2 \Pi^{\vec{\mu} \vec{\sigma}} \left( \delta_{\vec{\sigma}}^{\vec{\nu}} - \mathcal{M}_j {\delta  j_{\vec{\sigma}} \over \delta a_{\vec{\nu}}} \right) .
\end{aligned} \end{equation}
The functional derivatives ${\delta j^{\mu_1 ... \mu_p} \over \delta a_{\nu_1 ... \nu_p}}$ map to 2-point correlators $\braket{\mathcal{O}_j^{\mu_1 ... \mu_p} \mathcal{O}_j^{\nu_1 ... \nu_p}}$ under the holographic dictionary. Hence, the equation above becomes
\begin{equation} \begin{aligned} \label{vector_constraint}
	\braket{\mathcal{O}_j^{A} \mathcal{O}_j^{A}} \braket{\mathcal{O}_j^{\vec{\mu}} \mathcal{O}_j^{\vec{\nu}}} & = \Big( \vartheta + \mathcal{M}_j \braket{\mathcal{O}_j^{A} \mathcal{O}_j^{A}} \Big) k^2 \Pi^{\vec{\mu}} {}_{\vec{\sigma}} \left( \eta^{\vec{\sigma} \vec{\nu}} \vartheta + \mathcal{M}_j \braket{\mathcal{O}_j^{\vec{\sigma}} \mathcal{O}_j^{\vec{\nu}}} \right) .
\end{aligned} \end{equation}
where $\vartheta$ is the relevant coupling constant which, for our choice of holographic generating functionals, is given by $i \kappa^2 \equiv i / \mathcal{C} (N)$. 
Because in the case at hand the fundamental fields are vector valued, they are usually classified as transverse or longitudinal with respect to $k^\mu$. The former belong to the scalar sector, while the latter to the vector one. Hence, \cref{vector_constraint} constrains transverse and longitudinal correlators in terms of one another.

\paragraph{}All in all, the general \cref{general_constraint} is a constraint on holographic correlators that follows from self-duality of a massless bulk $p$-form when $d = 2 p + 1$. The particular case of \cref{vector_constraint} when $\mathcal{M}_j = 0$ had been derived in the seminal paper \cite{Herzog:2007ij}. Here, we extend it to arbitrary theories with a continuous higher-form symmetry and double-trace ($\ast \mathcal{O}_j \wedge \mathcal{O}_j$) deformations. Note that the case of magnetic quantisation has been left out of the discussion. Although it also possesses a constraint closely related to \eqref{general_constraint}, its presentation in a form similar to \eqref{vector_constraint} is slightly more cumbersome due to the choice of gauge-fixing for $\mathcal{O}_a$.

\section{Conclusion} \label{conclusion}

\paragraph{}In \cref{HologPathInt,WeaklyBroken_sec}, we developed a general holographic treatment of systems with higher-form symmetries and their low-energy effective descriptions, based on a detailed analysis of bulk path integrals --- including electric/magnetic quantisation and double-trace deformations --- and holographic renormalisation. These discussions were guided by symmetries and the patterns through which they are realised. Along the way, however, we imposed a few simplifying restrictions on our models, such as the truncation of topological contributions. Relaxing these provides natural directions for further investigation. \\ \hspace*{1.75ex}
The main result here is a precise holographic understanding of how a bulk antisymmetric tensor field with parametrically small mass realises an approximate higher-form symmetry, in the sense of \cref{boundarytheories}. Alternatively, under a suitable choice of quantisation, the same can be achieved (for arbitrary mass) via a strong double-trace deformation associated with Robin boundary conditions.
Since our analysis was obtained by specialising a more general treatment of explicitly broken symmetries, it also clarifies how the massive bulk field fits within broader holographic models. For instance, the former may arise as the linearised description of the latter. Furthermore, in the case of an electric symmetry, the use of a massive $p$-form is tied to the defect being realised through a St\"uckelberg field.

\paragraph{}In \cref{sec:duality}, we studied the holographic implications of bulk classical dualities, referred to as massless and massive Hodge dualities. These induce dualities between holographic theories that differ mainly by quantisation --- the precise mapping consists of \eqref{HodgeZ}.
For an exact higher-form symmetry of degree ${(d - 3) / 2}$, self-duality of a Maxwell ${d-1 \over 2}$-form produces strong constraints on the holographic current-current correlators. These stem from \cref{general_constraint} which is valid in the presence of the double-trace deformation \eqref{defM1} and for arbitrary spacetime dimension. 
In \cite{Herzog:2007ij}, the authors studied how constraints of this kind (in $d=3$) affected transport in the strongly coupled large-$N$ theory. One may ask what happens for explicitly broken symmetries of degree ${d - 2 \over 2}$, as this corresponds to the self-dual case under massive Hodge transformations, which was left out of \cref{self_duality}. (Real electrically charged matter breaks a higher-form symmetry of precisely the aforementioned degree in $d=4$). A preliminary inspection of equations \eqref{5.18} for $\lambda = 3$ suggests that `massive self-duality constraints' would not take the form of an inverse proportionality between correlators, as in \eqref{vector_constraint} without the deformation, but rather a direct proportionality. We postpone the treatment of the massive self-dual case, including an investigation along the lines of \cite{Pinheiro:2025fqg}, to future work. \\ \hspace*{1.75ex}
The holographic models that we have studied were used in \cite{Pinheiro:2025fqg} to compute the low-energy thermal spectra of systems with higher-form symmetries, in the low-charge density limit. These spectra, together with the correlation functions from which they were derived, provide a
check on the aforementioned dualities, as well as on the strong/weak-coupling duality from \cref{Stuckelberg}.

\paragraph{}It would be interesting to extend our approach to continuous higher-form symmetries with anomalies (including deformations, quantisation and explicit breaking, as well as the discussion of dualities). In particular, mixed anomalies and higher-group structures, in which multiple symmetries interact nontrivially, deserve a systematic treatment within holography. Progress in this direction was initiated in \cite{Hofman:2017vwr,DeWolfe:2020uzb,Iqbal:2020lrt} and applied to chiral magnetohydrodynamics in \cite{Das:2022auy}, but several questions remain open, one of which was first raised by the authors of \cite{Delacretaz:2019brr}. Here, superfluid hydrodynamics was recast in terms of a mixed ’t Hooft anomaly between a $0$-form and a $(d{-}2)$-form U(1) symmetries. This brings about the question of whether this structure is to any extent present in the conventional holographic approach to superfluids.

\acknowledgments
This work was supported by an EPSRC Doctoral Training Partnership Award.

\appendix

\section{Exterior calculus} \label{conventions_append}
\paragraph{}In this appendix, we collect the conventions we used, along with several useful relations. We start with exterior calculus on the bulk. For a differential form $\omega = \omega_{a_1 ... a_p} \mathrm{d} x^{a_1} \wedge \ldots \wedge \mathrm{d} x^{a_p} \in \Omega^p (\mathbb{B})$, the ``{components}" of $\omega$ are $ \omega_{a_1 ... a_p}$. The Hodge Star $\star$ map associated with the metric $G$ is such that the components of $\star \omega$ are given by
\begin{equation} \begin{aligned} 
	(\star \omega)_{a_0 ... a_{d-p}} & = {\varepsilon^{(G)}_{a_0 ... a_{d-p} b_1 ... b_p} \omega^{b_1 ... b_p} \over p!} \, ,
\end{aligned} \end{equation} 
where $\varepsilon^{(G)}$ is the volume form ($\varepsilon^{(G)}_{r 1 ... d} = \sqrt{|G|}$). Normalisation of the exterior derivative is such that $(\mathrm{d} \omega)_{a_0 ... a_p} = \partial_{[ a_0} \omega_{a_1 ... a_p]}$. We also define the adjoint exterior derivative $\mathrm{d}^\dagger$ according to
\begin{equation} \begin{aligned} 
	(\mathrm{d}^\dagger \omega)_{a_2 ... a_p} \vcentcolon = {(-1)^{p (d-p)} \over (d+1-p)!} ({\star \mathrm{d} \star} \omega)_{a_2 ... a_p} = \nabla^{(G)}_{a_1} \omega^{a_1} {}_{a_2 ... a_p} \, . 
\end{aligned} \end{equation} 
Raising indices we obtain $(\mathrm{d}^\dagger \omega)^{a_2 ... a_p} = \nabla^{(G)}_{a_1} \omega^{a_1 ... a_p} = \partial_{a_1} \big( \sqrt{|G|} \omega^{a_1 ... a_p} \big) / \sqrt{|G|}$.
Moving to the physical spacetime, we have the Hodge Star $\ast$ associated with $g$ such that
\begin{equation} \begin{aligned} 
	(\ast \omega)_{\mu_1 ... \mu_{d-p}} & = {\varepsilon_{\mu_1 ... \mu_{d-p} \nu_1 ... \nu_p} \omega^{\nu_1 ... \nu_p} \over p!} \, ,
\end{aligned} \end{equation} 
where $\omega \in \Omega^p$ and $\varepsilon$ is the volume form ($\varepsilon_{1 ... d} = \sqrt{|g|}$), and
\begin{equation} \begin{aligned} 
	\int {\ast Y \wedge Y' \over (d-m)!} = \int \mathrm{d}^d x \sqrt{|g|} Y^{\mu_1 ... \mu_m} Y'_{\mu_1 ... \mu_m} \equiv \int Y \cdot Y' \, ,
\end{aligned} \end{equation} 
where $Y , Y' \in \Omega^{m} (\partial \mathbb{B})$. The notation on the right-hand side was used in \cref{massiverenorm}. 
The volume forms are related to the Levi‑Civita tensor density $\epsilon$ by, for example, $\varepsilon^{(G)}_{a_1 ... a_{d+1}} = \sqrt{|G|} \epsilon_{a_1 ... a_{d+1}}$. We would like to bring to the reader’s attention:
\begin{equation} \begin{aligned}  
	|g| \epsilon^{a_1 ... a_r a_{r+1} ... a_{d+1}} \epsilon_{a_1 ... a_r b_{r+1} ... b_{d+1}} = - r! \delta^{[ a_{r+1}}_{b_{r+1}} \ldots \delta^{a_{d+1} ]}_{b_{d+1}}
	\qquad \text{and} \qquad 
	\epsilon_{a_1 ... a_{d+1}} = - |g| \epsilon^{a_1 ... a_{d+1}} .
\end{aligned} \end{equation} 

\section{Higher-form Maxwell equations} \label{MaxwellEOM}

\paragraph{}The equations of motion associated with the action \eqref{actionmassless} of a massless $p$-form are $\mathrm{d}^\dagger \bar{\mathcal{F}} = 0$ or, in components,
\begin{equation} \begin{aligned} \label{0.2a} 
	\partial_{a_0} \left( \sqrt{|G|} \bar{\mathcal{F}}^{a_0 ... a_p} \right) = 0 \, .
\end{aligned} \end{equation}
Instead of solving this equation for the gauge potential $\bar{\phi}$, we will solve it together with the Bianchi identity $\mathrm{d} \bar{\mathcal{F}} = (\mathrm{d}^2 \bar{\phi} =) 0$ for the gauge-invariant field strength $\bar{\mathcal{F}}$. In components, the Bianchi identity can be written as
\begin{equation} \begin{aligned} \label{0.2b} 
	\partial_{[ a_0} \bar{\mathcal{F}}_{a_1 ... a_{p+1} ]} = 0 \, .
\end{aligned} \end{equation}
Manipulating the equations above, we obtain the following equation for $\bar{\mathcal{F}}_{r \mu_1 ... \mu_p}$,
\begin{equation} \begin{aligned} \label{4.2}
	\left( {\bar{\lambda}} + r \partial_r \right) \left( 3 + r \partial_r \right) \bar{\mathcal{F}}_{r \mu_1 ... \mu_p} + r^{-2} \Box \bar{\mathcal{F}}_{r \mu_1 ... \mu_p} = 0 \, ,
\end{aligned} \end{equation}
After solving it, the remaining components $\bar{\mathcal{F}}_{\mu_0 ... \mu_p}$ can be found by integrating 
\begin{equation} \begin{aligned} \label{eq:4.2}
	(\mathrm{d} \bar{\mathcal{F}})_{r \mu_0 ... \mu_p} = p! \partial_{r} \bar{\mathcal{F}}_{\mu_0 ... \mu_p} + (-1)^{p+1} \partial_{[ \mu_0} \bar{\mathcal{F}}_{\mu_1 ... \mu_p ] r} = 0 \, .
\end{aligned} \end{equation}
Before explicitly solving these equations, we would like to make a few remarks regarding the general structure of solutions, which we assume to be given by a radial polyhomogeneous expansion:
\begin{equation} \begin{aligned} \label{poly}
	\bar{\mathcal{F}}_{a_0 ... a_p} = \sum_{l \in \mathbb{Z}} r^{-l} \left[ P^{l}_{a_0 ... a_p} (x^\mu) + \ln r L^{l}_{a_0 ... a_p} (x^\mu) \right] .
\end{aligned} \end{equation}
It is convenient to view the arbitrary constants that parametrise this ansatz as $p$ and $(p+1)$-forms living in $\partial \mathbb{B}_\varLambda$. 
Consider, for example, the $p$-form $P^{l}_r$ with components $P^{l}_{r \mu_1 ... \mu_p}$ and the $(p+1)$-form $P^{l}$ with components $P^{l}_{\mu_0 ... \mu_p}$. When these forms are closed (co-closed), their components are curl-free (divergenceless) and, when they are exact (co-exact), then their components are identically curl-free (divergenceless). \\ \hspace*{1.75ex}
Due to $(\mathrm{d}^\dagger \bar{\mathcal{F}})^{r \mu_2 ... \mu_p} = 0$, the coefficients $P^{l}_r$ and $L^{l}_r$ are co-closed. On top of this, $(\mathrm{d}^\dagger \bar{\mathcal{F}})^{\mu_1 ... \mu_p} = 0$ implies that all coefficients except $P^{{\bar{\lambda}}}_r$, with components,
\begin{equation} \begin{aligned}
	P^{{\bar{\lambda}}}_{r \mu_1 ... \mu_p} \equiv J^{{\bar{\lambda}}}_{\mu_1 ... \mu_p} \, ,
\end{aligned} \end{equation}
are in fact co-exact. This is due to $r^{-{\bar{\lambda}}}$ being annihilated by $({\bar{\lambda}} + r \partial_r)$. Note that $(\mathrm{d}^\dagger \bar{\mathcal{F}})^{r \mu_2 ... \mu_p} = 0$, as the radial component of the equations of motion, gives rise to the conservation equation of the boundary theory in the large-$N$ limit. Hence, we expect $J^{{\bar{\lambda}}}_{\mu_1 ... \mu_p}$ to be closely related to the conserved current. \\ \hspace*{1.75ex} 
Similarly, the coefficients $P^{l}$ and $L^{l}$ are closed due to $(\mathrm{d} \bar{\mathcal{F}})_{\mu_0 ... \mu_{p+1}} = 0$ but $(\mathrm{d} \bar{\mathcal{F}})_{r \mu_0 ... \mu_p} = 0$ additionally requires all coefficients except $P^{0}$, with components
\begin{equation} \begin{aligned}
	P^{0}_{\mu_0 ... \mu_p} \equiv \beta_{\mu_0 ... \mu_p} \, ,
\end{aligned} \end{equation} 
to be exact. This is due to $r^{0}$ being annihilated by $\partial_r$. When working with the action, rather than equations of motion, one needs to solve the definition of the field strength $\bar{\mathcal{F}}$ for the gauge field $\bar{\phi}$. At this level, $\beta_{\mu_0 ... \mu_p}$ also becomes identically curl-free, since it is given by the exterior derivative of a more fundamental field parametrising the solution $\bar{\phi}$. Such a boundary field will inevitably carry some gauge ambiguity such that $\ast \beta$ is a magnetic-current precursor.

\subsection{Near-boundary massless solutions} \label{masslessolutions}

\paragraph{}We are interested in perturbative solutions to \cref{4.2,eq:4.2} in a regime that can be defined in terms of eigenvalues $\delta_r$ and $\delta_\Box$ of operators $r \partial_r$ and $r^{-2} \Box$, respectively, acting on a suitable basis (e.g. Fourier) for $\bar{\mathcal{F}}_{r \mu_1 ... \mu_p}$. Specifically, we consider $\delta_r \sim 1$ and $\delta_\Box \ll 1$. In this regime, \cref{4.2} becomes effectively an ordinary (rather than partial) differential equation. We separate the cases of even ${\bar{\lambda}}$, odd ${\bar{\lambda}} \neq 3$ and ${\bar{\lambda}} = 3$, as was done in \cref{masslessrenorm_new}.

\paragraph{} \noindent $\boxed{\text{Even } {\bar{\lambda}}}$ \\[1ex] \indent
In order to solve \cref{4.2}, we use the polyhomogeneous ansatz and obtain
\begin{equation} \begin{aligned} \label{Fr1}
	\bar{\mathcal{F}}_{r \mu_1 ... \mu_p} & = r^{- \bar{\varDelta}_-} J^{\bar{\varDelta}_-}_{\mu_1 ... \mu_p} + \ldots + r^{- \bar{\varDelta}_+} J^{\bar{\varDelta}_+}_{\mu_1 ... \mu_p} + \ldots
\end{aligned} \end{equation}
where we have introduced $J^{\bar{\varDelta}_\mp} \in \Omega^{p}$ as arbitrary constants.\footnote{$(\bar{\varDelta}_+ , \bar{\varDelta}_-)$ are $({\bar{\lambda}} , 3)$ when ${\bar{\lambda}} \geq 3$ and  $(3 , {\bar{\lambda}})$ when ${\bar{\lambda}} \leq 3$, as introduced in \cref{masslessrenorm_new}.} Note that we are using the convention introduced in \cref{masslessrenorm_new} regarding the ellipses.
Now that we know $\bar{\mathcal{F}}_{r \mu_1 ... \mu_p}$, we can integrate $(\mathrm{d} \bar{\mathcal{F}})_{r \mu_0 ... \mu_p} = 0$ and obtain
\begin{equation} \begin{aligned} \label{A1}
	p! \bar{\mathcal{F}}_{\mu_0 ... \mu_p} & = p! \beta_{\mu_0 ... \mu_p} + {r^{1 - \bar{\varDelta}_-} \over 1 - \bar{\varDelta}_-} \partial_{[ \mu_0} J^{\bar{\varDelta}_-}_{\mu_1 ... \mu_p ]} + \ldots + {r^{1 - \bar{\varDelta}_+} \over 1 - \bar{\varDelta}_+} \partial_{[ \mu_0} J^{\bar{\varDelta}_+}_{\mu_1 ... \mu_p]} + \ldots
\end{aligned} \end{equation}
where we have introduced the constant of integration $\beta \in \Omega^{p+1}$. 

\paragraph{} \noindent $\boxed{\text{Odd } {\bar{\lambda}} \neq 3}$ \\[1ex] \indent
Solutions in this case take a more complex form. One can write $\bar{\mathcal{F}}_{r \mu_1 ... \mu_p}$ on-shell as
\begin{equation} \begin{aligned} \label{Fr2}
	\bar{\mathcal{F}}_{r \mu_1 ... \mu_p} = & r^{-\bar{\varDelta}_-} J^{\bar{\varDelta}_-}_{\mu_1 ... \mu_p} + \ldots + {r^{-\bar{\varDelta}_+} \ln r \over \bar{\varDelta}_- - \bar{\varDelta}_+} {(- \Box)^{\bar{\varDelta}_+ - \bar{\varDelta}_- \over 2} J^{\bar{\varDelta}_-}_{\mu_1 ... \mu_p} \over \varOmega_{\bar{\varDelta}_+ - \bar{\varDelta}_-}} + r^{-\bar{\varDelta}_+} J^{\bar{\varDelta}_+}_{\mu_1 ... \mu_p} + \ldots
\end{aligned} \end{equation}
where $\varOmega_w = \Pi^{{w \over 2} - 1}_{s=1} 2 s (\bar{\varDelta}_- + 2s - 3)$ for $w > 2$ and $\varOmega_2 = 1$. 
It follows that
\begin{equation} \begin{aligned} \label{A2}
	p! \bar{\mathcal{F}}_{\mu_0 ... \mu_p} & = p! \beta_{\mu_0 ... \mu_p} + {r^{1 - \bar{\varDelta}_-} \over 1 - \bar{\varDelta}_-} \partial_{[ \mu_0} J^{\bar{\varDelta}_-}_{\mu_1 ... \mu_p ]} + \ldots + \ln r {(- \Box)^{1 - \bar{\varDelta}_- \over 2} \partial_{[ \mu_0} J^{\bar{\varDelta}_-}_{\mu_1 ... \mu_p ]} \over \varOmega_{\bar{\varDelta}_+ - \bar{\varDelta}_-}} \\
	& + {r^{1-\bar{\varDelta}_+} \over 1-\bar{\varDelta}_+} \left( \partial_{[ \mu_0} J^{\bar{\varDelta}_+}_{\mu_1 ... \mu_p ]} + {(\bar{\varDelta}_+ - 1) \ln r + 1 \over \bar{\varDelta}_- - \bar{\varDelta}_+} {(- \Box)^{\bar{\varDelta}_+ - \bar{\varDelta}_- \over 2} \partial_{[ \mu_0} J^{\bar{\varDelta}_-}_{\mu_1 ... \mu_p ]} \over (\bar{\varDelta}_+ - 1) \varOmega_{\bar{\varDelta}_+ - \bar{\varDelta}_-}} \right) + \ldots
\end{aligned} \end{equation}
When ${\bar{\lambda}} > 3$, this equation should be read without the purely logarithmic term.

\paragraph{} \noindent $\boxed{{\bar{\lambda}} = 3}$ \\[1ex] \indent
Lastly, we have
\begin{subequations} \begin{align} 
	\label{Fr3} \bar{\mathcal{F}}_{r \mu_1 ... \mu_p} & = r^{-3} \left( \ln r \hat{J}^3_{\mu_1 ... \mu_p} + J^3_{\mu_1 ... \mu_p} \right) + \ldots \\
	\label{A3} p! \bar{\mathcal{F}}_{\mu_0 ... \mu_p} & = p! \beta_{\mu_0 ... \mu_p} - {r^{-2} \over 2} \left( {2 \ln r + 1 \over 2} \partial_{[ \mu_0} \hat{J}^3_{\mu_1 ... \mu_p ]} + \partial_{[ \mu_0} J^3_{\mu_1 ... \mu_p ]} \right) + \ldots
\end{align} \end{subequations}

\paragraph{} ~

\paragraph{}The $r$-constants $J^{3}$ (or $J^{\hat{3}}$ when ${\bar{\lambda}}=3$) are not independent from $\beta$. Due to $(\mathrm{d}^\dagger \bar{\mathcal{F}})^{\mu_1 ... \mu_p} = 0$ they are given according to \cref{0.12,0.13,0.14}, depending on the value of ${\bar{\lambda}}$. 
In agreement with the statements below \cref{poly}, $(\mathrm{d} \bar{\mathcal{F}})_{\mu_0 ... \mu_{p+1}} = 0$ only implies  
\begin{equation} \begin{aligned} \label{eq:3.25}
	\partial_{[ \mu_0} \beta_{\mu_1 ... \mu_{p+1} ]} = 0 \, ,
\end{aligned} \end{equation}
since the remaining terms in $\bar{\mathcal{F}}_{\mu_0 ... \mu_p}$ are identically curl-free, and $(\mathrm{d}^\dagger \bar{\mathcal{F}})^{r \mu_2 ... \mu_p} = 0$ only implies  
\begin{equation} \begin{aligned} \label{3.26}
	\partial^{\mu_1} J^{{\bar{\lambda}}}_{\mu_1 ... \mu_p} = 0 \, ,
\end{aligned} \end{equation}
since $\partial^{\mu_0} \beta_{\mu_0 ... \mu_p}$ is identically divergenceless.

\bibliographystyle{JHEP}
\bibliography{references}

\providecommand{\href}[2]{#2}\begingroup\raggedright\begin{thebibliography}{10}

\bibitem{Gaiotto:2014kfa}
D.~Gaiotto, A.~Kapustin, N.~Seiberg and B.~Willett, \emph{{Generalized Global
  Symmetries}}, \href{https://doi.org/10.1007/JHEP02(2015)172}{\emph{JHEP}
  {\bfseries 02} (2015) 172} [\href{https://arxiv.org/abs/1412.5148}{{\ttfamily
  1412.5148}}].

\bibitem{Cordova:2022ruw}
C.~Cordova, T.T.~Dumitrescu, K.~Intriligator and S.-H.~Shao, \emph{{Snowmass
  White Paper: Generalized Symmetries in Quantum Field Theory and Beyond}},  in
  \emph{{Snowmass 2021}}, 5, 2022
  [\href{https://arxiv.org/abs/2205.09545}{{\ttfamily 2205.09545}}].

\bibitem{Chen:2025uno}
X.~Chen, \emph{{Essay: Generalized Landau Paradigm for Quantum Phases and Phase
  Transitions}}, \href{https://doi.org/10.1103/tmvy-vsqd}{\emph{Phys. Rev.
  Lett.} {\bfseries 135} (2025) 250001}
  [\href{https://arxiv.org/abs/2511.19793}{{\ttfamily 2511.19793}}].

\bibitem{McGreevy:2022oyu}
J.~McGreevy, \emph{{Generalized Symmetries in Condensed Matter}},
  \href{https://doi.org/10.1146/annurev-conmatphys-040721-021029}{\emph{Ann.
  Rev. Condensed Matter Phys.} {\bfseries 14} (2023) 57}
  [\href{https://arxiv.org/abs/2204.03045}{{\ttfamily 2204.03045}}].

\bibitem{Wen:2018zux}
X.-G.~Wen, \emph{{Emergent anomalous higher symmetries from topological order
  and from dynamical electromagnetic field in condensed matter systems}},
  \href{https://doi.org/10.1103/PhysRevB.99.205139}{\emph{Phys. Rev. B}
  {\bfseries 99} (2019) 205139}
  [\href{https://arxiv.org/abs/1812.02517}{{\ttfamily 1812.02517}}].

\bibitem{Delacretaz:2019brr}
L.V.~Delacr{\'e}taz, D.M.~Hofman and G.~Mathys, \emph{{Superfluids as
  Higher-form Anomalies}},
  \href{https://doi.org/10.21468/SciPostPhys.8.3.047}{\emph{SciPost Phys.}
  {\bfseries 8} (2020) 047} [\href{https://arxiv.org/abs/1908.06977}{{\ttfamily
  1908.06977}}].

\bibitem{Grozdanov:2016tdf}
S.~Grozdanov, D.M.~Hofman and N.~Iqbal, \emph{{Generalized global symmetries
  and dissipative magnetohydrodynamics}},
  \href{https://doi.org/10.1103/PhysRevD.95.096003}{\emph{Phys. Rev. D}
  {\bfseries 95} (2017) 096003}
  [\href{https://arxiv.org/abs/1610.07392}{{\ttfamily 1610.07392}}].

\bibitem{Grozdanov:2018ewh}
S.~Grozdanov and N.~Poovuttikul, \emph{{Generalized global symmetries in states
  with dynamical defects: The case of the transverse sound in field theory and
  holography}}, \href{https://doi.org/10.1103/PhysRevD.97.106005}{\emph{Phys.
  Rev. D} {\bfseries 97} (2018) 106005}
  [\href{https://arxiv.org/abs/1801.03199}{{\ttfamily 1801.03199}}].

\bibitem{Armas:2018ibg}
J.~Armas, J.~Gath, A.~Jain and A.V.~Pedersen, \emph{{Dissipative hydrodynamics
  with higher-form symmetry}},
  \href{https://doi.org/10.1007/JHEP05(2018)192}{\emph{JHEP} {\bfseries 05}
  (2018) 192} [\href{https://arxiv.org/abs/1803.00991}{{\ttfamily
  1803.00991}}].

\bibitem{Armas:2018atq}
J.~Armas and A.~Jain, \emph{{Magnetohydrodynamics as superfluidity}},
  \href{https://doi.org/10.1103/PhysRevLett.122.141603}{\emph{Phys. Rev. Lett.}
  {\bfseries 122} (2019) 141603}
  [\href{https://arxiv.org/abs/1808.01939}{{\ttfamily 1808.01939}}].

\bibitem{Glorioso:2018kcp}
P.~Glorioso and D.T.~Son, \emph{{Effective field theory of magnetohydrodynamics
  from generalized global symmetries}},
  \href{https://arxiv.org/abs/1811.04879}{{\ttfamily 1811.04879}}.

\bibitem{Armas:2018zbe}
J.~Armas and A.~Jain, \emph{{One-form superfluids {\&} magnetohydrodynamics}},
  \href{https://doi.org/10.1007/JHEP01(2020)041}{\emph{JHEP} {\bfseries 01}
  (2020) 041} [\href{https://arxiv.org/abs/1811.04913}{{\ttfamily
  1811.04913}}].

\bibitem{Armas:2019sbe}
J.~Armas and A.~Jain, \emph{{Viscoelastic hydrodynamics and holography}},
  \href{https://doi.org/10.1007/JHEP01(2020)126}{\emph{JHEP} {\bfseries 01}
  (2020) 126} [\href{https://arxiv.org/abs/1908.01175}{{\ttfamily
  1908.01175}}].

\bibitem{Maldacena:1997re}
J.M.~Maldacena, \emph{{The Large $N$ limit of superconformal field theories and
  supergravity}}, \href{https://doi.org/10.4310/ATMP.1998.v2.n2.a1}{\emph{Adv.
  Theor. Math. Phys.} {\bfseries 2} (1998) 231}
  [\href{https://arxiv.org/abs/hep-th/9711200}{{\ttfamily hep-th/9711200}}].

\bibitem{Witten:1998qj}
E.~Witten, \emph{{Anti de Sitter space and holography}},
  \href{https://doi.org/10.4310/ATMP.1998.v2.n2.a2}{\emph{Adv. Theor. Math.
  Phys.} {\bfseries 2} (1998) 253}
  [\href{https://arxiv.org/abs/hep-th/9802150}{{\ttfamily hep-th/9802150}}].

\bibitem{Gubser:1998bc}
S.S.~Gubser, I.R.~Klebanov and A.M.~Polyakov, \emph{{Gauge theory correlators
  from noncritical string theory}},
  \href{https://doi.org/10.1016/S0370-2693(98)00377-3}{\emph{Phys. Lett. B}
  {\bfseries 428} (1998) 105}
  [\href{https://arxiv.org/abs/hep-th/9802109}{{\ttfamily hep-th/9802109}}].

\bibitem{Grozdanov:2017kyl}
S.~Grozdanov and N.~Poovuttikul, \emph{{Generalised global symmetries in
  holography: magnetohydrodynamic waves in a strongly interacting plasma}},
  \href{https://doi.org/10.1007/JHEP04(2019)141}{\emph{JHEP} {\bfseries 04}
  (2019) 141} [\href{https://arxiv.org/abs/1707.04182}{{\ttfamily
  1707.04182}}].

\bibitem{Hofman:2017vwr}
D.M.~Hofman and N.~Iqbal, \emph{{Generalized global symmetries and
  holography}},
  \href{https://doi.org/10.21468/SciPostPhys.4.1.005}{\emph{SciPost Phys.}
  {\bfseries 4} (2018) 005} [\href{https://arxiv.org/abs/1707.08577}{{\ttfamily
  1707.08577}}].

\bibitem{Bhattacharyya:2007vjd}
S.~Bhattacharyya, V.E.~Hubeny, S.~Minwalla and M.~Rangamani, \emph{{Nonlinear
  Fluid Dynamics from Gravity}},
  \href{https://doi.org/10.1088/1126-6708/2008/02/045}{\emph{JHEP} {\bfseries
  02} (2008) 045} [\href{https://arxiv.org/abs/0712.2456}{{\ttfamily
  0712.2456}}].

\bibitem{Rangamani:2009xk}
M.~Rangamani, \emph{{Gravity and Hydrodynamics: Lectures on the fluid-gravity
  correspondence}},
  \href{https://doi.org/10.1088/0264-9381/26/22/224003}{\emph{Class. Quant.
  Grav.} {\bfseries 26} (2009) 224003}
  [\href{https://arxiv.org/abs/0905.4352}{{\ttfamily 0905.4352}}].

\bibitem{Hubeny:2011hd}
V.E.~Hubeny, S.~Minwalla and M.~Rangamani, \emph{{The fluid/gravity
  correspondence}},  in \emph{{Theoretical Advanced Study Institute in
  Elementary Particle Physics}: {String theory and its Applications: From meV
  to the Planck Scale}}, pp.~348--383, 2012
  [\href{https://arxiv.org/abs/1107.5780}{{\ttfamily 1107.5780}}].

\bibitem{Davison:2025sze}
R.A.~Davison and A.O.~Pinheiro, \emph{{Viscoelastic hydrodynamics of charged
  black holes}},  \href{https://arxiv.org/abs/2512.10905}{{\ttfamily
  2512.10905}}.

\bibitem{Hofman:2018lfz}
D.M.~Hofman and N.~Iqbal, \emph{{Goldstone modes and photonization for higher
  form symmetries}},
  \href{https://doi.org/10.21468/SciPostPhys.6.1.006}{\emph{SciPost Phys.}
  {\bfseries 6} (2019) 006} [\href{https://arxiv.org/abs/1802.09512}{{\ttfamily
  1802.09512}}].

\bibitem{Lake:2018dqm}
E.~Lake, \emph{{Higher-form symmetries and spontaneous symmetry breaking}},
  \href{https://arxiv.org/abs/1802.07747}{{\ttfamily 1802.07747}}.

\bibitem{Armas:2023tyx}
J.~Armas and A.~Jain, \emph{{Approximate higher-form symmetries, topological
  defects, and dynamical phase transitions}},
  \href{https://doi.org/10.1103/PhysRevD.109.045019}{\emph{Phys. Rev. D}
  {\bfseries 109} (2024) 045019}
  [\href{https://arxiv.org/abs/2301.09628}{{\ttfamily 2301.09628}}].

\bibitem{Pinheiro:2025fqg}
A.O.~Pinheiro, \emph{{Higher-form (Quasi)Hydrodynamics from Holography:
  Deformations and Dualities}},
  \href{https://arxiv.org/abs/2509.26583}{{\ttfamily 2509.26583}}.

\bibitem{Hartnoll:2009sz}
S.A.~Hartnoll, \emph{{Lectures on holographic methods for condensed matter
  physics}}, \href{https://doi.org/10.1088/0264-9381/26/22/224002}{\emph{Class.
  Quant. Grav.} {\bfseries 26} (2009) 224002}
  [\href{https://arxiv.org/abs/0903.3246}{{\ttfamily 0903.3246}}].

\bibitem{Herzog:2009xv}
C.P.~Herzog, \emph{{Lectures on Holographic Superfluidity and
  Superconductivity}},
  \href{https://doi.org/10.1088/1751-8113/42/34/343001}{\emph{J. Phys. A}
  {\bfseries 42} (2009) 343001}
  [\href{https://arxiv.org/abs/0904.1975}{{\ttfamily 0904.1975}}].

\bibitem{McGreevy:2009xe}
J.~McGreevy, \emph{{Holographic duality with a view toward many-body physics}},
  \href{https://doi.org/10.1155/2010/723105}{\emph{Adv. High Energy Phys.}
  {\bfseries 2010} (2010) 723105}
  [\href{https://arxiv.org/abs/0909.0518}{{\ttfamily 0909.0518}}].

\bibitem{Denef:2009tp}
F.~Denef and S.A.~Hartnoll, \emph{{Landscape of superconducting membranes}},
  \href{https://doi.org/10.1103/PhysRevD.79.126008}{\emph{Phys. Rev. D}
  {\bfseries 79} (2009) 126008}
  [\href{https://arxiv.org/abs/0901.1160}{{\ttfamily 0901.1160}}].

\bibitem{Gubser:2009qm}
S.S.~Gubser, C.P.~Herzog, S.S.~Pufu and T.~Tesileanu, \emph{{Superconductors
  from Superstrings}},
  \href{https://doi.org/10.1103/PhysRevLett.103.141601}{\emph{Phys. Rev. Lett.}
  {\bfseries 103} (2009) 141601}
  [\href{https://arxiv.org/abs/0907.3510}{{\ttfamily 0907.3510}}].

\bibitem{Gauntlett:2009dn}
J.P.~Gauntlett, J.~Sonner and T.~Wiseman, \emph{{Holographic superconductivity
  in M-Theory}},
  \href{https://doi.org/10.1103/PhysRevLett.103.151601}{\emph{Phys. Rev. Lett.}
  {\bfseries 103} (2009) 151601}
  [\href{https://arxiv.org/abs/0907.3796}{{\ttfamily 0907.3796}}].

\bibitem{Gauntlett:2009bh}
J.P.~Gauntlett, J.~Sonner and T.~Wiseman, \emph{{Quantum Criticality and
  Holographic Superconductors in M-theory}},
  \href{https://doi.org/10.1007/JHEP02(2010)060}{\emph{JHEP} {\bfseries 02}
  (2010) 060} [\href{https://arxiv.org/abs/0912.0512}{{\ttfamily 0912.0512}}].

\bibitem{Witten:2001ua}
E.~Witten, \emph{{Multitrace operators, boundary conditions, and AdS / CFT
  correspondence}},  \href{https://arxiv.org/abs/hep-th/0112258}{{\ttfamily
  hep-th/0112258}}.

\bibitem{lYi:1998trg}
W.S.~l'Yi, \emph{{Correlators of currents corresponding to the massive p-form
  fields in AdS/CFT correspondence}},
  \href{https://doi.org/10.1016/S0370-2693(99)00009-X}{\emph{Phys. Lett. B}
  {\bfseries 448} (1999) 218}
  [\href{https://arxiv.org/abs/hep-th/9811097}{{\ttfamily hep-th/9811097}}].

\bibitem{Skenderis:2002wp}
K.~Skenderis, \emph{{Lecture notes on holographic renormalization}},
  \href{https://doi.org/10.1088/0264-9381/19/22/306}{\emph{Class. Quant. Grav.}
  {\bfseries 19} (2002) 5849}
  [\href{https://arxiv.org/abs/hep-th/0209067}{{\ttfamily hep-th/0209067}}].

\bibitem{Papadimitriou:2007sj}
I.~Papadimitriou, \emph{{Multi-Trace Deformations in AdS/CFT: Exploring the
  Vacuum Structure of the Deformed CFT}},
  \href{https://doi.org/10.1088/1126-6708/2007/05/075}{\emph{JHEP} {\bfseries
  05} (2007) 075} [\href{https://arxiv.org/abs/hep-th/0703152}{{\ttfamily
  hep-th/0703152}}].

\bibitem{Bilal:1999ph}
A.~Bilal and C.-S.~Chu, \emph{{A Note on the chiral anomaly in the AdS / CFT
  correspondence and 1 / N**2 correction}},
  \href{https://doi.org/10.1016/S0550-3213(99)00553-2}{\emph{Nucl. Phys. B}
  {\bfseries 562} (1999) 181}
  [\href{https://arxiv.org/abs/hep-th/9907106}{{\ttfamily hep-th/9907106}}].

\bibitem{Gynther:2010ed}
A.~Gynther, K.~Landsteiner, F.~Pena-Benitez and A.~Rebhan, \emph{{Holographic
  Anomalous Conductivities and the Chiral Magnetic Effect}},
  \href{https://doi.org/10.1007/JHEP02(2011)110}{\emph{JHEP} {\bfseries 02}
  (2011) 110} [\href{https://arxiv.org/abs/1005.2587}{{\ttfamily 1005.2587}}].

\bibitem{Amado:2011zx}
I.~Amado, K.~Landsteiner and F.~Pena-Benitez, \emph{{Anomalous transport
  coefficients from Kubo formulas in Holography}},
  \href{https://doi.org/10.1007/JHEP05(2011)081}{\emph{JHEP} {\bfseries 05}
  (2011) 081} [\href{https://arxiv.org/abs/1102.4577}{{\ttfamily 1102.4577}}].

\bibitem{Landsteiner:2011iq}
K.~Landsteiner, E.~Megias, L.~Melgar and F.~Pena-Benitez, \emph{{Holographic
  Gravitational Anomaly and Chiral Vortical Effect}},
  \href{https://doi.org/10.1007/JHEP09(2011)121}{\emph{JHEP} {\bfseries 09}
  (2011) 121} [\href{https://arxiv.org/abs/1107.0368}{{\ttfamily 1107.0368}}].

\bibitem{Landsteiner:2012kd}
K.~Landsteiner, E.~Megias and F.~Pena-Benitez, \emph{{Anomalous Transport from
  Kubo Formulae}},
  \href{https://doi.org/10.1007/978-3-642-37305-3_17}{\emph{Lect. Notes Phys.}
  {\bfseries 871} (2013) 433}
  [\href{https://arxiv.org/abs/1207.5808}{{\ttfamily 1207.5808}}].

\bibitem{Rivelles:2003ge}
V.O.~Rivelles, \emph{{Quantization in AdS and the AdS / CFT correspondence}},
  \href{https://doi.org/10.1142/S0217751X03015544}{\emph{Int. J. Mod. Phys. A}
  {\bfseries 18} (2003) 2099}
  [\href{https://arxiv.org/abs/hep-th/0301025}{{\ttfamily hep-th/0301025}}].

\bibitem{Berkooz:2002ug}
M.~Berkooz, A.~Sever and A.~Shomer, \emph{{'Double trace' deformations,
  boundary conditions and space-time singularities}},
  \href{https://doi.org/10.1088/1126-6708/2002/05/034}{\emph{JHEP} {\bfseries
  05} (2002) 034} [\href{https://arxiv.org/abs/hep-th/0112264}{{\ttfamily
  hep-th/0112264}}].

\bibitem{Minces:2001zy}
P.~Minces and V.O.~Rivelles, \emph{{Energy and the AdS / CFT correspondence}},
  \href{https://doi.org/10.1088/1126-6708/2001/12/010}{\emph{JHEP} {\bfseries
  12} (2001) 010} [\href{https://arxiv.org/abs/hep-th/0110189}{{\ttfamily
  hep-th/0110189}}].

\bibitem{Mueck:2002gm}
W.~Mueck, \emph{{An Improved correspondence formula for AdS / CFT with
  multitrace operators}},
  \href{https://doi.org/10.1016/S0370-2693(02)01487-9}{\emph{Phys. Lett. B}
  {\bfseries 531} (2002) 301}
  [\href{https://arxiv.org/abs/hep-th/0201100}{{\ttfamily hep-th/0201100}}].

\bibitem{Minces:2002wp}
P.~Minces, \emph{{Multitrace operators and the generalized AdS / CFT
  prescription}}, \href{https://doi.org/10.1103/PhysRevD.68.024027}{\emph{Phys.
  Rev. D} {\bfseries 68} (2003) 024027}
  [\href{https://arxiv.org/abs/hep-th/0201172}{{\ttfamily hep-th/0201172}}].

\bibitem{Sever:2002fk}
A.~Sever and A.~Shomer, \emph{{A Note on multitrace deformations and AdS/CFT}},
  \href{https://doi.org/10.1088/1126-6708/2002/07/027}{\emph{JHEP} {\bfseries
  07} (2002) 027} [\href{https://arxiv.org/abs/hep-th/0203168}{{\ttfamily
  hep-th/0203168}}].

\bibitem{Aharony:2005sh}
O.~Aharony, M.~Berkooz and B.~Katz, \emph{{Non-local effects of multi-trace
  deformations in the AdS/CFT correspondence}},
  \href{https://doi.org/10.1088/1126-6708/2005/10/097}{\emph{JHEP} {\bfseries
  10} (2005) 097} [\href{https://arxiv.org/abs/hep-th/0504177}{{\ttfamily
  hep-th/0504177}}].

\bibitem{Elitzur:2005kz}
S.~Elitzur, A.~Giveon, M.~Porrati and E.~Rabinovici, \emph{{Multitrace
  deformations of vector and adjoint theories and their holographic duals}},
  \href{https://doi.org/10.1088/1126-6708/2006/02/006}{\emph{JHEP} {\bfseries
  02} (2006) 006} [\href{https://arxiv.org/abs/hep-th/0511061}{{\ttfamily
  hep-th/0511061}}].

\bibitem{Gubser:2002vv}
S.S.~Gubser and I.R.~Klebanov, \emph{{A Universal result on central charges in
  the presence of double trace deformations}},
  \href{https://doi.org/10.1016/S0550-3213(03)00056-7}{\emph{Nucl. Phys. B}
  {\bfseries 656} (2003) 23}
  [\href{https://arxiv.org/abs/hep-th/0212138}{{\ttfamily hep-th/0212138}}].

\bibitem{Hartman:2006dy}
T.~Hartman and L.~Rastelli, \emph{{Double-trace deformations, mixed boundary
  conditions and functional determinants in AdS/CFT}},
  \href{https://doi.org/10.1088/1126-6708/2008/01/019}{\emph{JHEP} {\bfseries
  01} (2008) 019} [\href{https://arxiv.org/abs/hep-th/0602106}{{\ttfamily
  hep-th/0602106}}].

\bibitem{Diaz:2007an}
D.E.~Diaz and H.~Dorn, \emph{{Partition functions and double-trace deformations
  in AdS/CFT}},
  \href{https://doi.org/10.1088/1126-6708/2007/05/046}{\emph{JHEP} {\bfseries
  05} (2007) 046} [\href{https://arxiv.org/abs/hep-th/0702163}{{\ttfamily
  hep-th/0702163}}].

\bibitem{Klebanov:1999tb}
I.R.~Klebanov and E.~Witten, \emph{{AdS / CFT correspondence and symmetry
  breaking}}, \href{https://doi.org/10.1016/S0550-3213(99)00387-9}{\emph{Nucl.
  Phys. B} {\bfseries 556} (1999) 89}
  [\href{https://arxiv.org/abs/hep-th/9905104}{{\ttfamily hep-th/9905104}}].

\bibitem{Faulkner:2012gt}
T.~Faulkner and N.~Iqbal, \emph{{Friedel oscillations and horizon charge in 1D
  holographic liquids}},
  \href{https://doi.org/10.1007/JHEP07(2013)060}{\emph{JHEP} {\bfseries 07}
  (2013) 060} [\href{https://arxiv.org/abs/1207.4208}{{\ttfamily 1207.4208}}].

\bibitem{Englert:1964et}
F.~Englert and R.~Brout, \emph{{Broken Symmetry and the Mass of Gauge Vector
  Mesons}}, \href{https://doi.org/10.1103/PhysRevLett.13.321}{\emph{Phys. Rev.
  Lett.} {\bfseries 13} (1964) 321}.

\bibitem{Higgs:1964pj}
P.W.~Higgs, \emph{{Broken Symmetries and the Masses of Gauge Bosons}},
  \href{https://doi.org/10.1103/PhysRevLett.13.508}{\emph{Phys. Rev. Lett.}
  {\bfseries 13} (1964) 508}.

\bibitem{Guralnik:1964eu}
G.S.~Guralnik, C.R.~Hagen and T.W.B.~Kibble, \emph{{Global Conservation Laws
  and Massless Particles}},
  \href{https://doi.org/10.1103/PhysRevLett.13.585}{\emph{Phys. Rev. Lett.}
  {\bfseries 13} (1964) 585}.

\bibitem{Stueckelberg:1938hvi}
E.C.G.~Stueckelberg, \emph{{Interaction energy in electrodynamics and in the
  field theory of nuclear forces}},
  \href{https://doi.org/10.5169/seals-110852}{\emph{Helv. Phys. Acta}
  {\bfseries 11} (1938) 225}.

\bibitem{Ruegg:2003ps}
H.~Ruegg and M.~Ruiz-Altaba, \emph{{The Stueckelberg field}},
  \href{https://doi.org/10.1142/S0217751X04019755}{\emph{Int. J. Mod. Phys. A}
  {\bfseries 19} (2004) 3265}
  [\href{https://arxiv.org/abs/hep-th/0304245}{{\ttfamily hep-th/0304245}}].

\bibitem{Klebanov:2002gr}
I.R.~Klebanov, P.~Ouyang and E.~Witten, \emph{{A Gravity dual of the chiral
  anomaly}}, \href{https://doi.org/10.1103/PhysRevD.65.105007}{\emph{Phys. Rev.
  D} {\bfseries 65} (2002) 105007}
  [\href{https://arxiv.org/abs/hep-th/0202056}{{\ttfamily hep-th/0202056}}].

\bibitem{Jimenez-Alba:2014iia}
A.~Jimenez-Alba, K.~Landsteiner and L.~Melgar, \emph{{Anomalous magnetoresponse
  and the St{\"u}ckelberg axion in holography}},
  \href{https://doi.org/10.1103/PhysRevD.90.126004}{\emph{Phys. Rev. D}
  {\bfseries 90} (2014) 126004}
  [\href{https://arxiv.org/abs/1407.8162}{{\ttfamily 1407.8162}}].

\bibitem{Leigh:2003ez}
R.G.~Leigh and A.C.~Petkou, \emph{{SL(2,Z) action on three-dimensional CFTs and
  holography}},
  \href{https://doi.org/10.1088/1126-6708/2003/12/020}{\emph{JHEP} {\bfseries
  12} (2003) 020} [\href{https://arxiv.org/abs/hep-th/0309177}{{\ttfamily
  hep-th/0309177}}].

\bibitem{Karch:2025hof}
A.~Karch and M.~Youssef, \emph{{Dissipation in Open Holography}},
  \href{https://arxiv.org/abs/2509.14312}{{\ttfamily 2509.14312}}.

\bibitem{Grozdanov:2025ulc}
S.~Grozdanov and M.~Vrbica, \emph{{Thermal field theory correlators in the
  large-$N$ limit and the spectral duality relation}},
  \href{https://arxiv.org/abs/2509.18074}{{\ttfamily 2509.18074}}.

\bibitem{Aurilia:1980jz}
A.~Aurilia, Y.~Takahashi and P.K.~Townsend, \emph{{The U(1) Problem and the
  Higgs Mechanism in Two-dimensions and Four-dimensions}},
  \href{https://doi.org/10.1016/0370-2693(80)90484-0}{\emph{Phys. Lett. B}
  {\bfseries 95} (1980) 265}.

\bibitem{Rajeev:1986id}
S.G.~Rajeev, ``{Duality and gauge invariance}.'' MIT-CTP-1335, 1986.

\bibitem{Allen:1990gb}
T.J.~Allen, M.J.~Bowick and A.~Lahiri, \emph{{Topological mass generation in
  (3+1)-dimensions}},
  \href{https://doi.org/10.1142/S0217732391000580}{\emph{Mod. Phys. Lett. A}
  {\bfseries 6} (1991) 559}.

\bibitem{Bizdadea:1996np}
C.~Bizdadea and S.O.~Saliu, \emph{{The BRST quantization of massive Abelian two
  form gauge fields}},
  \href{https://doi.org/10.1016/0370-2693(96)89545-1}{\emph{Phys. Lett. B}
  {\bfseries 368} (1996) 202}.

\bibitem{Borsten:2024gox}
L.~Borsten, M.~Jalali~Farahani, B.~Jur{\v{c}}o, H.~Kim,
  J.~N{\'a}ro{\v{z}}n{\'y}, D.~Rist et~al., \emph{{Higher Gauge Theory}},
  \href{https://arxiv.org/abs/2401.05275}{{\ttfamily 2401.05275}}.

\bibitem{Kawai:1980qq}
H.~Kawai, \emph{{A Dual Transformation of the Nielsen-olesen Model}},
  \href{https://doi.org/10.1143/PTP.65.351}{\emph{Prog. Theor. Phys.}
  {\bfseries 65} (1981) 351}.

\bibitem{Quevedo:1996uu}
F.~Quevedo and C.A.~Trugenberger, \emph{{Phases of antisymmetric tensor field
  theories}}, \href{https://doi.org/10.1016/S0550-3213(97)00337-4}{\emph{Nucl.
  Phys. B} {\bfseries 501} (1997) 143}
  [\href{https://arxiv.org/abs/hep-th/9604196}{{\ttfamily hep-th/9604196}}].

\bibitem{Quevedo:1997jb}
F.~Quevedo, \emph{{Duality and global symmetries}},
  \href{https://doi.org/10.1016/S0920-5632(97)00517-3}{\emph{Nucl. Phys. B
  Proc. Suppl.} {\bfseries 61} (1998) 23}
  [\href{https://arxiv.org/abs/hep-th/9706210}{{\ttfamily hep-th/9706210}}].

\bibitem{Hell:2021wzm}
A.~Hell, \emph{{On the duality of massive Kalb-Ramond and Proca fields}},
  \href{https://doi.org/10.1088/1475-7516/2022/01/056}{\emph{JCAP} {\bfseries
  01} (2022) 056} [\href{https://arxiv.org/abs/2109.05030}{{\ttfamily
  2109.05030}}].

\bibitem{Burgess:2025geh}
C.P.~Burgess and F.~Quevedo, \emph{{Dual is Different: EFTs, Axions and
  Nonpropagating Form Fields in Cosmology}},
  \href{https://arxiv.org/abs/2509.11340}{{\ttfamily 2509.11340}}.

\bibitem{Witten:2003ya}
E.~Witten, \emph{{SL(2,Z) action on three-dimensional conformal field theories
  with Abelian symmetry}},  in \emph{{From Fields to Strings: Circumnavigating
  Theoretical Physics: A Conference in Tribute to Ian Kogan}}, pp.~1173--1200,
  7, 2003 [\href{https://arxiv.org/abs/hep-th/0307041}{{\ttfamily
  hep-th/0307041}}].

\bibitem{Yee:2004ju}
H.-U.~Yee, \emph{{A Note on AdS / CFT dual of SL(2,Z) action on 3-D conformal
  field theories with U(1) symmetry}},
  \href{https://doi.org/10.1016/j.physletb.2004.05.082}{\emph{Phys. Lett. B}
  {\bfseries 598} (2004) 139}
  [\href{https://arxiv.org/abs/hep-th/0402115}{{\ttfamily hep-th/0402115}}].

\bibitem{deHaro:2007eg}
S.~de~Haro and P.~Gao, \emph{{Electric-magnetic duality and deformations of
  three-dimensional CFT's}},
  \href{https://doi.org/10.1103/PhysRevD.76.106008}{\emph{Phys. Rev. D}
  {\bfseries 76} (2007) 106008}
  [\href{https://arxiv.org/abs/hep-th/0701144}{{\ttfamily hep-th/0701144}}].

\bibitem{deHaro:2007fg}
S.~de~Haro and A.C.~Petkou, \emph{{Holographic Aspects of Electric-Magnetic
  Dualities}}, \href{https://doi.org/10.1088/1742-6596/110/10/102003}{\emph{J.
  Phys. Conf. Ser.} {\bfseries 110} (2008) 102003}
  [\href{https://arxiv.org/abs/0710.0965}{{\ttfamily 0710.0965}}].

\bibitem{Herzog:2007ij}
C.P.~Herzog, P.~Kovtun, S.~Sachdev and D.T.~Son, \emph{{Quantum critical
  transport, duality, and M-theory}},
  \href{https://doi.org/10.1103/PhysRevD.75.085020}{\emph{Phys. Rev. D}
  {\bfseries 75} (2007) 085020}
  [\href{https://arxiv.org/abs/hep-th/0701036}{{\ttfamily hep-th/0701036}}].

\bibitem{Hartnoll:2007ip}
S.A.~Hartnoll and C.P.~Herzog, \emph{{Ohm's Law at strong coupling: S duality
  and the cyclotron resonance}},
  \href{https://doi.org/10.1103/PhysRevD.76.106012}{\emph{Phys. Rev. D}
  {\bfseries 76} (2007) 106012}
  [\href{https://arxiv.org/abs/0706.3228}{{\ttfamily 0706.3228}}].

\bibitem{Frankel_2011}
T.~Frankel, \emph{The Geometry of Physics: An Introduction}, Cambridge
  University Press, 3~ed. (2011).

\bibitem{DeWolfe:2020uzb}
O.~DeWolfe and K.~Higginbotham, \emph{{Generalized symmetries and 2-groups via
  electromagnetic duality in $AdS/CFT$}},
  \href{https://doi.org/10.1103/PhysRevD.103.026011}{\emph{Phys. Rev. D}
  {\bfseries 103} (2021) 026011}
  [\href{https://arxiv.org/abs/2010.06594}{{\ttfamily 2010.06594}}].

\bibitem{Iqbal:2020lrt}
N.~Iqbal and N.~Poovuttikul, \emph{{2-group global symmetries, hydrodynamics
  and holography}},
  \href{https://doi.org/10.21468/SciPostPhys.15.2.063}{\emph{SciPost Phys.}
  {\bfseries 15} (2023) 063}
  [\href{https://arxiv.org/abs/2010.00320}{{\ttfamily 2010.00320}}].

\bibitem{Das:2022auy}
A.~Das, R.~Gregory and N.~Iqbal, \emph{{Higher-form symmetries, anomalous
  magnetohydrodynamics, and holography}},
  \href{https://doi.org/10.21468/SciPostPhys.14.6.163}{\emph{SciPost Phys.}
  {\bfseries 14} (2023) 163}
  [\href{https://arxiv.org/abs/2205.03619}{{\ttfamily 2205.03619}}].

\end{thebibliography}\endgroup

\end{document}